---------------------------------------------------------------

\documentstyle[12pt]{article}
\advance \textheight \topskip

\begin{document}
\begin{flushright}
Preprint DFPD 95/TH/56
\end{flushright}
\centerline{\Large {\bf The Fermion Generations Problem In The GUST} }
\centerline{\Large {\bf In The Free World-Sheet Fermion Formulation} }

\vspace{24pt}
\begin{center}
{\Large {\bf A.A. Maslikov $^{b}$ , I.A. Naumov $^{b}$,
  G.G. Volkov $^{a,b}$ }}\\

\bigskip
$^a$ INFN Sezione di Padova and Dipartimento di Fisica \\
Universit\`a di Padova, \\
Via Marzolo 8, 35100 Padua, Italy \\

\bigskip
$^b$ Institute for  High Energy Physics\\
142284 Protvino, Moscow Region, Russia\\
\end{center}

\bigskip
\centerline{ABSTRACT}
In the framework of the four dimensional heterotic superstring with free
fermions
we present a revised version of the rank eight
Grand Unified String Theories (GUST) which contain
the $SU(3)_H$-gauge family symmetry. We also develop some methods for
building of corresponding string models.
We explicitly construct GUST with gauge symmetry
$ G = SU(5)
\times U(1)\times (SU(3) \times U(1))_H$ and
 $G = SO(10)\times (SU(3) \times U(1))_H $
or $E(6)\times SU(3)_H$ $\subset E(8)$
and consider the full massless spectrum for our string models.
 We consider for the observable gauge
symmetry the diagonal subgroup  $G^{symm}$ of the rank 16 group $G \times G$
$\subset SO(16) \times SO(16)$ or $\subset E(8) \times E(8)$.
We discuss the possible
fermion matter and Higgs sectors in these theories.
We study renormalizable and nonrenormolizable contributions to the
superpotential.
There has to exist "superweak" light chiral matter ($m_H^f
< M_W$) in GUST under consideration.
The understanding of quark and lepton mass spectra and family mixing leaves
a possibility for the existence of an unusually low mass breaking scale of
the $SU(3)_H$ family gauge symmetry (some TeV).

\newpage

\tableofcontents

\newpage

\section{Introduction}
There are no experimental indications which would impel one to go
beyond the framework of the $SU(3)_C\times SU(2)_L\times U(1)_Y$ Standard
Model
$(SM)$ with three generations of quarks and leptons. None of the up-to-date
 experiments contradicts, within the limits of accuracy, the validity of
the SM predictions for low energy phenomena.
However SM includes a large number of arbitrary parameters and
many fundamental questions remain unanswered in its framework.
The number of generations,
the fermion mass origin,
generation mixing, neutrino mass uniqueness, CP-violation problems are
among most exciting theoretical puzzles in SM.
All these questions stimulate the searching for more fundamental theories
that have the SM as an effective low energy limit and predict some new
particles and phenomena.

 One of the interesting steps towards possible explanation of generation
problem and others  phenomena is the including a non-abelian horizontal
factor like $SU(3)$ into gauge group of the model. The benefit of this
approach could be a clarifying of such a question as generation number,
family mixing nature, quark-lepton mass hierarchy.
It is interesting to  note that there are some horizontal extensions that
do not add too many extra arbitrary parameters. These models allow to
analyse the rare processes experimental data by the universal way and to
obtain experimentally testable predictions.
The constraints of horizontal model parameters followed from this approach
permits the existence of the interesting flavour-changing physics in the
TeV region. Also these models gives rise to a rather natural way of
the superweak-like CP-violation.

The superstring theories are appear to be the most perspective candidates
for the Theory of Everything. That is why we consider this approach to be
useful in the investigations of GUTs that include horizontal gauge
symmetry. For the heterotic string the groups $E_8\otimes E_8$ and
$spin(32)/Z_2$ are characteristic. Hence it is interesting to consider GUT
based on its various rank 8 and 16 subgroup.
String theories possess infinite dimensional symmetries that place many
specific constraints on the theory spectrum. These symmetries origin from
2 dimensional conformal invariance, modular invariance, and Virasoro and
Kac-Moody algebras.

In this paper we analyse several Grand Unification models based on 4
dimensional heterotic string in free fermions approach.

We consider possible ways of breaking the "string" gauge group
$E_8\otimes E_8$ down to low energy supersymmetric model that includes
Standard Model group and horizontal factor $SU(3)$ and naturally
describes $3$ or $3+1$ generations.

The paper is organized as follows. In Chapter 2 we consider supersymmetric
extension of the Standard Model including horizontal $SU(3)$ factor.
This model could be the low energy limit of the string GUT considered
further in Chapter 5. The details of the model structure are described in
Appendix~A.
For such an effective model we present estimates of horizontal gauge
symmetry breaking scale and of contributions of horizontal interactions to
various rare process.

When we demand that the quark-lepton generations mixing and the splitting
of the horizontal gauge boson mass spectrum have the same nature this
model gain minimal number of extra new parameters like $g_{3H}$, $M_{H_0}$
and Yukawa coupling. This allows us to carry complete  analysis of the
 experimental data of all known rare processes.
The horizontal gauge symmetry breaking in these models is seemed to  be
considerably low ($\sim$some TeV).

For estimating the experimental data of the rare processes in framework of
these models  Appendix B gives an justification of validity of our
$g_{3H}$ choosing basing on the renormalization group analysis of string
GUTs.

In Chapter 3 we discuss different ways of breaking of string rank 16 gauge
group $E_8\otimes E_8$ and its symmetric diagonal subgroup of rank 8. We
outline the perspective way including the symmetric subgroup on the
intermediate stage that does not involve higher level of Kac-Moody
algebra representations.  Namely in Chapter 4 we analyse
 restrictions on various unitary representations in the  superstring GUTs
 from the point of the level of the Kac-Moody algebras (KMA) and
conformal weights.  It is emphasized that the breaking of gauge group down
to symmetric diagonal subgroup effectively increases KMA level to 2 and
makes the existence of the representations for Higgs fields needed for GUT
symmetry breaking possible.

Chapter 5 presents various string models including Grand Unification group
$SO(10)$ or $U(5)$ along with horizontal gauge symmetry $U(3)$ and
describing $3+1$ generations.
For all these models  it is essential that gauge group is breaking down to
the symmetric diagonal subgroup. The rules of building the basis of spin
boundary conditions and GSO projections based on modular invariance
are given in Appendix C. We use
a computer program for calculating the full massless spectrum of states of
the considered models. The full spectra for the presented models are given
in corresponding tables.

The further analysis of root and weight lattices and their relations with
the GSO projection is given in Chapter 6. Based on the method proposed
there we build 4D superstring GUT model with ${\left[E_6\otimes
SU(3)\right]}^2$ and $SU(3)^{\otimes 4}$ gauge groups that seems to be
perspective in the way to obtain 3 generation model. Using that technique
we  build  these models with $N=2$ supersymmetry.

In the Chapter 7 the Model~1 is considered in details. Corresponding gauge
symmetry breaking is analysed, superpotential and nonrenormalizable
contribution permitted by string dynamics are presented. The construction
of electromagnetic charge for diagonal symmetric subgroup allows to avoid
states with exotic fractional charges that is a typical problem in string
GUTs on the level 1. The form of obtained superpotential implies that 2
generations remain massless comparing with the $m_W$ scale.
Using the condition of  $SU(3)_H$ anomaly cancellation  the theory
predicts the existence of the Standard Model singlet
"neutrino-like" particles that participate only in horizontal
interactions.  As following from the form of the superpotential some of
them   could be light (less than $m_W$) that will be very interesting in
sense of experimental searches. At last we calculate for model1
tne N=4,5 vertex non-renormalizable contributions to the superpotential.

\newpage
\section{ Towards a low energy gauge family symmetry with the small
                number of parameters}
\subsection{The family problems in SM, family mixing
            and quark/lepton mass hierarchy}

One has ten parameters in the quark sector of the SM with three generations:
six quark masses, three mixing angles and
the Kobayashi-Maskawa $(KM)$ CP violation phase $(0<\delta^{KM}<\pi)$.
The CKM ( Cabibbo-Kobayashi-Maskawa )
matrix in Wolfenstein parameterization is determined by the four
parameters --- Cabibbo angle $\lambda \approx 0.22$, A, $\rho$ and $\eta$:

\begin{equation}
{\bf V_{CKM}}=
\left( \begin{array}{ccc}
V_{ud}& V_{us} & V_{ub}\\
V_{cd} & V_{cs}& V_{cb} \\
V_{td}  &V_{ts} & V_{tb} \\
\end{array} \right) =
\left( \begin{array}{ccc}
1- 1/2 {\lambda}^2 & \lambda & A {\lambda}^3 (\rho - i \eta)\\
- \lambda & 1 - 1/2{\lambda}^2 & A {\lambda}^2 \\
A {\lambda}^3 (1 - \rho - i \eta)  & -A {\lambda}^2  & 1 \\
\end{array} \right).
\end{equation}

In the complex plane the point $(\rho, \eta)$
is  a vertex of the  unitarity triangle and describes the CP violation
in SM. The unitarity triangle is constructed from
the following unitarity condition of $V_{CKM}$ :
$V_{ub}^{*} + V_{td} \approx A {\lambda}^3$.

Recently, the interest in the CP-violation problem was excited again due
to the
data on the search for the direct CP-violation effects in neutral K-mesons
\cite{1'},\cite{2'} :
\begin{equation}
Re \biggl (\frac{\varepsilon'}{\varepsilon}\biggr )
=(7.4\pm 6)\times10^{-4},\,\,\,\,\,\,
\end{equation}

\begin{equation}
Re \biggl ( \frac{\varepsilon'}{\varepsilon}\biggr )
=(23\pm 7)\times10^{-4},\,\,\,\,\,\,
\end{equation}

The major contribution to the CP-violation parameters $\varepsilon_K$ and
$\varepsilon_K'$ ($K^0$-decays), as well as to the $B_d^0-\bar B_d^0$
mixing parameter $x_d=\frac{\Delta m_{(B_d)}}{\Gamma_{(B_d)}}$  is due
to the large t-quark mass contribution. The same statement holds also
for some amplitudes of K- and B-meson rare decays.
 The CDF collaboration gives the following region for the top quark
mass: $m_t=174\pm 25$ GeV \cite{3'}. The complete fit which is based on
the low energy data as well as the latest LEP and SLC data and comparing
with the mass indicated by CDF measurements gives $m_t=162\pm9$ GeV
\cite{4'}.

The main undrawbacks of SM now are going from our non-understanding
the generation problem, their mixing and hierarchy
of quark and lepton mass spectra. For example, for quark
masses $\mu \approx 1 GeV$
we can get  the following approximate relations \cite {9'}:

\begin{eqnarray}
m_{i_k} \approx ( q_H^u )^{2k} m_0,
\,\,\,\,k = 0, 1, 2;
\,\,\, i_0= u, \, i_1 = c, \, i_2 = t,
\nonumber\\
m_{i_k} \approx ( q_H^d )^{2k} m_0,
\,\,\,\,k = 0, 1, 2;
  \,\,\, i_0= d, \, i_1 = s, \, i_2 = b,
\end{eqnarray}

where  $q_H^u \approx (q_H^d)^2$,
$q_H^d \approx 4-5 \approx 1/{\lambda}$ and $\lambda \approx
\sin {\theta}_C$.

Here we used the conventional ratios  of  the  "running"  quark
masses \cite{10'}

\begin{eqnarray}
m_d/m_s\ =0.051\pm 0.004 , \,\,\,  m_u/m_c\ =0.0038\pm 0.0012 , \,\,\,
m_s/m_b\ =0.033\pm 0.011 ,  \nonumber\\
m_c(\mu =1GeV)= (1.35\pm 0.05 ) GeV \,\,\,
and  \,\,\, m_t^{phys} \approx 0.6m_t(\mu = 1GeV) \label{5.1}.
\end{eqnarray}

This phenomenological formula (6) predicts the following value for the
$t$-quark mass:
\begin{equation}
{m_t}^{phys} \approx 180 - 200 GeV.
\end{equation}

In SM these mass matrices and mixing come from the
Yukawa sector :

\begin{eqnarray}
L_Y= Q  Y_u \bar u {h^*} + Q Y_d \bar d h + L Y_e {\bar {l_e}} h
\,\, + \,\, H.C.,
\end{eqnarray}
where $Q_i$ and $L_i$ are three quark and lepton isodoublets,
  $\bar u_i,\bar d_i$ and ${l_e}_i$ are three right-handed antiquark and
antilepton
isosinglets respectively. $h$ is the ordinary Higgs doublet.
In SM the $3\times 3$ family Yukawa matrices, $(Y_u)_{ij}$ and $(Y_d)_{ij}$
have no any particular symmetry.
 Therefore it is necessary to find some additional mechanisms or symmetries
beyond the SM which
could diminish the  number of the independent parameters in Yukawa
sector $L_Y$. These new structures can be
used for the determination of the mass hierarchy and family mixing.

To understand the generation mixing origin and fermion mass hierarchy
several models beyond the SM suggest special forms for the mass matrix of
"up" and "down" quarks (Fritzsch ansatz, "improved" Fritzsch ansatz,
"Democratic" ansatz, etc.\cite {11'}).
These mass matrices have less than ten independent parameters
or they could
have some matrix elements equal to zero ("texture zeroes")\cite {12'}.
This allows us
to determine the diagonalizing matrix $U_L$ and $D_L$ in terms of quark
masses:
\begin{equation}
Y_d^{diag} = D_L Y_d D_R^+,\,\\,\,\,\,\,\,
Y_u^{diag} = U_L Y_d U_R^+.
\end{equation}
For simplicity the symmetric form of Yukawa matrices  has been taken ,
therefore:  $D_L=D_R^*$, $U_L=U_R^*$.
 These ansatzes or zero "textures" could be checked experimentally in
predictions for the mixing angles of the CKM matrix: $V_{CKM} = U_L D_L^+$.
For example, one can consider the following approximate form
at the scale $M_X$ for the symmetric "texture" used in paper \cite{12'}:

\begin{equation}
{\bf Y_u}=
\left( \begin{array}{ccc}
0 & {\lambda}^6 & 0 \\
 {\lambda}^6  & 0 &  {\lambda}^2 \\
0 &  {\lambda}^2 & 1 \\
\end{array} \right)
\,\,\,\,\,\,
{\bf Y_d}=
\left( \begin{array}{ccc}
0 & 2{\lambda}^4 & 0 \\
 2{\lambda}^4  &  2{\lambda}^3 &  2{\lambda}^3 \\
0 &  2{\lambda}^3 & 1 \\
\end{array} \right).
\end{equation}

Given these conditions it is possible to evolve down to low energies via
the renormalization group equations all quantities  including the matrix
elements of Yukawa couplings $Y_{u,d}$, the values of the quark masses (see
(4))
and the CKM matrix elements (see (1)).
Also, using these relations we may compute $U_L$ (or $D_L$) in terms of CKM
matrix and/or of quark masses.

In GUT extensions of the SM with the family gauge  symmetry embedded
Yukawa matrices can acquire particular symmetry or an ansatz,
depending on the Higgs multiplets to which they couple.

The  family gauge symmetry could help us to study
in an independent way  the origin of the up- ($U$) and down-
($D$) quark mixing matrices  and consequently  the structure of the
CKM matrix $V_{CKM}=UD^+$.

Let us see the example of Abelian gauge U(1) horizontal
symmetry in papers \cite {hiermass}. In this approach  the mechanism of
mass and mixing hierarchy follows from nonrenormalizable terms in
Yukawa potential \cite {f, hiermass}.
\begin{eqnarray}
L_Y^{eff}= Y_{d_{ij}} Q_i \phi_d \frac{S^{m_{ij}}}{M^{m_{ij}}}\bar d_j
\, +\,  Y_{u_{ij}}Q_i\phi_u \frac{S^{n_{ij}}}{M^{n_{ij}}}\bar u_j \,+\,H.C.,
\end{eqnarray}
where the spontaneous breaking of the horizontal symmetry by VEV of a
scalar field $S$, $S \approx \Lambda_H$, which is a singlet of the SM.
The mass scale $M$ is connected with a mass of the new massive particles .
The powers $m_{ij}$
 ($n_{ij}$) are the horizontal charge difference between $Q_i$ and
$\bar d_j$  ($Q_i$ and $\bar u_j$):$H(Q_i)+H(\bar d_j)=m_{ij}$,
$H(Q_i)+H(\bar u_j)=n_{ij}$ \cite {f, hiermass}. The hierarchy in mixing
angles and quark (charged lepton) masses appears due to a small parameter
$\epsilon= \Lambda_H / M << 1$ \cite {f, hiermass}.

Due to  the local gauge family symmetry  with a  low energy breaking
scale $\Lambda_H \approx 1 TeV$  gives us a chance
to define the quantum numbers of
quarks and leptons and thus establishes a link between them in families.
For the  mass fermion ansatz considered above  in the extensions of
SM there could exist
the following types of the  $SU(3)\times SU(2_L)$ Higgs multiplets:(1,2),
(3,1), (8,1),(3,2),(8,2),(1,1),...., which in turn could exist
in the spectra of the String Models.

In the framework of the rank eight Grand Unified String
Theories we will consider an extension of SM due to local
family gauge symmetry,
 $G_H =SU(3)_H$, $SU(3)_H \times U(1)_H$ models and its developments and
their possible Higgs sector.
Thus,  for  understanding the quark mass spectra and
 the difference between  the origins of the up- ( or down) quark
and charged lepton  mass matrices in GUSTs we have to study  the
Higgs content of the model, which we must use from the one hand for
 breaking the GUT , Quark-Lepton -,
$G_H = SU(3)_H ,...$-, $SU(2)_L \times U(1)$-
symmetries and from the other hand --- for Yukawa matrix constructions.

\subsection{The "bootstrap"  $SU(3_H)$ gauge family models.}

In the paper \cite{9'} we investigated the samples of different
scenarios of $SU(3)_H$ breakings down to
the $SU(2)_H \times U(1)_{3H}$, $U(1)_{3H} \times U(1)_{8H}$ and
$U(1)_{8H}$-
subgroups, as well as the mechanism of the complete breaking of
the base group $SU(3)_H$. We tried to realize the
 SUSY conserving program (see Appendix A) on  the  scales where
the relevant gauge symmetry is broken. In
the framework of these  versions of the gauge symmetry breaking, we
were searching for the spectra of  horizontal gauge bosons and gauginos and
calculated the  amplitudes of some typical rare processes. Theoretical
estimates for the branching  ratios of some
rare processes obtained from these  calculations have been compared
with the experimental data on the corresponding values  \cite {28',29',30',9'}.
Further  we
have got  some bounds on the masses of $H_{\mu}$-bosons and the appropriate
$H$-gauginos. Of particular interest was  the case of the
$SU(3)_H$ -group which   breaks completely
on  the scale $M_{H_0}$. We calculated the
splitting of eight $H$-boson masses in a model dependent  fashion.
This    splitting, depending on the quark mass spectrum, allows us
to reduce considerably  the predictive ambiguity of the  model
-"almost exactly solvable model".

We assume at first
that all of the  8 gauge bosons of $SU(3)_H$ group acquire the same mass
$M_{H_0}$. Such a breaking is not difficult to get by, say, introducing
the Higgs fields transforming in accordance with the triplet
representation of the $SU(3)_H$  group. These fields are singlet
under the  Standard Model symmetries : $(z^i \in (3,1,1,0)$ and
${\bar z}^i \in (\bar 3,1,1,0) $ ,
$ {\langle }{\bar z}^{i\,\alpha}{\rangle }_0=\delta^{i\,\alpha}V$ ,
${\langle }z_i^{\alpha}{\rangle }_0=\delta_i^{\alpha}V\ ,\ $,$ i,\, \alpha\,
=1,2,3 $,
where $V=M_{H_0})$.  In this case
the fermion mass matrices are proportional to a unit matrix
because the fermions have a global $SU(3)_H$ symmetry.

The degeneracy of  the
masses of 8 gauge horizontal vector bosons is eliminated by including
the VEV's of the Higgs fields violating the  electroweak
symmetry and determining the hierarchical structure of
up- and  down- quarks (leptons) mass matrices.
Thus,  there is a set  of  the  Higgs  fields
(see corresponding Table \ref{tabl0}):
 $H(8,\ 2)\ ,h(8,\ 2)\ $ or $ \ Y (\bar 3 ,\ 2),
\ X(3,\ 2)\ $ and $ \  {\kappa}_{1,2}(1,\ 2)$  which could determine  the
mass matrix of up-and down-quarks. On the other hand,
in order to  calculate
the splitting between the masses of horizontal gauge  bosons,  one
has to take into account the VEV's contributions  of
the Higgs fields $H, h$ or $X, Y$. For example, the VEV's of the
Higgs fields  $H$ or $X$ can give the corresponding contributions:

\begin{eqnarray}
(\Delta M_u^2)^{ab}_{\underline 8} =  { g_H^2}\, \sum_{d=1}^{8}\,
f^{adc} f^{bdc'} {\langle }H^c{\rangle }{\langle }H^{c'}{\rangle }^*
,\label{5.6}\\
(\Delta M_u^2)^{ab}_{\underline 3} =  { g_H^2} \sum_{k=1}^{3}\,
 \frac{{\lambda}_{ik}^a}{2}\frac{{\lambda}_{kj}^b}{2}
{\langle }X^{i}{\rangle }{{\langle }X^{j}{\rangle }}^* ,\label{5.7}
\end{eqnarray}

Now we can come to constructing the    horizontal
gauge boson mass  matrix  ${M^{ab}}^2$ \\
(a,b=1,2,...,8):
\begin{eqnarray}
(M_H^2)^{ab}=M_{H_0}^2\delta^{ab} +({\Delta}M_d^2)^{ab} +({\Delta}M_u^2)^{ab}.
\label{5.8}
\end{eqnarray}
Here $({\Delta}M_d^2)^{ab}$ and $({\Delta}M_u^2)^{ab}$ are the "known"
functions of heavy fermions,
$({\Delta}M_{u,d}^2)^{ab} = F^{ab}(m_t, m_b,...)$ , which mainly get the
contributions due to the vacuum expectations of   the Higgs  bosons
that  were  used  for  construction of  the mass matrix ansatzes for
d- (u-) quarks.

For example in the case of $N_g=\underline 3 + \underline 1$ families
with Fritzsch ansatz for quark mass matrices and using
$SU(3)_H \times SU(2)$ Higgs fields,
$(8,2)$, \cite {9'},
  we can  write down some rough relations
between the masses of horizontal gauge bosons ("bootstrap" solution):

\begin{eqnarray}
M_{H_1}^2 &\approx& M_{H_2}^2 \approx M_{H_3}^2 \approx M_{H_0}^2 +
\frac{g_{H}^2}{4} \Bigl[\frac{1}{ \lambda^2} {{m_c m_t}
\over {1 -{m_t}/{m_{t'}}}} \Bigr]+
 ...  ,\nonumber \\
M_{H_4}^2 &\approx&  M_{H_5}^2 \approx M_{H_6}^2 \approx M_{H_7}^2
\approx M_{H_0}^2 +
\frac{g_{H}^2}{4} \Bigl[\frac{1}{\tilde {\lambda}^2} m_t m_{t'}\Bigr] +...  ,
\nonumber\\
 M_{H_8}^2
&\approx& M_{H_0}^2 +
\frac{g_{H}^2}{3} \Bigl[\frac{1}{\tilde {\lambda}^2} m_t m_{t'}\Bigr]
+...,
\end{eqnarray}
where $\lambda$ and $\tilde {\lambda}$ are Yukawa couplings.

We were interested how does the unitary compensation
for the contributions of horizontal forces  to rare processes \cite{9'}
depend on  different versions of the $SU(3)_H$-symmetry breaking. The
investigation
of this dependence allows, firstly, to understand how low may the horizontal
symmetry  breaking scale $M_H$  be, and, secondly, how is this scale
determined by a particular  choice of a  mass matrix ansatz both for quarks
and leptons.

 We would like to stress  a possible existing
of the local family symmetry with a low energy symmetry breaking scale,
i.e. the existence of rather light H-bosons:
$m_H\geq (1-10)TeV$ \cite{9'}.
 We have analyzed, in the framework of the
"minimal" horizontal
supersymmetric gauge model, the possibilities of  obtaining
a satisfactory
hierarchy for  quark masses and of connecting it with the splitting of
horizontal gauge boson masses. We expect that due to  this approach
the horizontal model will become more definite since it will
allow to study the amplitudes of rare processes and
the CP-violation mechanism more thoroughly.
In this way  we hope to get a deeper insight into  the nature of
interdependence between  the generation mixing mechanism and
the local horizontal symmetry breaking scale.

\subsection{The flavor-changing  rare processes and
superweak-like source of CP-violation in GUST with
the non-Abelian gauge family symmetry.}

The existence of  horizontal interactions with low energy breaking scale
($\Lambda_H < 10 TeV$) might lead  to large flavor changing neutral currents
(FCNC).
 This  interaction  is  described  by  the
relevant part of the SUSY $SU(3)_H$-Lagrangian and has the form

\begin{eqnarray}
{\cal L}_H=g_H\bar \psi_d {\Gamma}_{\mu}\ (\ D\ \frac{  \lambda^a } {2}\
D^+\ ) \ \psi_d O^{ab} Z_{\mu}^b  \,\,. \label{5.9}
\end{eqnarray}
Here we have (a,b=1,2,...,8). The matrix $O_{ab}$ determines the  relationship
between the bare,  $H_{\mu}^{b}$, and physical, $Z_{\mu}^b$,  gauge
fields and  is  calculated  for
the  mass  matrix  $(M_H^2)_{ab}$   diagonalized;
$\psi_d  =(d\   ,\ s\  ,\ b\ )$
(similarly for "up"-quarks and charged leptons);
$g_H$  is  the  gauge  coupling   of  the $SU(3)_H$  group.

After the calculations in "bootstrap" model with the Higgs fields
${\langle }H{\rangle }=
({\lambda}^a{\varphi}^a)/2$, ${\langle }h{\rangle }=
({\lambda}^a \tilde {\varphi}^a)/2$
the expressions for
the ($K_L^0\,-\,K_S^0$), ($B_{dL}^0\, -\,B_{dS}^0$),
- ($B_{sL}^0\, -\, B_{sS}^0$),-
 ($D_L^0\, -\, D_S^0$),...
meson mass differences (pure quark processes)
at tree level take the following general forms:

\begin{eqnarray}
\biggl[\frac{(M_{12})_{ij}^K}{m_K}\biggr]_H &=&
\frac{1}{2}\frac{g_H^4}{M_{H_0}^4}  \Biggl\{
\biggl[\tilde{\varphi^a}(D\,\frac{\lambda^a}{2}\,D^+)_{ij}\biggr]^2
+\biggl[\varphi^a(D\,\frac{\lambda^a}{2}\,D^+)_{ij}\biggr]^2 \Biggr\}
{f_{K_{ij}}^2}R_{K_{ij}}, \nonumber\\
\biggl[\frac{(M_{12})_{ij}^D}{m_D}\biggr]_H &=&
\frac{1}{2}\frac{g_H^4}{M_{H_0}^4}  \Biggl\{
\biggl[\tilde{\varphi^a}(U\,\frac{\lambda^a}{2}\,U^+)_{ij}\biggr]^2
+\biggl[\varphi^a(U\,\frac{\lambda^a}{2}\,U^+)_{ij}\biggr]^2 \Biggr\}
{f_{D_{ij}}^2}R_{D_{ij}}. \label{6.7}
\end{eqnarray}

where (i,j) = {(1,2)},{(1,3)},{(2,3)} -- the $K$ or$ D$, $B_d$ or $T_u$,
$B_s$ or $T_c$ - meson systems.

The coefficients in formulas (\ref{6.7}) are calculated from (\ref{5.9})
using formula (\ref{5.6}).

For example, for $K$-meson systems we find the following contribution
(if $D_L=D_R=D$)
\begin{eqnarray}
&& g_H^2/4 (D\lambda^a O^{ab} D^+)_{12}\frac{1}{M_0^2+\Delta M_b^2}
(D\lambda^c O^{cb} D^+)_{12} =\nonumber \\
&&=\frac{g_H^2}{4M_0^2}(D\lambda^a O^{ab} D^+)_{12}(1-\frac{\Delta M_b^2}
{M_0^2})
(D\lambda^c O^{cb} D^+)_{12} =\nonumber \\
&&=-\frac{g_H^2}{4M_0^4}(D \lambda^a  D^+)_{12}{\Delta M^2}_{ac}
(D\lambda^c D^+)_{12} =\nonumber \\
&&=-\frac{g_H^4}{4M_0^4}(D\lambda^a  D^+)_{12}
f^{kal}f^{kcm} \varphi^l {\bar{\varphi}}^{m}
(D \lambda^c D^+)_{12} =\nonumber \\
&&=-\frac{g_H^4}{4M_0^4} (D[\lambda^k , \lambda^l]D^+)_{12}
\varphi^l{\bar{\varphi}}^{m}
(D[\lambda^k,\lambda^m] D^+)_{12} =\nonumber \\
&&=-\frac{g_H^4}{M_0^4}\ ((D \lambda^a D^+)_{12} \varphi^a )
( (D\lambda^b  D^+)_{12} \bar{\varphi}^b)
\label{vyvodf}
\end{eqnarray}
Here we consider a case of the complex value for VEVs, $\varphi^a$.
Also we used the next formula:
$(D \lambda^a /2 D^+)_{ij}(D \lambda^a /2 D^+)_{kl}=
1/2(\delta_{il} \delta_{jk} - 1/3\delta_{ij} \delta_{kl})$.

It is interesting, that if a difference between the gauge boson's masses
is generated
by Higgs fields in representation $(3,\ 2)$ (see (\ref{5.7})),
than the contribution
in $\biggl[\frac{\Delta m}{m}\biggr]_H$ equal to zero in considering order
(for case $D_L=D_R$),
since we will use Higgs fields $(8,\ 2)$ for these evaluations.
However, for processes including three equivalent index
(like $\mu\longrightarrow 3e$) Higgs fields $(3,\ 2)$ give
nonzero contribution $\sim (\varphi D^+)_i\cdot (D{\bar \varphi} )_j$.

Note, that formula (\ref{vyvodf}) is true for the case when
$D_L$ differs from $D_R$ by diagonal phase multiply too. For us the
case $D_L=-D_R$ which corresponds to axial-vector terms is important.
In general if $D_L\ne D_R$ (or $U_L\ne U_R$) than in formulas
(\ref{6.7}) there is a  quadratic term
$g_H^2/M_0^2\, (D_LD_R^+)_{ij}(D_RD_L^+)_{ij}$, $i\ne j$.

 So, we could analyse
the ratios (similar for $B_{d,s}$-meson system):
\begin{equation}
\biggl[\frac{\Delta m_K}{m_K}\biggr]_H=
\frac{g_H^2}{M_{H_0}^2}Re[C_K]f_K^2R_K < 7\cdot 10^{-15} \label{6.8}
\end{equation}
and
\begin{equation}
\biggl[\frac{Im{M}_{12}}{m_K}\biggr]_H=
\frac{1}{2} \frac{g_H^2}{M_{H_0}^2}Im[C_K]f_K^2R_K < 2\cdot
10^{-17}.\label{6.9}
\end{equation}

In these formulas in "bootstrap" models \cite{9'}  the expression for
$C_{K,D}$ , namely
\begin{equation}
      C_{K,D} = \frac{g_H^2}{2{{\lambda}_Y}^2}
\frac{m_t^2}{M_{H{0}}^2} \times f(\frac{m_i^{up}}{m_j^{up}};
\frac{m^{down}_k}{m^{down}_l}),
\end{equation}
contains  known complex functions  $f$'s and their forms  depend on
quark fermion mass ansatzes \cite{9'}.

Substituting in formula (\ref{6.7}) the expressions for $\varphi$,
$\tilde{\varphi}$
 and the elements $d_{ij}$ of the $D$ mixing matrix ("bootstrap" solution),
we can obtain the lower limit for the value $M_{H_0}$ \cite{9'}:\\
\begin{equation}
M_{H_0} < O(1 TeV).
\end{equation}

 Here noteworthy are the  following two
points:
 a)The appearance of the phase in the CKM mixing matrix may be
due to new dynamics  working at short distances $(r\ll {{1}\over {M_W}})$.
Horizontal forces may be the  source of this new dynamics \cite{9'}.
Using this approach,
 we might have the CP violation effects both
due to electroweak and  horizontal  interactions.

(b) The CP is conserved in the
electroweak sector $(\delta^{KM}=0)$, and its breaking is provided by the
structure of the horizontal interactions. Let us consider  the
situation when $\delta^{KM}=0$. In the  SM, such a case might be
realized just accidentally. The vanishing phase of the electroweak sector
$(\delta^{KM}=0)$ might arise spontaneously due to some additional
symmetry. Again, such a situation might occur within the
horizontal extension of the electroweak model.

In particular, this model gives rise to a rather natural mechanism of superweak
-like  CP-violation due to the  $(CP=-1)$ part of the effective
Lagrangian of  horizontal interactions $({\epsilon}^{\prime}/
{\epsilon})_K\leq \,{10}^{-4}$.
 That part of $\cal L\rm_{eff}$
includes the product of the $SU(3)_H$-currents $I_{\mu i}$~$I_{\mu j}$
(i=1,4,6,3,8; j=2,5,7 or, vice versa,  i$\longleftrightarrow$j ) \cite{9'}.
In the case of a vector-like $SU(3)_H$-gauge model the
CP violation could
be only due to the charge symmetry breaking.

The space-time structure of horizontal interactions depends on the $SU(3)_H$
quantum numbers of quark and lepton superfields and their C-conjugate
superfields. One can obtain vector
(axial)-like
horizontal interactions as far as
the $G_H$ particle quantum numbers are conjugate (equal)
 to those of antiparticles.
The question arising in these theories
is how such horizontal interactions are related with strong and
electroweak ones. All these interactions can be unified within
one gauge group, which would allow  to calculate the value of the coupling
constant of horizontal interactions.
Thus, an unification of horizontal, strong and
electroweak interactions might  rest  on the GUTs
$\tilde G \equiv G\times SU(3)_H$
 (where, for example, $\tilde G \equiv E(8)$,
$G\equiv SU(5),SO(10)$ or $E_6$), which may  be  further broken down to
$SU(3)_H\times SU(3)_C \times SU(2)_L \times U(1)_Y$.
For including "vector"-like horizontal gauge symmetry into GUT
 we have to introduce "mirror" superfields. Speaking more definitely,
if we want to construct GUTs of the
$\tilde G \equiv G\times SU(3)_H$ type,
  each generation must
encompass double  $G$-matter supermultiplets, mutually conjugate under the
$SU(3)_H$-group. In this approach the first supermultiplet consists of the
superfields $f$ and $f_m^c\in 3_H$, while the second is constructed with
the help of the supermultiplets $f^c$ and $f_m\in \bar 3_H$.
In this scheme, proton decays are only possible in the case of
 mixing between ordinary and "mirror" fermions. In its turn, this mixing
must, in particular, be related
with the $SU(3)_H$-symmetry breaking.

The GUSTs spectra also predict the existing of the new neutral
neutrino - like particles interacting with the matter only by
"superweak"-like coupling. It is possible to estimate the masses of
these particles, and, as will be shown further, some of them have to be light
(superlight) to be observed in modern experiment.

 A variant for unusual nonuniversal family gauge
interactions  of known quarks and leptons could be realized if for each
generation
we introduce new heavy quarks (F = U, D ),
and leptons (L, N) which are
 singlets ( it is possible to consider doublets also)
 under   SU(2)$_L$- and triplets under
$SU(3)_H$-groups.( This fermion matter could exist in string spectra.
See the all three models with $SU(3)_H \times SU(3)_H$ family gauge
symmetry).
Let us consider for concreteness a case of charged leptons:
${\Psi}_l= (e,\mu, \tau)$ and ${\Psi}_L$=(E, M,$\cal T$).
Primarily, for simplicity we suggest that  the ordinary
fermions do not take  part in  $SU(3)_H$-interactions ("white" color states).
Then the interaction is described by  the
relevant part of the SUSY $SU(3)_H$- Lagrangian and gets the form

\begin{eqnarray}
{\cal L}_H=g_H \bar{\Psi}_{\cal L} \gamma_{\mu}
\frac{ {\bf {\Lambda^a}_{6x6}} } {2}
 {\Psi}_{\cal L} O_{ab} Z_{\mu}^b  \,\,,
\end{eqnarray}
where

\begin{displaymath}
{\bf {\Lambda^a}_{6x6}} =
\left( \begin{array}{cc}
 S(L\lambda^a  L^+)S& -S(L\lambda^a L^+)C \\
-C(L\lambda^a L^+)S &  C(L\lambda^a L^+)C
\end{array} \right) .
\end{displaymath}

Here we have ${\Psi}_{\cal L} = ({\Psi}_l; {\Psi}_L )$.
The matrix $O_{ab}$  (a,b=1,2,3...8) determines the  relationship
between the bare,  $H_{\mu}^{b}$, and physical, $Z_{\mu}^b$,  gauge
fields. The diagonal $3\times3$ matrices  S=diag ($s_e, s_{\mu}, s_{\tau}$)
and C=diag ($c_e,c_{\mu},c_{\tau}$)
define the nonuniversal
character for lepton horizontal interactions, as the elements
$s_i$  depend on the lepton masses, like
$s_i \sim \sqrt{m_i/M_0}$ (i=e,$\mu$,$\tau$).
The same suggestion we might accept for local quark family
interactions.

For the family mixing  we might  suggest
the next scheme. The primary $3\times3$ mass matrix
for the light ordinary fermions is equal to zero :
$M_{ff}^0 \approx  0$.
The $3\times3$- mass matrix for heavy fermions is approximately
proportional to unit $M_{FF}^0 \approx M^Y_0 \times 1 $,
where $M^Y_0 \approx 0.5 - 1.0 TeV$ and might be different
for $F_{up}$- , $F_{down}$-  quarks and for $F_L$- leptons.
We assume that the splitting between new heavy fermions in
each class $F_Y$ (Y=up, down, L) is small and,at least in quark sector,
might be described
by the t- quark mass. Thus we think that at the first
approximation it is possible to neglect  the heavy fermion mixing.
The mixing in the
light sector is completely explained by the coupling of the
light fermions with the heavy fermions. As a result of this
coupling the $3\times3$- mass matrix $ M_{fF}^0 $ could be constructed
by "democratic" way which could lead to the well known mass family
hierarchy:

\begin{displaymath}
{\bf {M}_{6x6}^0 } =
\left( \begin{array}{cc}
    M_{ff}^0 & M_{fF}^0 \\
   M_{Ff}^0  &  M_{FF}^0
\end{array} \right) ,
\end{displaymath}
where
\begin{equation}
M_{fF}^0 \approx   M_{fF}^{dem}  \,+ \, M_{fF}^{corr}.
\end{equation}

The diagonalization of the $ M_{fF}^0$- mass matrix
$ X M_{fF}^0 X^+ $
(X = L-, D-, U- mixing matrices)
  gives us the eigenvalues, which define
the family mass
  hierarchy- $ n_1^Y << n_2^Y << n_3^Y $ and the following relations
between the masses of the known light fermions and a new heavy mass
scale:

\begin{eqnarray}
n_i^Y = \sqrt{ m_i M^Y_0},\,\,\, i=1_g,2_g,3_g;\,
\,\,Y=up-, down- fermions. \nonumber
\end{eqnarray}

 In this "see-saw" mechanism
the common mass scale of new heavy fermions  might be not very far
from the energy $\sim 1 TeV $, and as a consequence of it
the mixing angles $ s_i$- might be not too small.

\newpage
 \section{The Heterotic Superstring Theory with Rank 8 and 16
Grand Unified Gauge Groups.}

\subsection{Conformal symmetry in heterotic superstring.}

 In the heterotic string theory in left-moving
  (supersymmetric) sector there are $d-2$
 (in the light-cone gauge) real fermions $\psi^{\mu}$,
 their bosonic superpartners $X^{\mu}$, and $3(10-d)$ real
 fermions $\chi^I$. In the right-moving sector there are $d-2$
 bosons ${\bar X}^{\mu}$ and $2(26-d)$ real fermions.

In heterotic string theories \cite{13',14'} ${(N=1\: SUSY)}_{LEFT}$
${(N=0\: SUSY)}_{RIGHT}$ $\oplus$ ${\cal M}_{c_{L};c_{R}}$ with $d \leq 10$,
the conformal anomalies of the space-time sector are canceled by the conformal
anomalies of the internal sector ${\cal M}_{c_L;c_R}$, where $c_L=15-3d/2$ and
$c_R=26-d$ are the conformal anomalies in the left- and right--moving string
sectors respectively.

One can consider the operator product expansion between the
 energy-momentum tensor T(z):
\begin{eqnarray}\label{eq T}
T(w)T(z) \sim \frac{c/2}{(w-z)^4} +
 \frac{2}{(w-z)^2}T(z)+ \frac{1}{w-z} \partial_z T(z),
\end{eqnarray}
where c is a central charge or conformal anomaly.

If we take the moments of the energy-momentum operator T(z) we will get the
conformal generators with the following Virasoro algebra:

\begin{eqnarray}
[L_n,\,L_m]\,= (n-m)\,L_{n+m}\,+\,\frac{c}{12}\,n\,(n^2-1)\, {\delta}_{n,-m}.
\end{eqnarray}

Using Virasoro algebra we can construct representations of the conformal group
where highest weight state is specified by two quantum numbers,
conformal weight $h$ and central charge $c$, such that:

\begin{eqnarray}
L_0 |h,c\rangle =&& h |h,c\rangle \nonumber\\
L_n |h,c\rangle =&& 0,\,\,\,n=1,2,3,....
\end{eqnarray}
For massless state the conformal weight $h=1$.

 In the left supersymmetric sector world-sheet supersymmetry is non-linearly
 realized via the supercharge
 \begin{equation}
 T_F=\psi^{\mu}\partial X_{\mu} +f_{IJK}\chi^I\chi^J\chi^K\ ,\label{sch}
 \end{equation}
 where $f_{IJK}$ are the structure constants of a semi-simple
 Lie group $G$ of dimension $3(10-d)$. The operator product expansions
 T(z) and $T_F(z)$ give the N=1, (1,0), superconformal algebra:

\begin{eqnarray}
T(w)T(z)&&\sim \frac{3{\hat c}/4}{(w-z)^4} +
 \frac{2}{(w-z)^2}T(z)+ \frac{1}{w-z} \partial_z T(z),\nonumber\\
T(w)T_F(z) &&\sim \frac{3/2}{(w-z)^2}T_F +
  \frac{1}{w-z} \partial_z T(z),\nonumber\\
T_F(w)T_F(z) &&\sim \frac{{\hat c}/4}{(w-z)^4} +
 \frac{1/2}{(w-z)^2}T(z),
\end{eqnarray}
where $\hat c = 2/3c= 6$.
 The possible Lie algebras of dimension 18 for $d=4$
 are $SU(2)^6$, $SU(3)\times SO(5)$, and $SU(2)\times SU(4)$.

In papers \cite {bank'} it has been shown that the N=1  space-time SUSY
vacuum
of heterotic string with local (1,0) worldsheet supeconformal invariance
extends to a global N=2, (2,0), superconformal invariance:
\begin{eqnarray} \label {equation 2.0}
{T_F^+}(w) {T_F^{-}}(z) &&\sim \frac{\hat c}{(w-z)^3} + \frac{2\partial J}
{(w-z)^2} +
 \frac{2T + \partial J}{w-z}\nonumber\\
J(w) {T_F^{\pm}}(z) &&\sim \pm \frac{T_F^{\pm}}{w-z}\nonumber\\
J(w) J(z) &&\sim \frac{\hat c/2}{(w-z)^2}.
\end{eqnarray}
 A Sugawara- Sommerfeld
construction of the energy- momentum tensor T(z)
 algebra in terms of bilinears in the Kac-Moody generators   $J^a_n(z)$
\cite {21',21''},\cite {22'},
\begin{eqnarray}
T(z)=\sum_n L_n z^{-n-2}=
-\frac{1}{2k + Q_{\psi}}\sum_{n,m}:J^a_{n-m}J^a_m: z^{-n-2},
\end{eqnarray}
allows to get, commuting two generators of the Virasoro algebra,
 the following expression for the central Virasoro
"charge":
\begin{eqnarray}\label{eq 101}
c_g = \frac {2 k dim g}{2 k + Q_{\psi}}=
\frac{x dim g}{ x + \tilde h}.
\end{eqnarray}

In the fermionic formulation of the four-dimensional heterotic string theory in
addition to the two transverse bosonic coordinates $X_{\mu}$ ,$\bar{X}_{\mu}$
and their left-moving superpartners ${\psi}_{\mu}$, the internal sector
${\cal M}_{c_L;c_R}$ contains 44 right-moving ($c_R=22$) and 18 left-moving
($c_L=9$) real fermions (each real world- sheet fermion has
$c_f=1/2$).

For a couple of years superstring theories, and particularly the heterotic
string theory, have provided an efficient way to construct the Grand Unified
Superstring Theories ($GUST$) of all known interactions, despite the
fact that it is
still difficult to construct unique and fully realistic low energy models
resulting after decoupling of massive string modes. This is because
of the fact that only
10-dimensional space-time  allows existence of two consistent (invariant
under reparametrization, superconformal, modular, Lorentz and SUSY
transformations) theories with the gauge symmetries $E(8)\times E(8)$ or
${spin(32)}/{Z_2}$ \cite{13',14'} which after compactification of
the six extra
space coordinates (into the Calabi-Yau \cite{15',16'} manifolds or into the
orbifolds) can be used for constructing GUSTs. Unfortunately, the process of
compactification to four dimensions is not unique and the number of possible
low energy models is very large. On the other hand, constructing
the theory directly in $4$-dimensional space-time requires including a
considerable number of free bosons or fermions into the internal string sector
of the heterotic superstring \cite{17',18',19',20'}.
This leads to as large internal
symmetry group such as e.g. rank  $22$ group.
The way of breaking this primordial symmetry
is again not unique and leads to a huge number of possible models,
each of them
giving different low energy predictions.

 Because of the presence of the affine Kac-Moody algebra (KMA)
$\hat g$ (which is a 2-dimensional manifestation of gauge symmetries of the
string itself) on the world sheet, string constructions yield definite
predictions concerning representation of the symmetry group that can be used
for
low energy models building\cite {21',22'}.
Therefore the following longstanding questions
have a chance to be answered in this kind of unification schemes:

\begin{enumerate}
\item How are the chiral matter fermions assigned to the multiplets of the
unifying group?
\item How is the GUT gauge symmetry breaking realized?
\item What is the origin of the fermion mass hierarchy?
\end{enumerate}

The first of these problems is, of course, closely connected to the
quantization of the electromagnetic charge of matter fields. In addition,
string constructions can shed some light on the questions about the number of
generation and possible existence of mirror fermions which remain unanswered
in conventional GUTs \cite {23'}.

There are not so many GUSTs describing the observable sector of Standard
Models. They are well known: the SM gauge group, the Pati-Salam
($SU(4)\times SU(2)\times SU(2)$) gauge group, the flipped SU(5) gauge group
and SO(10) gauge group, which includes flipped SU(5) \cite{20',20''}.

There are good physical reasons for including the horizontal $SU(3)_H$ group
into the unification scheme. Firstly, this group naturally accommodates three
fermion families presently observed (explaining their origin) and, secondly,
can provide correct and economical description of the fermion mass spectrum
and mixing  without invoking high dimensional representation of conventional
$SU(5)$, $SO(10)$ or $E(6)$ gauge groups. Construction of a string model
(GUST)
containing the horizontal gauge symmetry provides additional strong
motivation
to this idea. Moreover, the fact that in GUSTs high dimensional
representations
are forbidden by the KMA is a very welcome feature in this context.

\subsection{The possible ways of  E(8)-GUST breaking
leading to the $N_G=3$ or $N_G=3+1$ families}

All this leads us naturally to consider possible forms for horizontal
symmetry
$G_H$, and $G_H$ quantum number assignments for quarks (anti-quarks) and
leptons (anti-leptons) which can be realized within GUST's framework.
To include the
horizontal interactions with three known generations in the ordinary
GUST it is
natural to consider rank eight gauge symmetry.
 We can consider
$SO(16)$ (or $E(6) \times SU(3)$) which is the maximal subgroup of $E(8)$
and which contains the rank eight
subgroup $SO(10)\times (U(1)\times SU(3))_H$ \cite {24'}. We will be,
therefore,
concerned with
the following chains (see Fig. \ref{fig1}):
\vskip 0.3cm
\small
$$
\begin{array}{cccc}
\:\:E(8) \longrightarrow SO(16) \longrightarrow
\underline {SO(10) \times (U(1) \times SU(3))_H} \longrightarrow  \\
\longrightarrow SU(5) \times U(1)_{Y_5}
\times (SU(3) \times U(1))_H  \\
\end{array}
$$
 \vskip 0.3cm
\small
or
$$
\begin{array}{cccc}
\:\:E(8) \longrightarrow E(6) \times SU(3) \longrightarrow
 (SU(3))^{ \times 4}.  \\
\end{array}
$$
 \vskip 0.3cm
\begin{figure}
\caption{The possible ways of E(8) gauge symmetry breaking leading to the
3+1 or 3 generations.}
\label{fig1}
\setlength{\unitlength}{1mm}
\begin{picture}(165,135)(10,0)
\setlength{\unitlength}{0.8mm}
{\tiny
\put(20,140){{\large E(8) } }
\put(110,140){{\large SO(16) }}
\put(6,80){{\large $ E(6)\times SU(3)_H $ }}
\put(85,80){{\large $ SO(10)\times SU(3)_H\times U(1)_H $ }}
\put(15,20){{\large $ SU(3)^{\otimes 4} $ }}
\put(73,20){{\large $ SU(5)\times U(1)\times SU(3)_H\times U(1)_H $ }}
\put(73,10){{\large $N_g = 3 $ , $N_g = 3 + 1 $ }}

\put(23,130){\vector(0,-1){33}}
\put(115,130){\vector(0,-1){33}}
\put(50,142){\vector(1,0){30}}
\put(23,70){\vector(0,-1){33}}
\put(115,70){\vector(0,-1){33}}
\put(50,82){\vector(1,0){30}}

\put(5,105){\shortstack[l]{ {$ 248 \longrightarrow $} \\ {$(78,1)\oplus $} \\
 {$ (1,8)\oplus $} \\ {$ (27,3)\oplus $} \\ {$ (\bar{27},\bar 3) $}  }}

\put(5,30){\shortstack[l]{ {$ 78 \longrightarrow $} \\ {$(8,1,1)\oplus $} \\
      {$ (1,8,1)\oplus $} \\ {$ (1,1,8)\oplus $} \\
          {$ (3,3,3)\oplus $} \\{$ (\bar 3,\bar 3,\bar 3) $} \\ \\
   {$ 27 \longrightarrow $} \\ {$(3,\bar 3,1)\oplus $} \\
      {$ (1,3,\bar 3)\oplus $} \\ {$ (\bar 3,1,3) $}    }}

\put(52,147){\shortstack[l]{ {$  248 \longrightarrow 120 \oplus 128 $} }}

\put(120,100){\shortstack[l]{ {$ 120 \longrightarrow (45,1)^0 \oplus (1,8)^0
 \oplus $} \\ {$ (1,1)^0 \oplus  (10,3)^2 \oplus (10,\bar 3)^{-2} \oplus $} \\
              {$ (1,3)^{-4} \oplus (1,3)^{+4} $} \\
 {$ 128 \longrightarrow (16,3)^{-1} \oplus (\bar{16},\bar 3)^{+1} \oplus $} \\
   {$ (16,1)^{+3} \oplus  (\bar{16},1)^{-3} $}    }}

\put(30,90){\shortstack[l]{
 {$ (78,1) \longrightarrow (45,1)^0 \oplus (1,1)^0
 \oplus  (16,1)^{+3} \oplus (\bar{16},1)^{-3} $} \\
 {$ (27,3) \longrightarrow (16,3)^{-1} \oplus (10,3)^{+2} \oplus (1,3)^{-4} $}
\\
 {$ (\bar{27},\bar{3}) \longrightarrow (\bar{16},\bar 3)^{+1} \oplus (10,\bar
3)^{-2}
            \oplus (1,\bar 3)^{+4} $} }}

\put(120,50){\shortstack[l]{
 {$ 45 \longrightarrow (24,1) \oplus (1,1) \oplus  (10,1)
                \oplus (\bar{10},1) $} \\
 {$ 16 \longrightarrow (1)_{+5/2} \oplus (\bar 5)_{-3/2}
                        \oplus (10)_{+1/2} $} \\
 {$ \bar{16} \longrightarrow (\bar 1)_{-5/2} \oplus (5)_{+3/2}
                        \oplus (\bar{10})_{-1/2} $} }}

}
\end{picture}

\end{figure}
\normalsize

According to this scheme one can get $SU(3)_H\times U(1)_H$ gauge family
symmetry with $N_g = 3 + 1 $ (there are also other possibilities as e.g.
 $E(6)\times SU(3)_H\subset E(8)$
 $N_g = 3 $ generations can be obtained due to the second way of $E(8)$
gauge symmetry breaking via $E(6)\times SU(3)_H$, see Fig.\ref{fig1}), where
the possible
additional fourth massive matter superfield could appear from
$\underline {78}$ as a singlet of
$SU(3)_H$ and transforms as $\underline {16}$ under the $SO(10)$ group.

In this note  starting from the rank 16 grand unified gauge group (which is
the minimal rank allowed in strings) of the form $G\times G$
\cite{25',26'}and making use of the KMA which select
the possible gauge group
representations we construct the string models based on the diagonal subgroup
$G^{symm}\subset G\times G\subset SO(16)\times SO(16)~(\subset E_8\times E_8)$
\cite {25'}.
We discuss and consider $G^{symm}=SU(5)\times U(1)\times (SU(3)\times U(1))_H$
$\subset SO(16)$ where the factor $(SU(3)\times U(1))_H$ is interpreted as the
horizontal gauge family symmetry. We explain how the unifying gauge symmetry
can be broken down to the Standard Model group. Furthermore, the horizontal
interaction predicted in our model can give an alternative description of the
fermion mass matrices without invoking high dimensional Higgs representations.
In contrast with other GUST constructions,
our model does not contain particles with exotic fractional electric charges
\cite {27',25'}.
This important virtue of the model is due to the symmetric construction
of the electromagnetic charge  $Q_{em}$ from $ Q^I$ and $Q^{II}$ -- the
two electric charges of each of the $U(5)$ groups\cite {25'}:

\begin{eqnarray}\label{eq102}
Q_{em} &=& Q^{II} \oplus  Q^I.
\end{eqnarray}

We consider the possible forms   of the
$G_H= SU(3)_H$ ,$SU(3)_H \times U(1)$, $G_{HL} \times G_{HR}$... - gauge family
symmetries  in the framework of Grand Unification Superstring Approach.
Also we will study the
 matter spectrum of these GUST, the possible Higgs sectors.
The form of the Higgs sector it is very important for GUST- , $G_H$- and SM
- gauge symmetries breaking
and for constructing Yukawa couplings.

\newpage

\section{World-Sheet Kac-Moody Algebra
         And Main Features of Rank Eight GUST}\label{sec2}

\subsection{ The representations of Kac-Moody Algebra}

Let us begin with a short review of the KMA results \cite{21',22'}.
In heterotic string the KMA is constructed by the operator product
expansion (OPE) of the fields $J^a$ of the conformal dimension $(0,1)$:
\begin{equation}
{J^a}(w) {J^b}(z)\sim {\frac{1}{{(w-z)}^2}} k {\delta}^{ab} +
{\frac{1}{w-z}} i f^{abc} J^c + ....
\end{equation}
 The structure constants $f^{abc}$ for the group $g$ are normalized so that
\begin{equation}
 f^{acd}f^{bcd} = Q_{\psi} {\delta}^{ab} =\tilde h {\psi}^2
{\delta}^{ab}
\end{equation}
 where $Q_{\psi}$ and $\psi$ are  the
quadratic Casimir and the highest weight of the adjoint representation and
$\tilde h$ is the dual Coxeter number.
The $\frac{\psi}{{\psi}^2}$ can be
expanded as in integer linear combination of the simple roots of $g$:
\begin{equation}
 \frac{\psi}{{\psi}^2} = \sum_{i=1}^{rank\,g} m_i {\alpha}_i.
\end{equation}
The dual Coxeter number can be expressed through the integers numbers $m_i$
\begin{equation}
\tilde h = 1 + \sum_{i=1}^{rank \,g} m_{i}
\end{equation}
and for the simply laced groups (all roots are equal and ${\psi}^2 =2$):
$A_n$, $D_n$, $E_6$, $E_7$, $E_8$
they are equal $n+1$, $2n-2$, $12$, $18$ and $30$, respectively.

The KMA $\hat g$ allows to grade the representations $R$ of the gauge group by
a level number $x$ (a non negative integer) and by a conformal weight $h(R)$.
An irreducible representation of the affine algebra $\hat g$ is characterized
by the vacuum representation of the algebra $g$ and the value of the central
term $k$, which is connected to  the level number by the relation
$x=2 k/{\psi}^2$.
The value of the level
number of the KMA determines the possible highest weight unitary
representations
which are present in the spectrum in the following way

\begin{equation}\label{eq1}
x=\frac{2 k}{{\psi}^2} \geq \sum_{i=1}^{rank \,g}n_{i} m_{i},
\end{equation}

\noindent where the sets of non-negative integers $\{m_i = m_1,..., m_r\}$
and $\{ n_i = n_1,...,n_r \}$ define the highest root and the highest weight
in terms of fundamental weights
of a representation $R$ respectively \cite{21',22'}:
\begin{equation}
{\mu}_0 = \sum_{i=1}^{rank \,g} n_{i} {\lambda}_{i}
\end{equation}

In fact, the KMA on the level one is realized in the 4-dimensional heterotic
superstring theories with free world sheet fermions which allow a
complex fermion description \cite{18',19',20'}. One can obtain KMA on a higher
level working with real fermions  and using some tricks \cite{33'}.
For these models the level of KMA coincides with the Dynkin index of
representation $M$ to which free fermions are assigned,

\begin{equation}\label{eq2}
x = x_M = \frac{Q_M}{{\psi}^2}\frac{dim M}{dim g}
\end{equation}
($Q_M$ is a quadratic Casimir eigenvalue of representation $M$) and equals one
in cases when real fermions form vector representation $M$ of $SO(2N)$, or when
the world sheet fermions are complex and $M$ is the fundamental representation
of $U(N)$ \cite{21',22'}.

Thus, in strings with KMA on the level one realized on the world-sheet, only
very restricted set of unitary representations can arise in the spectrum:

\begin{enumerate}
\item  singlet and totally antisymmetric tensor representations of $SU(N)$
groups, for which $m_i = (1,...,1) $;
\item singlet, vector, and spinor representations of $SO(2N)$ groups with\\
$m_i = (1,2,2,...2,1,1)$;
\item singlet, $\underline{27}$, and $\bar{\underline{27}}$-plets of $E(6)$
corresponding to $m_i = (1,2,2,3,2,1)$;
\item singlet  of $E(8)$ with $m_i = (2,3,4,6,5,4,3,2)$.
\end{enumerate}
Therefore only these representations can be used to incorporate matter
and Higgs fields in GUSTs with KMA on the level 1.

In principle it might be possible to construct explicitly an example of
a level 1 KMA-representation of the simply laced algebra $\hat g$
(A-, D-, E - types)
from the level one representations of the Cartan subalgebra of $g$.
This construction is achieved using the vertex operator of string, where
these operators are assigned to a set of lattice point corresponding to
the roots of a simply-laced Lie algebra $g$.

\subsection{The features of the level one KMA in matter and Higgs
representations in
rank 8 and 16 GUST Constructions}
For example, to describe chiral matter fermions in GUST with the gauge symmetry
group $SU(5)\times U(1)\subset SO(10)$ the following sum of the level-one
complex representations: $\underline {1}(-5/2) + \underline {\bar{5}}(+3/2) +
\underline {10}(-1/2) = \underline{16}$ can be used. On the other side, as real
representations of $SU(5)\times U(1)\subset SO(10)$, from which Higgs fields
can arise, one can take for example $\underline{5} + \underline{\bar 5}$
representations arising from real representation $\underline {10}$ of $SO(10)$.
Also, real Higgs representations like $\underline {10}$(-1/2)
+ $\underline{\bar{10}}$(+1/2) of $SU(5)\times U(1)$ originating
from $\underline{16}$+$\underline{\bar{16}}$ of $SO(10)$, which has been used
in ref. \cite{10'} for further symmetry breaking, are allowed.

Another example is provided by the decomposition of $SO(16)$ representations
under $SU(8)\times U(1)\subset SO(16)$. Here, only singlet,
$v=\underline{16}$,
$s = \underline{128}$, and $s^{\prime} = \underline{128}^{\prime}$
representations of $SO(16)$
are allowed by the KMA ($s = \underline{128}$ and
$s^{\prime}= \underline{128}^{\prime}$ are
the two nonequivalent, real spinor representations with the highest
weights
${\pi}_{7,8} =1/2 ({\epsilon}_1+{\epsilon}_2, + \dots +
{\epsilon}_7 \mp {\epsilon}_8)$,
${\epsilon}_{i}{\epsilon}_{j}= {\delta}_{ij}$).
 From the item 2. we can
obtain the following $SU(8)\times U(1)$ representations: singlet,
$\underline{8}$+$\underline{\bar 8}$ $(=\underline{16})$,
$\underline{8}+\underline{56}+\underline{\bar {56}}+\underline{\bar 8}$
$(=\underline{128})$, and $ \underline{1}+\underline{28}+\underline{70} +
+\underline{\bar{28}} + \underline{\bar {1}}$ $(=\underline{128}^{'})$.
The highest weights of $SU(8)$ representations
 ${\pi}_1 ={\pi}( \underline{8})$, ${\pi}_7 = ={\pi}(\underline{\bar 8})$
and
${\pi}_3 =={\pi}(\underline{56})$,
${\pi}_5 =={\pi}(\underline{\bar {56}})$  are:
\begin{eqnarray}
{\pi}_{1}& = &1/8 ( \underbrace{7{\epsilon}_1 - {\epsilon}_2
- {\epsilon}_3 - {\epsilon}_4 - {\epsilon}_5}  \underbrace{- {\epsilon}_6
-{\epsilon}_7 -  {\epsilon}_8}),\nonumber \\
{\pi}_{7}& = & 1/8(  \underbrace{{\epsilon}_1 + {\epsilon}_2
+ {\epsilon}_3 + {\epsilon}_4 + {\epsilon}_5}  \underbrace{+ {\epsilon}_6
+{\epsilon}_7 - 7 {\epsilon}_8}),\nonumber \\
{\pi}_{3}& = &1/8( \underbrace{{5\epsilon}_1+ 5{\epsilon}_2 + 5{\epsilon}_3
- 3{\epsilon}_4 - 3{\epsilon}_5}  \underbrace{- 3{\epsilon}_6 -
3{\epsilon}_7 - 3{\epsilon}_8}),\nonumber\\
 {\pi}_{5}& = &1/8( \underbrace{{3\epsilon}_1+ 3{\epsilon}_2 + 3 {\epsilon}_3
+ 3{\epsilon}_4 + 3{\epsilon}_5}  \underbrace{ - 5{\epsilon}_6 -
5{\epsilon}_7  - 5{\epsilon}_8}).
\end{eqnarray}
Similarly,the highest weights of $SU(8)$ representations
 ${\pi}_2 =={\pi}( \underline{28})$,  ${\pi}_6 =={\pi}(\underline{\bar {28}})$
and
${\pi}_4 =={\pi}(\underline{70})$ are:
\begin{eqnarray}
{\pi}_{2}& = &1/4 ( \underbrace{3{\epsilon}_1 + 3{\epsilon}_2 -{\epsilon}_3
-{\epsilon}_4 -{\epsilon}_5}  \underbrace{ -{\epsilon}_6
-{\epsilon}_7 -  {\epsilon}_8}),\nonumber \\
{\pi}_{6}& = &1/4 ( \underbrace{{\epsilon}_1 + {\epsilon}_2
+ {\epsilon}_3 + {\epsilon}_4 + {\epsilon}_5}  \underbrace{
+  {\epsilon}_6 -
3{\epsilon}_7 - 3 {\epsilon}_8 }),\nonumber \\
{\pi}_{4}& = &1/2( \underbrace{{\epsilon}_1+ {\epsilon}_2 + {\epsilon}_3
+ {\epsilon}_4 - {\epsilon}_5}  \underbrace{ - {\epsilon}_6 -
{\epsilon}_7 - {\epsilon}_8}).
\end{eqnarray}

However, as we will demonstrate, in each of the string sectors
the generalized Gliozzi--Scherk--Olive projection (the $GSO$ projection
in particular guarantees the modular invariance and supersymmetry of the
theory and also give some nontrivial restrictions on gauge
groups and its representations) necessarily eliminates either
$\underline{128}$ or $\underline{128}^{\prime}$. It is therefore important
that, in order to incorporate chiral matter in the model, only one spinor
representation is sufficient. Moreover, if one wants to solve the
chirality problem applying further $GSO$ projections (which break the
gauge symmetry) the representation $\underline {\bar{10}}$ which
otherwise, together with $\underline{10}$, could form real Higgs
representation, also disappears from this sector. Therefore, the existence of
$\underline{\bar{10}}_{-1/2}$+$\underline{10}_{1/2}$, needed for breaking
$SU(5)\times U(1)$ is incompatible (by our opinion) with the possible
solution of the chirality problem for the family matter fields.

Thus, in the rank eight group $SU(8) \times U(1) \subset SO(16)$ with Higgs
representations from the level-one KMA only, one cannot arrange for further
symmetry breaking. Moreover, construction of the realistic fermion mass
matrices
seems to be impossible. In old-fashioned GUTs (see e.g.\cite{23'}), not
originating from strings, the representations of the level two were commonly
used to solve these problems.

The way out from this difficulty is based on the following important
observations. Firstly, all higher-dimensional representations of (simple laced)
groups like $SU(N)$, $SO(2N)$ or $E(6)$, which belong to the level two
representation of the KMA (according to equation \ref{eq1}),
appear in the direct product of the level one representations:

\begin{eqnarray}\label{eq3}
R_G(x=2) \subset R_G(x=1) \times {R_G}^{'}(x=1).
\end{eqnarray}

For example, the level-two representations of $SU(5)$
 will appear in the corresponding direct products of
\begin{eqnarray}
 \underline{15},
\underline{24}, \underline{40}, \underline{45}, \underline{50},
\underline{75}
\subset \underline{5}\times
\underline{5}, \underline{5}\times\underline{\bar 5}, \underline{5}\times
\underline{10}, etc.
\end{eqnarray}

In the case of $SO(10)$ the level two
representations   can be obtained by the
suitable direct products:
\begin{eqnarray}
 \underline{45}, \underline{54}, \underline{120},
\underline{126}, \underline{210}, \underline{144} \subset
\underline{10}\times\underline{10},
\underline{\bar{16}}\times\underline{10},
\underline{\bar{10}}\times\underline{16}, \underline{16}\times
\underline{16}, \underline{\bar{16}}\times\underline{16}.
\end{eqnarray}

The level-two
representations of $E(6)$ are the corresponding factors of the
decomposition of the direct
products:
\begin{eqnarray}
 \underline{78}, \underline{351}, \underline {351}^{'},
\underline{650}
\subset
\underline{\bar{27}}\times\underline{27}\mbox{ or }\underline{27}
\times\underline{27}.
\end{eqnarray}

The only exception from this rule is the $E(8)$
group, two level-two representations ($\underline {248}$ and
$\underline{3875}$) of which cannot be constructed as a product of level-one
representations \cite{24'}.

Secondly, the diagonal (symmetric) subgroup $G^{symm}$ of $G \times G $
effectively corresponds to the level-two KMA
$g(x=1)\oplus g(x=1)$ \cite{25',26'} because taking the $G\times G$
representations in the form $(R_G,R^{'}_G)$ of the $G\times G$,
where $R_G$ and $R_G^{'}$ belong to the level-one of G,
one obtains representations of the form $R_G\times R^{'}_G$
when one considers only the diagonal subgroup of $G\times G$.
This observation is crucial, because such a construction
allows one to obtain level-two representations. (This construction
has implicitly been used in \cite{26'} (see also \cite{25'}
where we have  constructed some examples
of GUST with gauge symmetry realized
as a diagonal subgroup of direct product
of two rank eight groups $U(8) \times U(8) \subset SO(16) \times SO(16)$.)

In strings, however, not all level-two representations can be obtained in that
way because, as we will demonstrate, some of them become massive (with masses
of order of the Planck scale). The condition ensuring that states in the string
spectrum transforming as a representation $R$ are massless reads:
\begin{equation}\label{eq4}
h(R) = \frac{Q_R}{2 k + Q_{ADJ}} =
\frac{Q_R}{2 Q_M} \leq 1 ,
\end{equation}
where $Q_i$ is the quadratic Casimir invariant of the corresponding
representations, and M has been already defined before (see eq. \ref{eq2}).
Here the conformal weight is defined by $L_0 |0{\rangle } = h(R)|0{\rangle }$,
\begin{eqnarray}\label{eq 100}
L_0 =\frac{1}
{2k + Q_{\psi}}\times  \biggl (\sum_{a=1}^{dim \,g}(T_0^aT_0^a +
2 \sum_{n=1}^{\infty}{T_{-n}^aT_n^a})\biggr),
\end{eqnarray}
where $T_n^a|0{\rangle }=0$ for $n>0$.
The condition (\ref {eq4}), when combined with (\ref{eq1}), gives a
restriction on the rank of
GUT's group  ($r \leq 8$), whose representations  can accommodate chiral matter
fields. For example, for antisymmetric representations of $SU(n=l+1)$ we have
the following values correspondingly : $h = p (n-p) /(2 n )$.
More exactly , for SU(8) group: $h(\underline 8)=7/16$,
$h(\underline{28})=3/4$, $h(\underline{56})=15/16$, $h(\underline {70})=1 $;
for SU(5), correspondingly $h(\underline 5)=2/5$ and $h(\underline{10})=3/5$;
for SU(3) group
$h(\underline 3)=1/3$ although  for adjoint representation of SU(3) -
$h(\underline 8)=3/4$;
for SU(2) doublet representation we have $h(\underline 2)=1/4$.
For vector representation of orthogonal series $D_l$- $h= 1/2$,
and, respectively, for spinor - $h(spinor)= l/8$.

 There are some another important cases. The values of conformal weights
for $ G =SO(16)$ or $E(6) \times SU(3)$, representations
$\underline{128}$, $(\underline{27}, \underline {3})$ ($h(\underline{128})= 1$,
$h(\underline{27}, \underline{3}) = 1$) respectively, satisfy both conditions.
Obviously, these (important for incorporation of chiral matter) representations
will exist in the level-two KMA of the symmetric subgroup of the group
$G\times G$.

In general, condition (\ref{eq4}) severely constrains  massless string states
transforming as $(R_G(x = 1), {R_G}^{'}(x = 1))$ of the direct product
$G\times G$. For example, for $SU(8)\times SU(8)$ and for $SU(5)\times SU(5)$
constructed from $SU(8)\times SU(8)$ only representations of the form
\begin{eqnarray}\label{eq5}
R_{N,N} =  \bigl((\underline {N},\underline {N}) +
h.c.\bigr ),\,\,\,\,\,
\bigl((\underline {N},\underline {\bar N}) + h.c.\bigr);\,\,\,
\end{eqnarray}
with $h(R_{N,N}) = (N-1)/N$, where $N\,=\,8$ or $5$ respectively can be
massless. For $SO(2N) \times SO(2N)$ massless states are contained
only in representations
\begin{eqnarray}\label{eq6}
R_{v,v} = (\underline{2N} ,\underline{2N})
\end{eqnarray}
with $h(R_{v,v}) = 1$. Thus, for the GUSTs based on a diagonal subgroup
$G^{symm}\subset G\times G$, $G^{symm}$ - high dimensional representations,
which are embedded in $R_G(x = 1) \times {R^{'}}_G(x = 1)$ are also severely
constrained by  condition (\ref{eq4}).

For spontaneous breaking of $G\times G$ gauge symmetry down to $G^{symm}$
(rank $G^{symm}$ = rank $G$) one can use the direct product of representations
$R_G(x=1)\times R_G(x=1)$, where $R_G(x = 1)$ is the fundamental representation
of $G = SU(N)$ or vector representation of $G = SO(2N)$. Furthemore,
$G^{symm}\subset G\times G$ can subsequently be broken down to a smaller
dimension gauge group (of the same rank as $G^{symm}$) through the VEVs of the
adjoint representations which can appear as a result of $G\times G$ breaking.
Alternatively, the real Higgs superfields (\ref{eq5}) or (\ref{eq6}) can
directly break the
$G\times G $ gauge symmetry down to a $G_1^{symm}\subset G^{symm}$ (rank
$G_1^{symm}\leq$ rank $G^{symm}$). For example when $G = SU(5)\times U(1)$ or
$SO(10)\times U(1)$,  $G\times G$ can directly be broken in this way down to
$SU(3^c)\times G^I_{EW}\times G^{II}_{EW}\times...$.

The above examples show clearly, that within the framework of GUSTs with the
KMA one can get interesting gauge symmetry breaking chains including the
realistic ones when $G\times G $ gauge symmetry group is considered.
However the lack of the higher dimension representations (which are forbidden
by \ref{eq4}) on the level-two KMA prevents the construction of the realistic
fermion mass matrices.
That is why we consider an extended grand unified string model of rank eight
$SO(16)$ or $E(6) \times SU(3)$ of $E(8)$.

The full chiral $SO(10)\times SU(3)\times U(1)$ matter multiplets can be
constructed from $SU(8)\times U(1)$--multiplets
\begin{eqnarray}\label{eq7}
(\underline {8} + \underline {56} + \underline {\bar{8}}
+ \underline {\bar{56}})~ = ~\underline {128}
\end{eqnarray}
of $SO(16)$. In the 4-dimensional heterotic superstring with free complex
world sheet fermions, in the spectrum of the Ramond sector there can appear
also
representations which are factors in the decomposition of
$\underline{128^{'}}$,
in particular, $SU(5)$-decouplets $(\underline {10} + \underline {\bar {10}})$
from $(\underline{28}+\underline{\bar{28}})$ of $SU(8)$. However their
$U(1)_5$ hypercharge does not allow to use
them for $SU(5)\times U(1)_5$--symmetry
breaking. Thus, in this approach we have only singlet and
$(\underline{5} + \underline{\bar5})$ Higgs fields which can break the grand
unified $SU(5)\times U(1)$ gauge symmetry. Therefore it is necessary (as we
already explained) to construct rank eight GUST based on a diagonal subgroup
$G^{symm}\subset G\times G$ primordial symmetry group, where in each
rank eight group $G$ the Higgs fields will appear only in singlets and in the
fundamental representations as in (see \ref{eq5}).

A comment concerning $U(1)$ factors can be made here. Since the available
$SU(5)\times U(1)$ decouplets have non-zero hypercharges with respect
to $U(1)_5$ and $U(1)_H$, these $U(1)$ factors may remain unbroken down to
the low energies in the model considered which seems to be very interesting.

\newpage

\section{ Modular Invariance in GUST Construction with
Non-Abelian Gauge Family Symmetry}\label{sec4}
\subsection{Spin-basis in free world-sheet fermion sector.}\label{sbsec41}

 The GUST model is completely defined by a set $\Xi$ of spin
boundary conditions for all these world-sheet fermions (see Appendix C).
In a diagonal basis the
vectors of $\Xi$ are determined by the values of phases $\alpha(f)$ $\in$(-1,1]
fermions $f$ acquire ~($f\longrightarrow -\exp({i\pi\alpha(f)}) f $) ~when
parallel transported around the string.
[5~
To construct the GUST according to the
scheme outlined at the end of the previous section we consider
three different bases each of them  with six elements
$B= {b_1, b_2, b_3, b_4 \equiv S, b_5, b_6 }$. (See Tables \ref{tabl1},
\ref{basis2}, \ref{tabl3}.)

Following \cite{19'} (see Appendix C) we  construct the canonical
basis in such a way that the
vector $\bar1$, which belongs to $\Xi$, is the first element $b_1$ of the
basis.
The basis vector $b_4=S$ is the generator of supersymmetry \cite{20'}
responsible for the conservation of the space-time $SUSY$.

In this chapter we have chosen a basis in which all left movers
$({\psi}_{\mu}; {\chi}_i, y_i,
{\omega}_i; i=1,...6)$ (on which the world sheet supersymmetry is realized
nonlinear\-ly) as well as 12 right movers $({\bar\varphi}_k; k=1,...12)$ are
real whereas (8 + 8) right movers $\bar{\Psi}_A$, $\bar{\Phi}_M$ are complex.
Such a construction corresponds to $SU(2)^6$ group of automorphisms of the
left supersymmetric sector of a string. Right- and left-moving real
fermions can be used for breaking $G^{comp}$ symmetry \cite{20'}. In
order to have a possibility to reduce the rank of the compactified group
$G^{comp}$, we have to select the spin boundary conditions for the maximal
possible number, $N_{LR}$ = 12, of left-moving, ${\chi}_{3,4,5,6}$,
$y_{1,2,5,6}$, ${\omega}_{1,2,3,4}$, and right-moving,
{}~${\bar{\phi}}^{1,...12}$~
(${\bar{\phi}}^p~=~{\bar{\varphi}}_p$,~$p=1,...12$) real fermions.
The KMA based on $16$ complex right moving fermions gives rise to the
"observable" gauge group $G^{obs}$ with:
\begin{equation}\label{eq8}
rank (G^{obs})  \leq 16.
\end{equation}

The study of the Hilbert spaces of the string theories
is connected to the problem of finding all possible choices
of the GSO coefficients ${\cal C}
\left[
\begin{array}{c}
{\alpha} \\
{\beta}
\end{array}\right]$
(see Appendix C),
such that the one--loop partition function
\begin{equation}
Z=\sum_{ \alpha , \beta} {\cal C}
\left[
\begin{array}{c}
{\alpha} \\
{\beta}
\end{array}\right] \prod_f Z
\left[
\begin{array}{c}
{\alpha}_f \\
{\beta}_f
\end{array}\right]
\label{Z}
\end{equation}
and its multiloop conterparts are all modular invariant.
In this formula ${\cal C}
\left[
\begin{array}{c}
{\alpha} \\
{\beta}
\end{array}\right]$ are GSO coefficients,
$\alpha$ and $\beta$ are $(k+l)$--component spin--vectors
$\alpha=[\alpha(f_1^r), ... , \alpha(f_k^r);
\alpha(f_1^c), ... , \alpha(f_l^c)]$,
the components $\alpha_f$, $\beta_f$ specify the spin
structure of the $f$th fermion and $Z[...]$ -- corresponding one-fermion
partition functions on torus: $Z[...]=\mbox{Tr exp}[2\pi iH_{(sect.)}]$.

The physical states in the Hilbert space of a given sector $\alpha$ are
obtained
acting on the vacuum ${|0{\rangle }}_{\alpha}$ with the bosonic and fermionic
operators
with frequencies
\begin{eqnarray}\label{eq9}
n(f) = 1/2 + 1/2 \alpha(f),\:\:\:   n(f^*) = 1/2  -1/2 \alpha(f^*)
\end{eqnarray}
and subsequently applying the generalized GSO projections. The physical states
satisfy the Virasoro condition:
\begin{eqnarray}\label{eq10}
M_L^2 = - 1/2 + 1/8 \:({\alpha}_L \cdot {\alpha}_L) + N_L =
-1 +1/8\: ({\alpha}_R \cdot {\alpha}_R) +N_R = M_R^2,
\end{eqnarray}
where $\alpha=({\alpha}_L,{\alpha}_R)$ is  a sector in the set $\Xi$,
$N_L = {\sum }_{L} (frequencies)$ and  $N_R = {\sum}_{R} (freq.)$.

We keep the same sign convention for the fermion number operator  $F$
as in \cite{20'}. For complex fermions we have $F_{\alpha}(f)~=~1$,
{}~$F_{\alpha}(f^*)~=~-1$ with the exception of the periodic fermions
for which we get $F_{{\alpha}=1}(f) ~=~-1/2(1-{\gamma}_{5f})$, where
{}~${\gamma}_{5f}|\Omega{\rangle }~=~|\Omega{\rangle }$,
{}~${\gamma}_{5f}b^+_o|\Omega{\rangle }~=~-b^+_o|\Omega{\rangle }$.

The full Hilbert space of the string theory is constructed as a direct sum of
different sectors ${\sum}_{i} {m_ib_i}$, ($m_i=0,1,..,N_i$), where the integers
$N_i$ define additive groups $Z(b_i)$ of the basis vectors  $b_i$.
The generalized GSO projection leaves in sectors $\alpha$ those states, whose
$b_i$-fermion number satisfies:
\begin{equation}\label{eq11}
\exp(i \pi b_i F_{\alpha})
= {\delta}_{\alpha}
{\cal C}^*
\left[
\begin{array}{c}
{\alpha} \\
b_i
\end{array}\right],
\end{equation}
where the space-time phase ${\delta}_{\alpha}=\exp(i\pi{\alpha}({\psi}_{\mu}))$
is equal $-1$ for the Ramond sector and +1 for the Neveu-Schwarz sector.

\subsection{$SU(5)\times U(1)\times SU(3)\times U(1)$- Model~1.}\label{sbsec42}
Model~1 is defined by 6 basis vectors given in Table \ref{tabl1} which
generates the $Z_2\times Z_4\times Z_2\times Z_2\times Z_8\times Z_2$
group under addition.

\begin{table}[h]
\caption{\bf Basis of the boundary conditions for all world-sheet fermions.
Model~1.}
\label{tabl1}
\footnotesize
\begin{center}
\begin{tabular}{|c||c|ccc||c|cc|}
\hline
Vectors &${\psi}_{1,2} $ & ${\chi}_{1,..,6}$ & ${y}_{1,...,6}$ &
${\omega}_{1,...,6}$
& ${\bar \varphi}_{1,...,12}$ &
${\Psi}_{1,...,8} $ &
${\Phi}_{1,...,8}$ \\ \hline
\hline
$b_1$ & $1 1 $ & $1 1 1 1 1 1$ &
$1 1 1 1 1 1 $ & $1 1 1 1 1 1$ &  $1^{12} $ & $1^8 $ & $1^8$ \\
$b_2$ & $1 1$ & $1 1 1 1 1 1$ &
$0 0 0 0 0 0$ &  $0 0 0 0 0 0 $ &  $ 0^{12} $ &
${1/2}^8 $ & $0^8$  \\
$b_3$ & $1 1$ & $ 1 1 1 1 0 0 $ & $0 0 0 0 1 1 $ & $ 0 0 0 0 0 0 $ &
$0^4  1^8 $ & $ 0^8 $ &  $ 1^8$ \\
$b_4 = S $ & $1 1$ &
$1 1 0 0 0 0 $ & $ 0 0 1 1 0 0 $ & $ 0 0 0 0 1 1 $ &
$ 0^{12} $ & $ 0^8 $ & $ 0^8 $ \\
$  b_5 $ & $ 1 1 $ & $ 0 0  1 1 0 0 $ &
$0 0 0 0 0 0 $ &  $1 1  0 0 1 1$ &  $ 1^{12} $ &
${ 1/4}^5  {-3/4}^3 $ & $ {-1/4}^5\ {3/4}^3  $  \\
$  b_6 $ & $ 1 1 $ & $ 1 1  0 0  0 0 $ &
$ 0 0  0 0 1 1 $ &  $0 0  1 1 0 0$ &  $ 1^2 0^4 1^6 $ &
$ 1^8  $ & $ 0^8  $  \\
\hline \hline
\end {tabular}
\end{center}
\normalsize
\end{table}

In our approach the basis vector $b_2$ is constructed as a complex vector
with the $1/2$ spin-boundary conditions for the right-moving fermions
${\Psi}_A$, $A = 1,...8$. Initially  it generates chiral  matter fields
in the $\underline{8}+\underline{56}+\underline{\bar{56}}+\underline{\bar{8}}$
representations of $SU(8)\times U(1)$, which subsequently are decomposed under
$SU(5)\times U(1)\times SU(3)\times U(1)$ to which $SU(8)\times U(1)$ gets
broken by applying the $b_5$ $GSO$ projection.

Generalized GSO projection coefficients are originally  defined up to fifteen
signs  but some of them are fixed by the supersymmetry conditions.
Below, in Table \ref{tabl2}, we present a set of numbers
$$
\gamma\left[\begin{array}{c}b_i\\b_j\end{array}\right]=\frac{1}{i \pi}
\log{\cal C}\left[\begin{array}{c}b_i\\b_j\end{array}\right].
$$
which we use as a basis for our GSO projections.

\begin{table}[h]
\caption{\bf The choice of the GSO basis $\gamma [b_i, b_j]$. Model~1.
($i$ numbers rows and $j$ -- columns)}
\label{tabl2}
\footnotesize
\begin{center}
\begin{tabular}{|c||c|c|c|c|c|c|}
\hline
& $b_1$ & $b_2$ & $b_3$ & $b_4$ & $b_5$ & $b_6$\\ \hline
\hline
$b_1$ & $0$ &    $1$ & $1$ & $1$ &    $1$ & $0$\\
$b_2$ & $1$ &  $1/2$ & $0$ & $0$ &  $1/4$ & $1$\\
$b_3$ & $1$ & $-1/2$ & $0$ & $0$ &  $1/2$ & $0$\\
$b_4$ & $1$ &    $1$ & $1$ & $1$ &    $1$ & $1$\\
$b_5$ & $0$ &    $1$ & $0$ & $0$ & $-1/2$ & $0$\\
$b_6$ & $0$ &    $0$ & $0$ & $0$ &    $1$ & $1$\\
\hline \hline
\end {tabular}
\end{center}
\normalsize
\end{table}

In our case of the ~${Z_2}^4\times {Z_4}\times {Z_8}$ ~model, we initially have
$256\times2$ sectors. After applying the GSO-projections we
get only $49\times2$ sectors containing massless states, which depending on the
vacuum energy values, $E^{vac}_L$ and $E^{vac}_R$, can be naturally divided
into some classes  and which determine the GUST representations.

Generally RNS (Ramond -- Neveu-Schwarz) sector (built on vectors $b_1$
and $S=b_4$) has high symmetry including $N=4$ supergravity and gauge
$SO(44)$ symmetry. Corresponding gauge bosons are constructed as follows:
\begin{eqnarray}\label{eq12}
&{\psi}_{1/2}^{\mu}{|0{\rangle }}_L \otimes  {\Psi}_{1/2}^I
 {\Psi}_{1/2}^{J} |0{\rangle }_R ,\nonumber\\
 &{\psi}_{1/2}^{\mu}{|0{\rangle }}_L \otimes  {\Psi}_{1/2}^I
 {\Psi}_{1/2}^{*J} |0{\rangle }_R ,\:\:I,~J~=1,~\dots,22 &
\end{eqnarray}
 While $U(1)_J$ charges for Cartan subgroups is given by formula
 $Y=\frac{\alpha}{2}+F$ (where $F$ --- fermion number, see (\ref{eq11})),
 it is obvious that states (\ref{eq12}) generate root lattice for
 $SO(44)$:
 \begin{eqnarray}
\pm \varepsilon_I \pm \varepsilon_J \ \ (I\neq J);\qquad
  \pm \varepsilon_I \mp \varepsilon_J
\end{eqnarray}
 The others vectors breakes $N=4$ SUSY to $N=1$ and gauge group $SO(44)$
 to $SO(2)^3_{1,2,3}\times SO(6)_4\times {\left[ SU(5)\times U(1)
 \times SU(3)_H\times U(1)_H\right]}^2$, see Figure \ref{fig1}.

 Generally, additional basis vectors can generate extra vector bosons and
 extend gauge group that remains after applying GSO-projection to
 RNS-sector. In our case dangerous sectors are: $2b_2+nb_5,\ n=0,2,4,6;
 \  2b_5;6b_5$. But our choice of GSO coefficients cancels all the vector
 states in these sectors. Thus gauge bosons in this model  appear
 only from RNS-sector.

\begin{table}[t]
\caption{\bf The list of quantum numbers of the states. Model~1.}
\label{tabl3'}
\footnotesize
\noindent \begin{tabular}{|c|c||c|cccc|cccc|} \hline
N$^o$ &$ b_1 , b_2 , b_3 , b_4 , b_5 , b_6 $&
$ SO_{hid}$&$ U(5)^I $&$ U(3)^I $&$ U(5)^{II} $&$
U(3)^{II} $&$ {\tilde Y}_5^I $&$ {\tilde Y}_3^I $&$ {\tilde Y}_5^{II} $&$
{\tilde Y}_3^{II}$ \\ \hline \hline
1 & RNS &&5&$\bar 3$&1&1&--1&--1&0&0 \\
  &     &&1&1&5&$\bar 3$&0&0&--1&--1 \\
  &0\ 2\ 0\ 1\ 2(6)\ 0&&5&1&5&1&--1&0&--1&0 \\
${\hat \Phi }$
  &&&1&3&1&3&0&1&0&1 \\
  &&&5&1&1&3&--1&0&0&1 \\
  &&&1&3&5&1&0&1&--1&0 \\ \hline \hline
2 &0\ 1\ 0\ 0\ 0\ 0&&1&3&1&1&5/2&--1/2&0&0 \\
  &&&$\bar 5$&3&1&1&--3/2&--1/2&0&0 \\
  &&&10&1&1&1&1/2&3/2&0&0 \\
${\hat \Psi }$
  &0\ 3\ 0\ 0\ 0\ 0&&1&1&1&1&5/2&3/2&0&0 \\
  &&&$\bar 5$&1&1&1&--3/2&3/2&0&0 \\
  &&&10&3&1&1&1/2&--1/2&0&0 \\ \hline
3 &0\ 0\ 1\ 1\ 3\ 0&$-_1\ \pm_2$&1&1&1&3&0&--3/2&0&--1/2 \\
  &0\ 0\ 1\ 1\ 7\ 0&$-_1\ \pm_2$&1&$\bar 3$&1&1&0&1/2&0&3/2 \\
${\hat \Psi }^H$
  &0\ 2\ 1\ 1\ 3\ 0&$+_1\ \pm_2$&1&$\bar 3$&1&3&0&1/2&0&--1/2 \\
  &0\ 2\ 1\ 1\ 7\ 0&$+_1\ \pm_2$&1&1&1&1&0&--3/2&0& 3/2 \\ \hline
4 &1\ 1\ 1\ 0\ 1\ 1&$\mp_1\ \pm_3$&1&1&1&$\bar 3$&0&--3/2&0&1/2 \\
  &1\ 1\ 1\ 0\ 5\ 1&$\mp_1\ \pm_3$&1&$\bar 3$&1&1&0&1/2&0&--3/2 \\
${\hat \Phi }^H$
  &1\ 3\ 1\ 0\ 1\ 1&$\pm_1\ \pm_3$&1&$\bar 3$&1&$\bar 3$&0&1/2&0&1/2 \\
  &1\ 3\ 1\ 0\ 5\ 1&$\pm_1\ \pm_3$&1&1&1&1&0&--3/2&0&--3/2 \\ \hline
5 &0\ 1(3)\ 1\ 0\ 2(6)\ 1&$-_1\ \pm_3$&1&3($\bar 3$)&1&1&$\pm$5/4&$\pm$1/4
&$\pm$5/4&$\mp$3/4 \\
  &&$+_1\ \pm_3$&5($\bar 5$)&1&1&1&$\pm$1/4&$\mp$3/4&$\pm$5/4&$\mp$3/4 \\
${\hat \phi }$
  &0\ 1(3)\ 1\ 0\ 4\ 1&$-_1\ \pm_3$&1&1&1&3($\bar 3$)&$\pm$5/4&$\mp$3/4
&$\pm$5/4&$\pm$1/4 \\
  &&$+_1\ \pm_3$&1&1&5($\bar 5$)&1&$\pm$5/4&$\mp$3/4&$\pm$1/4&$\mp$3/4 \\
\hline
6 &1\ 2\ 0\ 0\ 3(5)\ 1&$\pm_1\ -_4$&1&1&1&1&$\pm$5/4&$\pm$3/4
&$\mp$5/4&$\mp$3/4 \\
  &1\ 1(3)\ 0\ 1\ 5(3)\ 1&$+_1\ \mp_4$&1&1&1&1&$\pm$5/4&$\pm$3/4
&$\pm$5/4&$\pm$3/4 \\
${\hat \sigma }$
  &0\ 0\ 1\ 0\ 2(6)\ 0&$\mp_3\ +_4$&1&1&1&1&$\pm$5/4&$\mp$3/4
&$\pm$5/4&$\mp$3/4 \\ \hline
\end{tabular}
\normalsize
\end{table}

In NS sector the $b_3$ GSO projection leaves $(5,\bar{3})+(\bar{5},3)$ Higgs
superfields (see Figure \ref{fig2}):
\begin{equation}\label{eq14}
\chi^{1,2}_{1/2}|\Omega{\rangle }_L\otimes {\Psi}_{1/2}^a
{\Psi}_{1/2}^{i*};\,\, {\Psi}_{1/2}^{a*}
{\Psi}_{1/2}^{i} |\Omega{\rangle }_R\ \ \mbox{and exchange}\  \Psi
 \longrightarrow\Phi, \end{equation}
where $a,\:b=1,\dots,\:5,\ \ i,\:j=1,2,3$.

Four $(3_H + 1_H)$ generations of chiral matter fields from
$~({SU(5)\times SU(3)})_I$ group forming $SO(10)$--multiplets
$(\underline 1, \underline 3) + (\underline {\bar 5},\underline 3) +
(\underline {10}, \underline 3)$ ; $( \underline 1,\underline 1) +
(\underline {\bar 5},\underline 1) + (\underline { {10}},\underline 1)$
are contained in $b_2$ and $3b_2$ sectors. Applying $b_3$ $GSO$
projection to the $3b_2$ sector yields the following massless states:

\begin{eqnarray}\label{eq15}
b_{\psi_{12}}^+ b_{{\chi}_{34}}^{+} b_{{\chi}_{56}}^+
|\Omega {\rangle }_L& \otimes &  \Biggl \{   {\Psi}_{3/4}^{i*}~,~~
  {\Psi}_{1/4}^{a}   {\Psi}_{1/4}^{b}   {\Psi}_{1/4}^{c},~
  {\Psi}_{1/4}^{a}   {\Psi}_{1/4}^{i}   {\Psi}_{1/4}^{j}~
\Biggr \} ~|\Omega{\rangle }_R, \nonumber \\
b_{{\chi}_{12}}^{+} b_{{\chi}_{34}}^{+} b_{{\chi}_{56}}^{+}
|\Omega {\rangle }_L& \otimes &  \Biggl \{   {\Psi}_{3/4}^{a*}~,~~
  {\Psi}_{1/4}^{a}   {\Psi}_{1/4}^{b}   {\Psi}_{1/4}^{i},~~
  {\Psi}_{1/4}^{i}   {\Psi}_{1/4}^{j}   {\Psi}_{1/4}^{k}~
\Biggr \}~|\Omega{\rangle }_R
\end{eqnarray}
with the space-time chirality  ${\gamma}_{5 {\psi}_{12}}~=-~1$
and  ${\gamma}_{5 {\psi}_{12}}~=~1$, respectively.
In these formulae the Ramond creation operators
$b_{\psi_{1,2}}^+$ and $b_{{\chi}_{\alpha, \beta}}^+$ of the zero modes
are built of a pair of real fermions (as indicated by double indices):
${\chi}_{\alpha, \beta}$,~ $(\alpha,\beta)$~= ~$(1,2)$, ~$(3,4)$, ~$(5,6)$.
Here, as in (\ref{eq14}) indices take values $a,b$~=~1,...,5 ~and
$i,j$~=~1,2,3 respectively.

We stress that without using the ~$b_3$~ projection we would get matter
supermultiplets belonging to real representations only i.e. "mirror"
particles would remain in the spectrum. The ~$b_6$~ projection instead,
eliminates all chiral matter superfields from $U(8)^{II}$ group.
It is interesting, that without $b_6$-vector the Model~1 is
fictitious U(1)-anomaly \cite{anomaly} fully free.

Since the matter fields form the chiral multiplets of $SO(10)$, it is possible
to write down  $U(1)_{Y_5}$--hypercharges of massless states. In order to
construct the right electromagnetic charges for matter fields we must define
the hypercharges operators for the observable $U(8)^{I}$ group as

\begin{equation}\label{eq16}
Y_5=\int^\pi_0 d\sigma\sum_a \Psi^{*a}\Psi^a ,\,\,\,\,\,
Y_3=\int^\pi_0 d\sigma\sum_i \Psi^{*i}\Psi^i
\end{equation}
and analogously for the $U(8)^{II}$ group.

Then the orthogonal combinations
\begin{equation}\label{eq17}
 \tilde Y_5 = {1\over 4}(Y_5 + 5Y_3), \,\,\,\,\,
 \tilde Y_3 = {1\over 4}(Y_3 - 3Y_5),
\end{equation}
play the role of the hypercharge operators of $U(1)_{Y_5}$ and
$U(1)_{Y_H}$ groups,
respectively. In  Table \ref{tabl3'} we give the hypercharges
 $\tilde Y_5^{I},\tilde Y_3^{I}, \tilde Y_5^{II},\tilde Y_3^{II}$.

 The full list of states in this model is given in  Table \ref{tabl3'}.
 For fermion states only sectors with positive (left) chirality are
written. Superpartners arise from sectors with $S=b_4$-component
changed by 1. Chirality under hidden $SO(2)^3_{1,2,3}\times SO(6)_4$ is
defined as $\pm_1,\ \pm_2,\ \pm_3,\ \pm_4$ respectively. Lower signs in item
5 and 6 correspond to sectors with components given in brackets.

 In the next section we discuss the problem of rank
eight GUST gauge symmetry breaking.  The matter is that according to the
results of section \ref{sec2} the Higgs fields
$(\underline{10}_{1/2}+\underline{\bar{10}}_{-1/2})$ do not appear.

\subsection{ $SU(5)\times U(1)\times SU(3)\times U(1)$ -- Model~2.}
Consider then another ${\left[U(5)\times U(3)\right]}^2$ model which after
breaking gauge symmetry by Higgs mechanism leads to the spectrum similar
to Model~1.

This model is defined by basis vectors given in  Table \ref{basis2}
with the $Z^4_2\times Z_6\times Z_{12}$ group under addition.

\begin{table}[h]
\caption{ Basis of the boundary conditions for Model~2.}
\label{basis2}
\begin{center}
\begin{tabular}{|c||c|ccc||c|cc|}
\hline
Vectors &${\psi}_{1,2} $ & ${\chi}_{1,..,6}$ & ${y}_{1,...,6}$ &
${\omega}_{1,...,6}$
& ${\bar \varphi}_{1,...,12}$ &
${\Psi}_{1,...,8} $ &
${\Phi}_{1,...,8}$ \\ \hline
\hline
$b_1$      & $1 1$ & $1^6$   & $1^6$   & $1^6$   &   $1^{12}$       & $1^8$
     & $1^8$ \\
$b_2$      & $1 1$ &  $1^6$  & $0^6$   & $0^6$   &   $0^{12}$       & $1^5$
$1/3^3$  & $0^8$  \\
$b_3$      & $1 1$ & $1^2 0^2 1^2$ & $0^6$   & $0^2 1^2 0^2$ &  $0^8\  1^4 $ &
$1/2^5\ 1/6^3$ & $-1/2^5\ 1/6^3 $ \\
$b_4 = S $ & $1 1$ & $1^2\ 0^4$ & $0^2 1^2 0^2$ & $0^4\ 1^2$ &  $0^{12}$
& $ 0^8 $        & $ 0^8 $ \\
$b_5 $     & $1 1$ & $1^4\ 0^2$ & $0^4\ 1^2$ & $0^6$   &  $1^8\ 0^4$   & $1^5\
0^3$    & $ 0^5\ 1^3  $  \\
$b_6 $     & $1 1$ & $0^2 1^2 0^2$ & $1^2\ 0^4$ & $0^4\ 1^2$ &  $1^2 0^2 1^6
0^2$     & $ 1^8  $       & $ 0^8  $  \\
\hline \hline
\end {tabular}
\end{center}
\end{table}
GSO coefficients are given in Table \ref{GSO2}.

\begin{table}[h]
\caption{The choice of the GSO basis $\gamma [b_i, b_j]$. Model~2.
($i$ numbers rows and $j$ -- columns)}
\label{GSO2}
\begin{center}
\begin{tabular}{|c||c|c|c|c|c|c|}
\hline
& $b_1$ & $b_2$ & $b_3$ & $b_4$ & $b_5$ & $b_6$\\ \hline
\hline
$b_1$ & $0$ &    $1$         & $1/2$        & $0$ &  $0$ & $0$\\
$b_2$ & $0$ &  $2/3$ & $-1/6$       & $1$ &  $0$ & $1$\\
$b_3$ & $0$ & $1/3$          &$5/6$ & $1$ &  $0$ & $0$\\
$b_4$ & $0$ &    $0$         & $0$          & $0$ &  $0$ & $0$\\
$b_5$ & $0$ &    $1$         & $-1/2$       & $1$ &  $1$ & $1$\\
$b_6$ & $0$ &    $1$         & $1/2$        & $1$ &  $0$ & $1$\\
\hline \hline
\end {tabular}
\end{center}
\end{table}

The given model corresponds to the following chain of the gauge
symmetry breaking:
$$E^2_8\longrightarrow SO(16)^2\longrightarrow U(8)^2
\longrightarrow [U(5)\times U(3)]^2\ . $$
Here the breaking of $U(8)^2-$group to $[U(5)\times U(3)]^2$
is determined by basis vector $b_5$, and the breaking
of N=2 SUSY$\longrightarrow$N=1 SUSY
is determined by basis vector $b_6$.

It is interesting to note that in the abcense of vector $b_5$
$U(8)^2$ gauge group is restored by sectors $4b_3,\ 8b_3,\ 2b_2+c.c.$
and $4b_2+c.c.$

\begin{table}[p]
\caption{The list of quantum numbers of the states. Model~2.}
\label{sost2}
\noindent \begin{tabular}{|c|c||c|cccc|cccc|} \hline
N$^o$ &$ b_1 , b_2 , b_3 , b_4 , b_5 , b_6 $&$SO_{hid}$&$ U(5)^I $&$ U(3)^I $&$
U(5)^{II} $&$
U(3)^{II} $&$  Y_5^I $&$ Y_3^I $&$ Y_5^{II} $&$
Y_3^{II}$ \\ \hline \hline
1 & RNS & $6_1\ 2_2$ & 1&1&1&1&0&0&0&0 \\
  && $2_3\ 2_4$ & 1&1&1&1&0&0&0&0 \\
  &&& 5&1&$\bar 5$&1&1&0&--1&0 \\
  &0\ 0\ 4\ 1\ 0\ 0&&1&3&1&3&0&--1&0&--1 \\
  &0\ 0\ 8\ 1\ 0\ 0&&1&$\bar 3$&1&$\bar 3$&0&1&0&1 \\ \hline\hline
2 &0\ 1\ 0\ 0\ 0\ 0&&5&$\bar 3$&1&1&--3/2&--1/2&0&0 \\
  &&&1&$\bar 3$&1&1&5/2&--1/2&0&0 \\
  &0\ 3\ 0\ 0\ 0\ 0&&$\bar {10}$&1&1&1&1/2&3/2&0&0 \\ \hline
3 &0\ 1\ 10\ 0\ 0\ 0&&1&1&$\bar {10}$&3&0&0&1/2&1/2 \\
  &0\ 3\ 6\ 0\ 0\ 0&&1&1&5&1&0&0&--3/2&--3/2 \\
  &&&1&1&1&1&0&0&5/2&--3/2 \\ \hline
4 &0\ 2\ 3\ 0\ 0\ 0&$-_3\ \pm_4$&1&3&1&1&--5/4&--1/4&5/4&3/4 \\ \hline
5 &0\ 0\ 3\ 0\ 0\ 0&$+_3\ \pm_4$&1&1&$\bar 5$&1&--5/4&3/4&1/4&3/4 \\ \hline
6 &0\ 0\ 9\ 0\ 0\ 0&$+_3\ \pm_4$&1&1&5&1&5/4&--3/4&--1/4&--3/4 \\ \hline
7 &0\ 4\ 9\ 0\ 0\ 0&$-_3\ \pm_4$&1&$\bar 3$&1&1&5/4&1/4&--5/4&--3/4 \\ \hline
8,9 &0\ 5\ 0\ 1\ 0\ 1&$-_1\ \pm_3$&1&3&1&1&0&--1&0&0 \\
  &0\ 3\ 0\ 1\ 0\ 1&$+_1\ +_3$&5&1&1&1&1&0&0&0 \\
  &&$+_1\ -_3$&$\bar 5$&1&1&1&--1&0&0&0 \\
  &&$-_1\ +_3$&1&1&5&1&0&0&1&0 \\
  &&$-_1\ -_3$&1&1&$\bar 5$&1&0&0&--1&0 \\
  &0\ 5\ 8\ 1\ 0\ 1&$+_1\ +_3$&1&1&1&$\bar 3$&0&0&0&1 \\ \hline
10 &0\ 3\ 3\ 0\ 0\ 1&$+_1\ \pm_4$&1&1&1&1&--5/4&3/4&5/4&3/4 \\ \hline
11 &1\ 0\ 3\ 0\ 0\ 1&$\pm_2\ -_3$&1&1&5&1&--1/4&3/4&--5/4&--3/4 \\
   &1\ 2\ 11\ 0\ 0\ 1&$\pm_2\ -_3$&1&1&1&$\bar 3$&--5/4&3/4&--5/4&1/4 \\ \hline
12 &1\ 0\ 9\ 0\ 0\ 1&$\pm_2\ +_3$&$\bar 5$&1&1&1&1/4&--3/4&5/4&3/4 \\
   &1\ 4\ 9\ 0\ 0\ 1&$\pm_2\ +_3$&1&$\bar 3$&1&1&5/4&1/4&5/4&3/4 \\ \hline
13 &0\ 0\ 0\ 1\ 1\ 1&$\pm_2\ +_3$&1&1&1&1&0&--3/2&0&3/2 \\
   &0\ 2\ 0\ 1\ 1\ 1&$\pm_2\ -_3$&1&3&1&1&0&1/2&0&3/2 \\
   &0\ 2\ 8\ 1\ 1\ 1&$\pm_2\ -_3$&1&1&1&$\bar 3$&0&--3/2&0&--1/2 \\
   &0\ 4\ 8\ 1\ 1\ 1&$\pm_2\ +_3$&1&3&1&$\bar 3$&0&1/2&0&--1/2 \\
   &1\ 0\ 3\ 1\ 1\ 1&$+_1\ +_3$&1&1&1&1&5/4&3/4&--5/4&3/4 \\
   &1\ 0\ 9\ 1\ 1\ 1&$+_1\ +_3$&1&1&1&1&--5/4&--3/4&5/4&--3/4 \\
   &1\ 3\ 3\ 0\ 1\ 1&$-_1\ -_3$&1&1&1&1&--5/4&--3/4&--5/4&3/4 \\
   &1\ 3\ 9\ 0\ 1\ 1&$-_1\ +_3$&1&1&1&1&5/4&3/4&5/4&--3/4 \\ \hline
\end{tabular}
\end{table}

The full massless spectrum  for the given model is given in Table \ref{sost2}.
By analogy with Table \ref{tabl3'}
only fermion states with positive chirality
are written and obviously vector supermultiplets are absent.
Hypercharges are determined by formula:
$$ Y_n=\sum_{k=1}^{n}(\alpha_k/2 + F_k)\ . $$

The given model possesses  the hidden gauge symmetry
$SO(6)_1\times SO(2)^3_{2, 3, 4}$.
The corresponding chirality is given in column $SO_{hid.}$.
The sectors  are divided by horizontal lines and
without including the $b_5-$vector form $SU(8)-$multiplets.

For example, let us consider row No 2.
In sectors $b_2$, $5b_2$ in addition to states $(1, \bar{3})$ and
$(5, \bar{3})$ the (10, 3)--state appears, and in the sector $3b_2$
besides the $(\bar{10}, 1)-$ the states (1, 1) and $(\bar{5}, 1)$
survive too. All these states form $\bar{8}+56$ representation
of the $SU(8)^I$ group.

Analogically we can get the full structure of the theory according
to the $U(8)^I\times U(8)^{II}-$group.
(For correct restoration of the $SU(8)^{II}-$group we must invert
3 and $\bar{3}$ representations.)

In Model~2 matter fields appear both in $U(8)^I$ and $U(8)^{II}$ groups.
This is the main difference between this model and Model~1.
However, note that in the
Model~2 similary to the Model~1 all gauge fields appear in RNS--sector only
and $10 +\bar{10}$ representation (which can be the Higgs field
for gauge symmetry breaking) is absent.

\subsection{ $SO(10) \times SU(4)$ -- Model~3.}
 As an illustration we can consider the GUST construction involving $SO(10)$ as
GUT gauge group.  We consider the set  consisting of seven vectors $B=
{b_1, b_2, b_3, b_4 \equiv S, b_5, b_6,b_7 }$ given in Table \ref{tabl3}.

\begin{table}[h]
\caption{Basis of the boundary conditions for the Model~3.}
\label{tabl3}
\begin{center}
\begin{tabular}{|c||c|ccc||c|cc|}
\hline
Vectors &${\psi}_{1,2} $ & ${\chi}_{1,..,6}$ & ${y}_{1,...,6}$ &
${\omega}_{1,...,6}$
& ${\bar \varphi}_{1,...,12}$ &
${\Psi}_{1,...,8} $ &
${\Phi}_{1,...,8}$ \\ \hline
\hline
$b_1$ & $1 1 $ & $1 1 1 1 1 1$ &
$1 1 1 1 1 1 $ & $1 1 1 1 1 1$ &  $1^{12} $ & $1^8 $ & $1^8$ \\
$b_2$ & $1 1$ & $1 1 1 1 1 1$ &
$0 0 0 0 0 0$ &  $0 0 0 0 0 0 $ &  $ 0^{12} $ &
$1^5 {1/3}^3 $ & $0^8$  \\
$b_3$ & $1 1$ & $ 0 0 0 0 0 0 $ & $1 1 1 1 1 1 $ & $ 0 0 0 0 0 0 $ &
$0^8 1^4$ & $ 0^5 1^3 $ &  $ 0^5 1^3$ \\
$b_4 = S $ & $1 1$ &
$1 1 0 0 0 0 $ & $ 0 0 1 1 0 0 $ & $ 0 0 0 0 1 1 $ &
$ 0^{12} $ & $ 0^8 $ & $ 0^8 $ \\
$  b_5 $ & $ 1 1 $ & $ 1 1 1 1 1 1 $ &
$0 0 0 0 0 0 $ &  $0 0 0 0 0 0$ &  $ 0^{12} $ &
$ 0^8 $ & $ 1^5 {1/3}^3  $  \\
$  b_6 $ & $ 1 1 $ & $ 0 0  1 1  0 0 $ &
$ 1 1  0 0 0 0 $ &  $0 0  0 0 1 1$ &  $ 1^2 0^2 1^6 0^2 $ &
$ 1^8  $ & $ 0^8  $  \\
$  b_7 $ & $ 1 1 $ & $ 0 0  1 1  0 0 $ &
$ 1 0  0 0 0 0 $ &  $1 0  0 0 1 1$ &  $ 1^2 1 0 1^2 1^2 1 0 0^2 $ &
$ 0^8  $ & $ 1^8  $  \\
\hline \hline
\end {tabular}
\end{center}
\end{table}
GSO projections are given in Table \ref{tabl4}.

It is interesting to note that in this model the horizontal gauge symmetry
$U(3)$ extends to $SU(4)$. Vector bosons which needed for this appear
in sectors $2b_2\ (4b_2)$ and $2b_5\ (4b_5)$. For further breaking $SU(4)$
to $SU(3) \times U(1)$ we need an additional basis spin-vector.

Of course for getting a realistic model we must add some basis vectors
which give additional GSO--projections.\\
\begin{table}[h]
\caption{The choice of the GSO basis $\gamma [b_i, b_j]$. Model~3.
($i$ numbers rows and $j$ -- columns)}
\label{tabl4}
\begin{center}
\begin{tabular}{|c||c|c|c|c|c|c|c|}
\hline
& $b_1$ & $b_2$ & $b_3$ & $b_4$ & $b_5$ & $b_6$ & $b_7$\\ \hline
\hline
$b_1$ & $0$ &    $1$ & $0$ & $0$ &    $1$ & $0$ & $0$\\
$b_2$ & $0$ &  $2/3$ & $1$ & $1$ &    $1$ & $1$ & $0$\\
$b_3$ & $0$ &    $1$ & $0$ & $1$ &    $1$ & $1$ & $0$\\
$b_4$ & $0$ &    $0$ & $0$ & $0$ &    $0$ & $0$ & $0$\\
$b_5$ & $0$ &    $1$ & $1$ & $1$ &  $2/3$ & $0$ & $0$\\
$b_6$ & $0$ &    $1$ & $0$ & $1$ &    $1$ & $1$ & $0$ \\
$b_7$ & $0$ &    $1$ & $1$ & $1$ &    $0$ & $1$ & $1$ \\
\hline \hline
\end {tabular}
\end{center}
\end{table}

The spectrum of the Model 3 is the next:

1.there is the $[U(1)\times SO(6)]_{Hid.}\times [SO(10)\times SU(4)]^2$  gauge
group, the $U(1)$ group is anomaly free;

2. the matter fields,  $(16,\ 4;\ 1,\ 1)$,
 are from $3b_2$ and $5b_2$ sectors;

3. there are  Higgs fields from RNS-sectors -
$(\pm 1)_1 (6)_6(1,\ 1;\ 1,\ 1)$ , $(10,\ 1;\ 10,\ 1)$
and 2 total singlets;

4.there also are some  Higgs fields from $mb_2+nb_5$ sectors,
where $m, n =2- 4$ : $(1,\ 6;\ 1,\ 6)$;

5. another additional fields are $(-)_6(10,\ 1;\ 1,\ 1)$ ,
$(+)_6(1,\ 1;\ 10,\ 1)$ ,
$(-)_6(1,\ 1;\ 1,\ 6)$ , $(+)_6(1,\ 6;\ 1,\ 1)$  \\
$(\pm1/2)_1(1,\ \bar 4;\ 1,\ \bar 4)$ , $(\pm1/2)_1(1,\ \bar 4;\ 1,\ 4)$ ,
$2\times (+)_6(1,\ 1;\ 1,\ 1)$ , $(\pm 1)_1(-)_6(1,\ 1;\ 1,\ 1)$.

The condition of generation chirality in this model results in the choice
of Higgs fields as vector representations of SO(10)
 ($\underline{16}+\underline{\bar{16}}$ are absent). According to
conclusion (\ref{eq6}) the only Higgs fields $(\underline{10}, \underline{1};
\underline{10}, \underline{1})$ of $(SO(10)\times SU(4))^{\times 2}$
appear in the model (from RNS--sector) which can be used for
GUT gauge symmetry.

\newpage

\section{More explicit methods of model building. Self dual charge lattice.}
  In a previous models we had to guess how to obtain certain algebra
  representation and select boundary conditions vectors and GSO
coefficients basing only on basis building rules. Below we will develop
 some methods that help to build models for more complicated cases
 such as $E_6\times SU(3)$ and  $SU(3)\times SU(3)\times SU(3)\times SU(3)$.

As it is known,
square of a root represented by state in sector
$\alpha$ is $\sum_i (\alpha_i/2+F_i)$.

Consider then a mass condition. It reads ( for right mass only )
$$ M^2_R=-1+\frac{1}{8}(\alpha_R\cdot\alpha_R)+N_R=0,$$
In general we can write $n_f$ as
$$ n(f)=F^2\frac{1+F\alpha(f)}{2}=\frac{F^2}{2}+F\frac{\alpha}{2}$$
for any $F=0,\pm1$ ( $F^3=F$ for that values ).

Now $M^2_R$ formulae reads
$$ M^2_R=-1+\frac{1}{8}\sum^{22}_{i=1}(\alpha^2_i)+
\sum^{22}_{i=1}(\frac{F_i^2}{2}+F_i\frac{\alpha_i}{2}) $$
Hence
$$ 2=\sum^{22}_{i=1}(\frac{\alpha}{2}+F)^2 $$
Clearly it is the square of algebra root and it equals to 2 for any
massless state. Obviously for massive states normalization will differ
from that.

\subsection{Building GSO-projectors for a given algebra}
As we follow certain breaking chain of $E_8$ then it is very naturally
to take $E_8$ construction as a starting point. Note that root lattice
of $E_8$ arises from two sectors: NS sector gives {\underline{120}} of
$SO(16)$ while sector with $1^8$ gives {\underline{128}}   of $SO(16)$.
This corresponds to the following choice of simple roots
$$ \pi_1=-e_1+e_2 $$
$$ \pi_2=-e_2+e_3 $$
$$ \pi_3=-e_3+e_4 $$
$$ \pi_4=-e_4+e_5 $$
$$ \pi_5=-e_5+e_6 $$
$$ \pi_6=-e_6+e_7 $$
$$ \pi_7=-e_7+e_8 $$
$$ \pi_8=\frac{1}{2}(e_1+e_2+e_3+e_4+e_5-e_6-e_7-e_8)$$
Basing on this choice of roots it is very clear how to build basis
of simple roots
for any subalgebra of $E_8$. One can just find out appropriate
vectors $\pi_i$ of the form as in $E_8$ with needed scalar products
or build weight diagram and break it in a desirable fashion to find
roots corresponding to certain representation in terms of $E_8$
roots.

After the basis of simple roots is written down one can build
GSO-projectors in a following way.

GSO-projection is defined by operator $(b_i\cdot F)$ acting on given state.
The goal is to find those $b_i$ that allow only states from algebra
lattice to survive. Note that $F_i=\gamma_i-\alpha_i/2$ ($\gamma_i$ ---
components of a root in basis of $e_i$),
so value of GSO-projector for sector $\alpha$ depends on $\gamma_i$ only.
So, if scalar products of all simple roots that arise from a given sector
 with vector $b_i$ is equal mod 2 then they surely will survive
 GSO-projection. Taking several such vectors $b_i$ one can eliminate
 all extra states that do not belong to a given algebra.

Suppose that simple roots of the algebra are in the form
$$ \pi_i=\frac{1}{2}(\pm e_1\pm e_2\pm e_3\pm e_4\pm e_5\pm e_6\pm e_7\pm
e_8)$$
$$ \pi_j=(\pm e_k\pm e_m)$$
In this choice we have to find vectors $b$ which gives 0 or 1 in a scalar
 product with all simple roots. Note that
$(b\cdot\pi_i)=(b\cdot\pi_j)\ \mbox{mod}2$ for
 all $i,j$ so $c_i=(b\cdot\pi_i)$ either all equal 0 mod 2 or equal 1 mod 2.
 ( for $\pi_j=(\pm e_k\pm e_m)$ it should be 0 mod 2 because they are arise
   from NS sector )
 Value 0 or 1 is taken because if root $\pi \in $ algebra lattice then
 $-\pi$ is a root also. With such choice of simple roots and scalar products
 with $b$ all states from sector like $1^8$ will have the same
 projector value. Roots like $\pm e_i\pm e_j$ rise from NS sector
 and are sum of roots like
$ \pi_i=\frac{1}{2}(\pm e_1\pm e_2\pm e_3\pm e_4\pm e_5\pm e_6\pm e_7\pm e_8)$
and therefore have scalar products equal to 0 mod 2 as is needed for NS sector.

Now vectors $b$ are obtained very simple. Consider
$$ c_i=(b\cdot \pi_i)=b_jA_{ji} $$
where $A_{ji}=(\pi_i)_j$ --- matrix of roots component in $e_j$ basis.
Hence $b=A^{-1}\cdot c$ where either all $c_j=0 \bmod 2$   or $c_j=1 \bmod 2$.
One has to try some combination of $c_j$ to obtain appropriate set of $b$.
The next task is to combine those $b_i$ that satisfy modular invariance
rules and do not give extra states to the spectrum.

\subsection{Breaking given algebra using GSO-projectors}
 It appears that this method of constructing GSO-projectors allows
to break a given algebra down to its subalgebra.

Consider root system of a simple Lie algebra. It is well known that if
$\pi_1,\,\pi_2\in\Delta$, where $\Delta$ is a set of positive roots then
$(\pi_1-\pi_2(\pi_1\cdot\pi_2))\in\Delta$ also. For simply laced
algebras it means that if $\pi_i,\,\pi\in\Delta$ and $(\pi_i\cdot\pi)=-1$
where $\pi_i$ is a simple root then $\pi+\pi_i$ is a root also.
This rule is hold automatically in string construction: if a sector
gives some simple roots then  all roots of algebra and only them also exist
(but part of them may be found in another sector). Because square of every
root represented by a state is 2 then if $(\pi_i\cdot\pi)\neq-1$ then
$(\pi+\pi_i)^2\neq2$. So one must construct GSO-projectors checking
 only simple roots.
On the other hand if one cut out some of simple roots then algebra
will be broken. For example if a vector $b$ has non-integer scalar product
with simple root $\pi_1$ of $E_6$ then we will obtain algebra
$SO(10)\times U(1)$ ( $(b\cdot\pi_1)$ even could be 1 if others products
are equals 0 mod 2).

More complicated examples are $E_6\times SU(3)$ and
 $SU(3)\times SU(3)\times SU(3)\times SU(3)$. For the former we must
forbid the $\pi_2$ root but permit it to form $SU(3)$ algebra.
Note that in $E_8$ root system there are two roots with $3\pi_2$.
We will use them for $SU(3)$. So the product $(b\cdot\pi_2)$ must be 2/3
while  others must be  0 mod 2.

We can also get GSO-projectors for all interesting subgroups of
$E_8$ in such a way but so far choosing of constant for scalar products
( $c_i$ in a previous subsection ) is rather experimental so
it is more convenient to follow certain breaking chain.

Below we will give some results for $E_6\times SU(3)$,
 $SU(3)\times SU(3)\times SU(3)\times SU(3)$ and
 $SO(10)\times U(1)\times SU(3)$.
We will give algebra basis and vectors that give GSO-projection needed
for  obtaining this algebra.

$E_6\times SU(3)$. This case follow from $E_8$ using root basis from a
previous subsection and choosing
$$c_i=(-2,-\frac{2}{3}, 0,2,-2,-2,2,0)$$
This gives GSO-projector of the form
$$b_1=(1,1,\frac{1}{3},\frac{1}{3},\frac{1}{3},\frac{1}{3},
\frac{1}{3},\frac{1}{3})$$
Basis of simple roots arises from sector with $1^8$ in right part and reads
\begin{eqnarray}
 \pi_1&=&\frac{1}{2}(+ e_1+ e_2+ e_3 +e_4 +e_5 +e_6 +e_7 +e_8)\nonumber\\
 \pi_2&=&\frac{1}{2}( +e_1 +e_2 -e_3 -e_4 -e_5 -e_6 -e_7 -e_8)\nonumber\\
 \pi_3&=&\frac{1}{2}( +e_1 -e_2 -e_3 -e_4 +e_5 +e_6 -e_7 +e_8)\nonumber\\
 \pi_4&=&\frac{1}{2}(- e_1 +e_2 +e_3 -e_4 -e_5 +e_6 +e_7 -e_8)\nonumber\\
 \pi_5&=&\frac{1}{2}( +e_1 -e_2 +e_3 +e_4 +e_5 -e_6 -e_7 -e_8)\nonumber\\
 \pi_6&=&\frac{1}{2}( -e_1 +e_2 -e_3 +e_4 -e_5 +e_6 -e_7 +e_8)\nonumber\\
 \pi_7&=&\frac{1}{2}( +e_1 -e_2 +e_3 -e_4 -e_5 -e_6 +e_7 +e_8)\nonumber\\
 \pi_8&=&\frac{1}{2}( -e_1 +e_2 -e_3 -e_4 +e_5 -e_6 +e_7 +e_8)\nonumber\\
\end{eqnarray}

$SO(10)\times U(1)\times SU(3)$. This case follow from $E_6\times SU(3)$.
In addition to $b_1$ we must find a vector that cut out $\pi_3$. Using
$$c_i=(0,0,1,0,0,0,0,0)$$
and inverse matrix of $E_6\times SU(3)$ basis we get GSO-projector of the form
$$b_2=(0,0,\frac{1}{3},-\frac{2}{3},\frac{1}{3},\frac{1}{3},-\frac{2}{3},\frac{1}{3})$$
Basis of simple roots is the same as for $E_6\times SU(3)$ excluding $\pi_3$.

 $SU(3)\times SU(3)\times SU(3)\times SU(3)$.
Using  $E_6\times SU(3)$ basis inverse matrix with
$$c_i=(1,-1,-1,\frac{1}{3},1,\frac{1}{3},-1,-1)$$
We get GSO-projector of the form
$$b_2=(-\frac{1}{3},\frac{1}{3},1,1,\frac{1}{3},\frac{1}{3},-\frac{1}{3},-\frac{1}{3})$$
Easy to see that such a $c_i$ cut out $\pi_4$ and $\pi_6$ roots but due to
appropriate combination in $E_6$ root system two $SU(3)$ groups will remain.
Basis of simple roots is
\begin{eqnarray}
 \pi_1&=&\frac{1}{2}(+ e_1+ e_2+ e_3 +e_4 +e_5 +e_6 +e_7 +e_8)\nonumber\\
 \pi_2&=&\frac{1}{2}( +e_1 +e_2 -e_3 -e_4 -e_5 -e_6 -e_7 -e_8)\nonumber\\
 \pi_3&=&\frac{1}{2}( +e_1 -e_2 +e_3 -e_4 -e_5 -e_6 +e_7 +e_8)\nonumber\\
 \pi_4&=&\frac{1}{2}(- e_1 +e_2 +e_3 -e_4 +e_5 +e_6 -e_7 -e_8)\nonumber\\
 \pi_5&=&\frac{1}{2}( +e_1 -e_2 +e_3 +e_4 +e_5 -e_6 -e_7 -e_8)\nonumber\\
 \pi_6&=&\frac{1}{2}( -e_1 +e_2 -e_3 -e_4 +e_5 -e_6 +e_7 +e_8)\nonumber\\
 \pi_7&=&\frac{1}{2}( -e_1 +e_2 +e_3 +e_4 -e_5 -e_6 +e_7 -e_8)\nonumber\\
 \pi_8&=&\frac{1}{2}( +e_1 -e_2 -e_3 -e_4 +e_5 +e_6 +e_7 -e_8)\nonumber\\
\end{eqnarray}

Using all this methods we could construct a model described in the next
section.

\newcommand{\thr}{\frac{1}{3}}
\newcommand{\tthr}{\frac{2}{3}}
\subsection{ $E_6 \times SU(3)$ three generations model -- Model~4.}
 This model illustrates a branch of $E_8$ breaking
$E_8\rightarrow E_6\times SU(3)$ and is an interesting result on a way to
obtain three generations with gauge horizontal symmetry. Basis of the boundary
conditions (see Table \ref{basis4}) is rather simple
but there are some subtle points.
In \cite{Lop} the possible left parts of basis vectors were worked out,
see it for details. We just use the notation given in \cite{Lop}
(~hat on left part means complex fermion, other fermions on the left
sector are real, all of the right movers are complex)
 and an example of commuting set of vectors.

\begin{table}[h]
\caption{ Basis of the boundary conditions for the Model~4.}
\label{basis4}
\footnotesize
\begin{center}
\begin{tabular}{|c||c|cc||c|cc|}
\hline
Vectors &${\psi}_{1,2} $ & ${\chi}_{1,..,9}$ &
${\omega}_{1,...,9}$
& ${\bar \varphi}_{1,...,6}$ &
${\Psi}_{1,...,8} $ &
${\Phi}_{1,...,8}$ \\ \hline
\hline
$b_1$ & $1 1 $ & $1^9$ &
$1^9$ & $1^{6} $ & $1^8 $ & $1^8$ \\
$b_2$ & $1 1$ & $\widehat\thr,1;-\widehat\tthr,0,0,\widehat\tthr$ &
$\widehat\thr,1;-\widehat\tthr,0,0,\widehat\tthr$ &  $ \tthr^3\:-\tthr^3$ &
$0^2\:-\tthr^6 $ & $1^2\:\thr^6$  \\
$b_3$ & $0 0$ & $0^9$ &
$0^9$ & $ 0^6$ &
$1^8$&$0^8$  \\
$b_4$ & $1 1$ & $\widehat{1},1;\widehat0,0,0,\widehat0$ &
$\widehat1,1;\widehat0,0,0,\widehat0$ &$0^6$ & $ 0^8 $ & $0^8$ \\
\hline \hline
\end {tabular}
\end{center}
\normalsize
\end{table}

A construction of an $E_6\times SU(3)$ group caused us to use rational
for left boundary conditions. It seems that it is the only way to
obtain such a gauge group with appropriate matter contents.

The model has $N=2$ SUSY. We can also construct model with
$N=0$ but according to \cite{Lop} using vectors that can give rise
to $E_6\times SU(3)$ (with realistic matter fields)
one cannot obtain $N=1$ SUSY.

\begin{table}[h]
\caption{ The choice of the GSO basis $\gamma [b_i, b_j]$. Model~4.
($i$ numbers rows and $j$ -- columns).}
\label{GSO4}
\footnotesize
\begin{center}
\begin{tabular}{|c||c|c|c|c|}
\hline
& $b_1$ & $b_2$ & $b_3$ & $b_4$   \\ \hline
\hline
$b_1$ & $0$ &  $1/3$ & $1  $ & $1$ \\
$b_2$ & $1$ &  $  1$ & $1  $ & $1$ \\
$b_3$ & $1$ &    $1$ & $0  $ & $1$ \\
$b_4$ & $1$ &   $1/3$ & $1  $ & $1$ \\
\hline \hline
\end {tabular}
\end{center}
\normalsize
\end{table}

Let us give a brief review of the model contents. First notice
that all superpartners of states in sector $\alpha$ are found in
sector $\alpha+b_4$ as in all previous models. Although the same sector
may contain, say, matter fields and gauginos simultaneously.

The observable gauge group
 $(SU(3)^I_H\times E^I_6)\times (SU(3)^{II}_H\times E^{II}_6)$ and hidden
group $SU(6)\times U(1)$ are rising up from sectors NS, $b_3$ and $3b_2+b_4$.
Matter fields in representations $({\bf 3},{\bf 27})+
(\overline{\bf 3},\overline{\bf 27})$
for each $SU(3)_H\times E_6$ group are found in sectors $3b_2$, $b_3+b_4$
and $b_4$. Also there are some interesting states in sectors $b_2,\:b_2+b_3,\:
2b_2+b_3+b_4,\:2b_2+b_4$ and $5b_2,\:5b_2+b_3,\:4b_2+b_3+b_4,\:4b_2+b_4$
that form representations $(\overline{\bf 3},{\bf 3})$
and $({\bf 3},\overline{\bf 3})$
of the $SU(3)^I_H\times SU(3)^{II}_H$ group. This states are singlets
under both $E_6$ groups but have nonzero $U(1)_{hidden}$ charge.

We suppose that the model permits further breaking of $E_6$ down to other
grand unification groups, but problem with breaking supersymmetry ${N=2}
\rightarrow N=1$ is a great obstacle on this way.

\section{Self-duality of the  charge lattice and possible gauge groups.}
Here we present some results based on the important feature of the  charge
lattice that is self-duality.

As was shown in \cite{Kawai} the  charge lattice ${\bf Q}$ is an odd
self-dual lorentzian lattice shifted by a constant vector ${\bf S}$
constituted by 32-components vectors with components
$$
Q_i=\frac{\alpha_i}{2}+F_i,
$$
where $\alpha_i$ is a boundary condition for
$i$th fermion and $F_i$ is the  corresponding fermion number in a
particular string state. Vector $S$ takes care of the space-time
spin-statistic and in  the case of heterotic string is $(1,0,0,\dots,0)$.

As we will see below this feature apply serious restriction on the
possible gauge group and matter spectrum of the GUST. In this section we
will consider only models that permit bosonization which means that
we write all fermions in terms of complex fermions and consequently can
construct fermionic charge. Also we will restrict ourself considering only
models that have only periodic or antiperiodic boundary conditions for
left moving fermions (supersymmetric sector). Other possible forms of the
left sector can be treated by the similar way but our case is more
convenient in sense of building a $N=1$ SUSY model.

Before analyzing particular GUST with appropriate gauge group we will
consider some common features of a class of the lattices that we define
above. Firstly notice that since we take all $\alpha_i$ in the left sector
to be 0 or 1 then all of the scalar products of the left parts of the
lattice vectors will be $\frac{n}{4},\ \ n\in Z$. So the scalar product of
the right parts must has the same form in order to obtain the integer
scalar product.

Secondly with accounting shifting vector $S$ the states that represent
space-time gauge bosons will have all 0 in the left part ($\alpha_i=0,\
F_1=1,\ F_j=0,\ j>1$). So the whole scalar product of such a vector is
determined by the scalar product of the right part that must be integer
in this case.  On the other hand if we construct a right part that give an
integer scalar product with every vector then we will have a gauge boson
with this right part.

Thirdly let us consider the problem of chirality. Suppose we have a state
with right part ${\bf Q}_r$ giving integer or half-integer scalar product
with all other vectors right parts (left part scalar products must be
correspondingly integer or half-integer). Consider then a vector ${\bf
Q' }$ equals $Q$ with conjugated (i.e. multiplied by -1) right part.
Obviously ${\bf Q' }$ also will have integer scalar products with all
other vectors hence we finally have a space-time boson (scalar particle).
The same way it is easy to prove that weight of chiral
fermion must have a scalar
product of the right part with some other vector's right part
equals to $\frac{n}{2}+\frac{k}{4},\ k\neq 0
\bmod 4$ .

Concluding we have a classification of charge vectors in a sense of their
space-time type. If we have a set of vectors that correspond to the
content of a particular model (the vectors of the set must have integer
scalar product with each other) then we can easily distinguish gauge
bosons, chiral fermions and scalars by calculating scalar product of the
right part with every vector in the set. The following list present
classification.
\begin{itemize}
\item All of the right part scalar products are integer. Then we have
gauge boson with this charges in the spectrum.
\item All of the right part scalar products are integer or half-integer.
Then we have scalar particle with this quantum numbers in the spectrum.
\item Other vectors (that have right part scalar product
with some of the vectors in a set in a form
$\frac{n}{2}+\frac{k}{4},\ k\neq 0 \bmod 4$ ) represent chiral fermions.
\end{itemize}

Finally notice that the scalar products of the right part is the  scalar
products of the weight vectors of particular gauge group representation.
The structure of a representation is well known and for particular gauge
group we can determine the representation that give appropriate model
spectrum and necessary scalar products.

As a first simple example that illustrates all above discussion of this
section we consider a model with $E_6\times SU(3)$  gauge group. Notice
that both $E_6$ and $SU(3)$ groups have only representations with scalar
products of the weight vectors of the form $\frac{n}{3}$. If we wish to
obtain a model which  has representation for matter then we have to
include representation $({\bf 27,3})$ in the spectrum so that this states
are space-time chiral fermions. But to make this representation to be the
chiral fermions one has to include another weight vector  that give scalar
product of the right part equal to $\frac{k}{4},\ k\neq 0 \bmod 2$. Since
there is no such weight vector among weight vector of $E_6$ and $SU(3)$
representations then it is impossible to build a model with $E_6\times SU(3)$
gauge group and chiral fermions in the representations appropriate for
the GUST.
However if one take spin structure vector with $\frac23$ in the
left part then it will be possible to build a model but the model will
have $N=2$ space-time supersymmetry which means that there are no chiral
fermions.

Now we will present a more complicated example, namely the $SO(10)\times
U(3)$ gauge group. We demand that the model spectrum includes
space-time chiral fermions in the representation $({\bf 16,3})$ of  the
$SO(10)\times U(3)$ gauge group. Note that $U(1)$ charge of this
representation is defined up to the sign by the massless condition.

Thus we have a weight vector that we have to include in the lattice  in
addition to the root lattice of $SO(10)\times U(3)$. Then we can find all
the weights (actually the components of the weights that have
nonvanishing scalar products with root vector and weight of
matter representation and that correspond to the $SO(10)\times U(3)$
representation) that give appropriate scalar product with weight of matter
representation.

Actually one can obtain a formula that gives $U(1)$ hypercharge for
arbitrary representation of  $SO(10)\times U(3)$ so that scalar product of
this weight with weight of matter representation has the form $\frac{n}4$.
But it appears that representation $({\bf 1,3})$ has integer scalar
product with all other vectors that are allowed by weight of $({\bf
16,3})$. It means that the  representation  $({\bf 1,3})$ will be a
space-time gauge bosons that extend the initial $SO(10)\times U(3)$ gauge
group to the $SO(10)\times SU(4)$. Now we can see why all attempts to
build a model with $SO(10)\times U(3)$ gauge group and chiral matter in
$({\bf 16,3})$ representation failed. All models obviously must have
$SO(10)\times SU(4)$ gauge group with corresponding matter
representations.

\newpage

\section{GUST Spectrum (Model~1)}
\subsection{Gauge Symmetry Breaking}
Let us consider Model~1 in detail.
In Model~1 there exists a
possibility to break the GUST group $(U(5)\times U(3))^I\times (U(5)\times
U(3))^{II}$
down to the symmetric group by the ordinary Higgs mechanism \cite{Li'}:
\begin{equation}\label{eq18}
G^I\times G^{II} \rightarrow {G}^{symm}
\rightarrow ...
\end{equation}
To achieve such breaking one can use nonzero vacuum expectation values of the
tensor Higgs fields (see Table \ref{tabl3'}, row No 1),
contained in the $2b_2 + 2(6)b_5 (+S)$ sectors which transform
under the $(SU(5)\times U(1)\times SU(3)\times U(1))^{symm}$ group in the
following way:
\begin{eqnarray}\label{eq19}
\begin{array}{lll}
(\underline 5,\underline 1;\underline 5,\underline 1)_{(-1,0;-1,0)}
{}~&\rightarrow~~&  (\underline {24},\underline 1)_{(0,0)}~~+~~
(\underline 1,\underline 1)_{(0,0)}; \\
(\underline 1,\underline 3;\underline 1,\underline 3)_{(0,1;0,1)}\,
{}~&\rightarrow~~&
(\underline 1,\underline 8)_{(0,0)}~~+~~(\underline 1,\underline 1)_{(0,0)},
\end{array}
\end{eqnarray}

\begin{eqnarray}\label{eq20}
\begin{array}{lll}
(\underline 5,\underline 1;\underline 1,\underline 3)_{(-1,0;0,1)}
{}~&\rightarrow~~&
 (\underline {\bar 5},\underline 3)_{(1,1)}; \\
 (\underline 1,\underline 3;\underline 5,\underline 1)_{(0,1;-1,0)}
{}~&\rightarrow~~&
(\underline 5,\underline {\bar3})_{(-1,-1)}.
\end{array}
\end{eqnarray}

The diagonal vacuum expectation values for Higgs fields (\ref{eq19}) break
the GUST group $(U(5)\times U(3))^I \times (U(5)\times U(3))^{II}$ down to
the "skew"-symmetric group with the generators $\triangle_{symm}$ of the form:
\begin{equation}\label{eq21}
\triangle_{symm} (t) = -t^* \times 1 + 1\times t,
\end{equation}
The corresponding  hypercharge of the symmetric group reads:
\begin{equation}\label{eq22}
\bar Y = \tilde {Y}^{II} - \tilde {Y}^{I}.
\end{equation}
Similarily, for the electromagnetic charge we get:
\begin{eqnarray}\label{eq23}
Q_{em} &=& Q^{II} - Q^I = \nonumber\\
&=& (T^{II}_5 - T^{I}_5) + \frac{2}{5}(\tilde Y^{II}_5 - \tilde Y^{I}_5) =
\bar T_5 + \frac{2}{5}\bar Y_5,
\end{eqnarray}
where $ T_5 = diag (\frac{1}{15},\frac{1}{15},\frac{1}{15}, \frac{2}{5},
-\frac{3}{5})$. Note, that this charge quantization does not lead to exotic
states with fractional electromagnetic charges \\(e.g. $Q_{em} =\pm 1/2,
\pm 1/6$).

Thus, in breaking scheme (\ref{eq21}) it is possible to avoid colour singlet
states with fractional electromagnetic charges, to achieve
desired GUT breaking and moreover to get the usual value for the weak mixing
angle at the unification scale (see (\ref{sin})).

Adjoint  representations which appear on the $rhs$ of (\ref{eq19}) can be used
for
further breaking of the symmetric group. This can lead to the final physical
symmetry
\begin{equation}\label{eq24}
(SU(3^c)\times SU(2_{EW})\times U(1_Y)\times U(1)^{'})
\times (SU(3)_H\times U(1)_H)
\end{equation}
with  low-energy gauge symmetry of the quark -- lepton generations with an
additional $U(1)^{'}$--factor.

Note, that when we use the same  Higgs fields as in (\ref{eq19}),
there exists also
another interesting way of breaking the $G^I \times G^{II}$ gauge symmetry:
\begin{eqnarray}\label{eq25}
 G^I\times G^{II} \rightarrow SU(3^c) \times SU(2)^I_{EW}
\times SU(2)^{II}_{EW} \times U(1_{\bar Y}) \times \nonumber\\
\times SU(3)_H^I \times SU(3)_H^{II} \times U(1_{{\bar Y}_H})
\rightarrow ....
\end{eqnarray}

In turn, the Higgs fields ${\hat h}_{(\Gamma, N)}$ from the NS sector
\begin{eqnarray}\label{eq26}
(\underline 5, \underline {\bar 3})_{(-1,-1)} +
(\underline {\bar 5}, \underline 3)_{(1,1)}
\end{eqnarray}
are obtained from N=2 SUSY vector representation ~$\underline{63}$
of $SU(8)^{I}$ (or $SU(8)^{II}$) by applying the $b_5$ GSO projection
 (see Fig. \ref{fig2} and Appendix B,C).
These Higgs fields (and fields (\ref{eq20})) can be used for constructing
chiral fermion (see Table \ref{tabl3'}, row No 2) mass matrices.

The $b$ spin boundary conditions (Tabl.\ref{tabl1}) generate chiral matter and
Higgs
fields with the $GUST$ gauge symmetry $G_{comp}\times (G^I\times G^{II})_{obs}$
(where $G_{comp} = {U(1)}^3\times SO(6) $ and $G^{I,II}$ have been already
defined). The chiral matter spectrum, which we denote as
${\hat \Psi}_{(\Gamma, N)}$ ~~with ($\Gamma = \underline 1,\underline {\bar 5},
\underline{10}; ~~N=\underline3, \underline1$),~~consists of ~~
$N_g = 3_H + 1_H $~~ families. See Table \ref{tabl3'}, row No 2 for
the $((SU(5)\times U(1))\times (SU(3)\times U(1))_H)^{symm}$ quantum numbers.

The $SU(3)_H$ anomalies of the matter fields (row No 2) are
naturally canceled by
the chiral "horizontal" superfields  forming two sets:
${\hat \Psi}^H_{(1,N;1,N)}$ and ${\hat \Phi}^H_{(1, N;1, N)}$,
{}~$\Gamma = \underline 1$,~~$N = \underline 1, \,  \underline 3$,~
(with both ${SO(2)}$ chiralities, see Table \ref{tabl3'}, row No 3, 4
respectively).

The horizontal fields (No 3, 4) cancel all  $SU(3)^{I}$ anomalies
introduced by the chiral matter spectrum (No 2) of the  $(U(5)\times U(3))^{I}$
group (due to $b_6$ GSO  projection the chiral fields of
the $(U(5)\times U(3))^{II}$ group disappear from the final string spectrum).
Performing the decomposition of fields (No 3) under $(SU(5)\times
SU(3))^{symm}$
we get (among other) three "horizontal" fields ${\hat \Psi }^H$:
\begin{eqnarray}\label{eq27}
2\times (\underline 1,\underline {\bar 3})_{(0,-1)}\quad ,\qquad
(\underline 1,\underline 1)_{(0,-3)}\quad ,\qquad
(\underline 1,\underline {\bar 6})_{(0,1)}\quad ,
\end{eqnarray}
coming from
${\hat \Psi}^H_{(\underline1,\bar{\underline 3};\underline1,\underline1)}$,
(and
${\hat \Psi}^H_{(\underline1,\underline1;\underline1,\underline {3})}$),
${\hat \Psi}^H_{(\underline1,\underline1;\underline1,\underline1)}$ and
${\hat \Psi}^H_{(\underline1,\underline {\bar3};\underline1,\underline 3)}$
respectively  which
make the low energy spectrum of the resulting model (\ref{eq25})
${SU(3)_H}^{symm}$-
anomaly free. The other fields ${\hat \Phi }^H$ arising from
rows No 4, Table \ref{tabl3'}  form
anomaly-free representations of $(SU(3)_H \times U(1)_H)^{symm}$:
\begin{eqnarray}\label{eq28}
2\times (\underline 1,\underline 1)_{(0,0)}\quad ,\qquad
(\underline 1,\underline {\bar3})_{(0,2)}~+~
(\underline 1,\underline 3)_{(0,-2)}\quad ,\qquad
(\underline 1,\underline {8})_{(0,0)}\quad.
\end{eqnarray}

The  superfields ~${\hat \phi}_{(\Gamma, N)} + h.c.$, where
($\Gamma = \underline 1, \underline 5$; $N = \underline 1,\underline 3$), from
the Table \ref{tabl3'}, row No 5
forming
representations of $(U(5) \times U(3))^{I.II}$ have either $Q^I$ or $Q^{II}$
exotic fractional charges.
Because of the strong $G^{comp}$ gauge forces these fields may
develop the double scalar condensate $ {{\langle }\hat \phi \hat \phi{\rangle
}}$, which can also
serve for $U(5)\times U(5)$ gauge symmetry breaking. For example, the composite
condensate ${{\langle }{\hat \phi}_{(5,1;1,1)} {\hat \phi}_{(1,1;\bar
5,1)}{\rangle }}$ can
break the $U(5)\times U(5)$ gauge symmetry down to the symmetric diagonal
subgroup with generators of the form
\begin{eqnarray}\label{eq29}
{\triangle}_{symm} (t) = t \times 1 + 1 \times t,
\end{eqnarray}
so for the electromagnetic charges  we would have the form
\begin{eqnarray}\label{eq30}
Q_{em} = Q^{II} + Q^I.
\end{eqnarray}
leading again to no exotic, fractionally charged states
in the low-energy string spectrum.

The superfields which transform nontrivially under the compactified group
$G^{comp} = SO(6)\times {SO(2)}^{\times 3}$,~~(denoted as $\hat{\sigma}$),
and which are singlets of $(SU(5)\times SU(3))\times (SU(5)\times SU(3))$,
arise
in three sectors, see Table \ref{tabl3'}, row No 6.
The superfields $\hat\sigma$ form the spinor representations $\underline4+
\underline {\bar4}$ of $SO(6)$ and they are also spinors of one of the $SO(2)$
groups.
With respect to the diagonal $G^{symm}$ group with generators given by
(\ref{eq21}) or
(\ref{eq29}), some $\hat {\sigma}$-fields
are of zero hypercharges and can, therefore, be used for breaking the
$SO(6)\times {SO(2)}^{\times 3}$ group.

Note, that for the fields $\hat {\phi}$ and for the fields $\hat {\sigma}$
any other electromagnetic charge quantization different from (\ref{eq23}) or
(\ref{eq30}) would lead to "quarks" and "leptons" with the exotic fractional
charges, for example,
for the $\underline 5$- and $\underline 1$- multiplets according to the values
of hypercharges (see Table \ref{tabl3'}, row No 6) the generator $Q^{II}$  (or
$Q^{I}$) has the
eigenvalues \\
$(\pm 1/6,\pm 1/6,\pm 1/6,\pm 1/2,\mp1/2)$ or $\pm 1/2$, respectively.

Scheme of the breaking of the gauge group to the symmetric subgroup,
which is similar to the scheme of Model~1, works for Model~2 too.
In this case vector-like multiplets
$(\underline5,\ \underline1;\ \underline{\bar 5},\ \underline1)$
from RNS--sector and
$(\underline1,\ \underline3;\ \underline1,\ \underline3)$
from $4b_3$ $(8b_3)$ play the role of Higgs fields.
Then generators of the symmetric subgroup and electromagnetic
charges of particles are determined by formulas:
\begin{eqnarray}
\Delta^{(5)}_{sym}&=&t^{(5)}\times 1\ \oplus\ 1\times t^{(5)} \nonumber \\
\Delta^{(3)}_{sym}&=&(-t^{(3)})\times 1\ \oplus\ 1\times t^{(3)} \nonumber \\
Q_{em}=t^{(5)}_5-2/5\,Y^5&,&\mbox{where}
\ t^{(5)}_5=(1/15,\ 1/15,\ 1/15,\ 2/5,\ -3/5) \label{eq}
\end{eqnarray}

After this symmetry breaking matter fields (see Table \ref{sost2}
rows No 2, 3) as usual for flip models take place in representations
of the $U(5)-$group and form four generations
$(\underline1 +\underline5 +\underline{\bar{10}};\ \underline{\bar3}
+\underline1)_{sym}$.
And Higgs fields form adjoint representation of the symmetric group,
similar to Model~1, which is necessary for breaking of the gauge
group to the Standard group. Besides, due to quantization of the
electromagnetic charge according to the formula (\ref{eq})
 sates with exotic charges in lowenergy
spectrum also do not appear in this model.
In this model U(1)-group in hidden sector has anomalies which is broken by
Dine-Seiberg-Witten mechanism \cite{Dine'}. The corresponding D- term
could break supersymmetry at very large scale however  it is possible to
show that there exist D-flat directions with
respect to some non-anomalous gauge symmetries, canceling the anomalous D-
term and restoring supersymmetry. As result the corresponding VEVs imply that
the final symmetry will be less than origine, which includes the
$$(SU(5)\times U(1))^{sym}\times SU(3_H)\times U(1_H)^{sym}$$
 observable gauge symmetry. Let us note that
the models 5 and 3 do not contain the anomalous U(1)-groups.

\subsection{On the problem of states with exotic fractional charges.}

Almost all experimental data points to the fact that all particles we can
observe have only integer electromagnetic  charge. Quarks are assumed to
[5~have charges $\frac13$ and $\frac23$ but there are no indications that
there are particles with charges other than that.

Unfortunately  many string models include states with fractional charges
(e.g. $\frac12\, , \frac16$).
We consider electromagnetic charge $q$ of a particular state to be exotic if
$q\not\in \frac13\mbox{\bf Z}$.
Our models are free from these states due to
symmetric construction of the electromagnetic charge and interesting
features of GSO projection.  This statement holds for all states in the
model (not only massless).

1. Remind that we define electromagnetic charge as follows
$$ Q = T^{II}_5-T^{I}_5+\frac25({\tilde Y}^{II}_5-{\tilde Y}^{I}_5)
=T^{II}_5-T^{I}_5+\frac1{10}(Y^{II}_5-Y^{I}_5)
		 +\frac12(Y^{II}_3-Y^{I}_3),$$
$$ T_5=\lbrace \frac1{15},\frac1{15},\frac1{15},\frac25,-\frac35\rbrace$$
Rewrite $T_5$ in a following way
$$ T_5=\lbrace -1,-1,-1,-\frac23,-\frac53\rbrace+\frac{16}{15}Y_5
= t_5+\frac{16}{15}Y_5.$$
It is easy to see that eigenvalues of $t_5$ on all the states will be
proportional to $\frac13$, so $t_5$ does not contribute to fractional
charge.
Now we have
$$ Q=Q'+t^{II}_5-t^I_5,\mbox{ where }Q'=\frac76(Y^{II}_5-Y^{I}_5)
		 +\frac12(Y^{II}_3-Y^{I}_3),$$

Now note that electromagnetic charge is defined by scalar product of
vector $Q$ and weight $\Lambda$ of a state, $\Lambda_i=\frac{\alpha_i}2 +
F_i$. Otherwise GSO projection is defined by scalar product of a weight
vector and basis vector. So we can express the charge vector $Q$ via basis
vectors. The difference will be only in left part of scalar product and
in the hidden part of basis vectors. But as we will see this additional
contributions do not make fractional charge.

Denote ${\tilde b}_i$ is the part of basis vector that forms observable
group (for the Model 1 it is the last 16 components). Now we can rewrite
$Q'$ as
$$ Q'=-4{\tilde b}_2+{\tilde b}_1-\frac23{\tilde b}_5$$
The GSO projection reads
$$(b_i\cdot \Lambda)=\bar\delta_\alpha+ c\left[{b_i \atop \alpha}\right]
$$
where we take a logarithm of a usual expression and make corresponding
redefinitions.

Now it is obvious that all of the $Q$ eigenvalues will $\in \frac13${\bf
Z}. Indeed if we now express scalar product $(Q'\cdot\Lambda)$ via GSO
coefficients and remaining parts of basis vectors we will see that most of
them contribute integers (e.g. $ c\left[{b_1 \atop b_i}\right],\,
 4c\left[{b_2 \atop b_i}\right],\, (b^L_1\cdot\Lambda), \,
4(b^L_2\cdot\Lambda)$, etc.) while the others contribute $\frac13$ or
$\frac23$ (e.g.  $-\frac23(b^L_5\cdot\Lambda),\,
-\frac23(b^{hid}_5\cdot\Lambda)$, etc.).

We see that states with fractional charges are forbidden by GSO
projections on all mass levels. Using exactly the same method one cane
prove analogical statement for Model 2.

\subsection{Superpotential. Vertex operators. Nonrenormalizable terms.}

The ability of making a correct descrition of the fermion masses
and mixings will, of
course, constitute the decisive criterion for selection of a model of this
kind. Therefore, within our approach one has to
\begin{enumerate}
\item study the possible nature of the $G_H$ horizontal gauge symmetry
($N_g=3_H$ or $3_H+1_H$),
\item investigate the possible cases for
$G_H$-quantum numbers for quarks (anti-quarks) and leptons (anti-leptons),
i.e. whether one can obtain vector-like or axial-like structure (or even
chiral $G_{HL}\times G_{HR}$ structure) for the horizontal interactions.
\item find the structure of the sector of the matter fields which
are needed for
the $SU(3)_H$ anomaly cancelation (chiral neutral "horizontal" or "mirror"
fermions),
\item write out all possible renormalizable and relevant non-renormalizable
contributions to the superpotential $W$ and their consequences for fermion mass
matrices.
\end{enumerate}
All these questions are currently under investigation.
Here we restrict ourselves to some general remarks only.

With the chiral matter and "horizontal" Higgs fields available in Model~1
constructed in this paper, the possible form of the renormalizable (trilinear)
part of the superpotential responsible for fermion mass matrices is well
restricted by the gauge symmetry:
\begin{eqnarray}\label{eq32}
W_1&=& g\sqrt{2} \biggl[ {\hat \Psi}_{(1,3)}
{\hat \Psi}_{({\bar 5},1)}
{\hat h}_{(5,{\bar 3})} +
 {\hat \Psi}_{(1,1)} {\hat \Psi}_{({\bar 5},3)}
{\hat h}_{(5,{\bar 3})} + \nonumber\\
&+& {\hat \Psi}_{(10,3)} {\hat \Psi}_{({\bar 5},3)}
{\hat h}_{({\bar5},3)} +
{\hat \Psi}_{(10,3)} {\hat \Psi}_{(10,1)}
{\hat h}_{(5,{\bar 3})} \biggr]
\end{eqnarray}

Another strong constraint, which we used for construction superpotential,
 comes from the interesting observation that a modular invariant,
N=1 space-time supersymmetric theory also extends to a global N=2
world sheet superconformal symmetry  \cite{bank'}, which contents
now two distinct fermionic components to the energy- momentum tensor,
 $T_F^+$ and  $T_F^-$ \cite {20'}, and
there is also the  $U_J(1)$ current $J$. This conserved $U(1)$ current of
this N=2 superalgebra may play a key role in
constructing of realistic phenomenology. So, all vertex operators
have the definite $U(1)$ charge.
For $J(z)$ we have
\begin{eqnarray}
J(z) \,=\,i \partial _z \,(S_{12}\,+\,S_{34}\,+\,S_{56}\,),
\end{eqnarray}
where $S_{ij}$ are the bosonized components of superspin generator
$T_F(z)$ \cite {20'}.
Let,s consider on an example in  Models 1,5 the three point
fermionic- fermionic - bosonic function for the case
 $$\Psi^1_{(1,3)} \Psi^2_{(\bar 5 ,1)} h^3_{(5,\bar 3 )}, $$
 where the fields 1 and 2 are from Ramond-sector and 3 is from NS -sector.
In canonical picture  for the fermionic (bosonic) vertex operator
the $U_J(1)$ charge equal to $-1/2$($+1$).
In  boson sector in $\Psi^1_{(1,3)}$,
$ \Psi^2_{(\bar 5 ,1)}$  the left Ramond fermions  are with the numbers
3, 4, 6, 10. The field $ h^3_{(5,\bar 3 )} $ from  NS-sector
has in boson sector the exitation by world sheet left fermion No 2.
The nonvanishing $U_J(1)$ \cite {20'} of these three vertex operators
are {: $\beta_1=\gamma_1=\beta_2=\gamma_2=1/2$
and $\alpha_3=1$.
So for the corresponding vertex operators one can obtain :

\begin{equation}
V^f_{1(-1/2)}=e^{-c/2}S_{\alpha}e^{-i/2\ H_2}\Sigma^{cc}_3
\Sigma^{cc}_4 e^{i/2\ K_1X_1} \bar G_1 e^{i/2\ K_1\bar X_1}
\end{equation}
\begin{equation}
V^f_{2(-1/2)}=e^{-c/2}S_{\alpha}e^{-i/2\ H_2}\Sigma^{cc}_3
\Sigma^{cc}_4 e^{i/2\ K_2X_2} \bar G_2 e^{i/2\ K_2\bar X_2}
\end{equation}
\begin{equation}
V^b_{3(-1)}=e^{-c}e^{i H_2} e^{i/2\ K_3X_3} \bar G_3 e^{i/2\ K_3\bar X_3}
\end{equation}
We see that the correlator of these   vertex operators is not equal to zero
 (for details see Kalara et al in \cite{20'}).
Following to the paper \cite{20'} one can easily  get
that the corresponding coefficient including to the superpotential is
equal to $g\sqrt{2}$.

{}From the above form of the Yukawa couplings
follows that two (chiral) generations
have to be very light (comparing to $M_W$ scale).
The construction of realistic quarks and leptons mass matrices depends, of
course, on the nature of the horizontal interactions.
To construct the realistic fermion mass matrices one has to also use
Higgs fields  (\ref{eq19}, \ref{eq20}) and (Table \ref{tabl3'}, No 5) and
also to take into account
all relevant non-renormalizable contributions \cite{f,d,20'}.

Higgs fields (\ref{eq19}) can be used for constructing Yukawa couplings
of the horizontal superfields (No 3 and 4). The
superpotential, $W_2$, consists of the next $R^2 NS$-terms:
\begin{eqnarray}\label{eq34}
W_2&=& g\sqrt{2} \biggl[
{\hat \Phi}^H_{(1,1;1,\bar3)}{\hat \Phi}^H_{(1,\bar 3;1,1)}
{\hat \Phi}_{(1,3;1,3)} +
{\hat \Phi}^H_{(1,1;1,1)}{\hat \Phi}^H_{(1,\bar 3;1,\bar 3)}
{\hat \Phi}_{(1,3;1,3)} + \nonumber\\
&+& {\hat \Phi}^H_{(1,\bar 3;1,\bar 3)}{\hat \Phi}^H_{(1,\bar 3;1,\bar 3)}
{\hat \Phi}_{(1,\bar 3;1,\bar 3)} +
 {\hat \Psi}^H_{(1,{\bar 3};1,1)}{\hat \Psi}^H_{(1,{\bar 3};1, 3)}
{\hat \Phi}_{(1,{\bar 3};1,{\bar 3})} +\nonumber\\
&+&{\hat \Psi}^H_{(1,1;1,3)}
{\hat \Psi}^H_{(1,{\bar 3};1,3)}
{\hat \Phi}_{(1, 3;1, 3)}
+{\hat \sigma }_{(-_1 -_4 )} {\hat \sigma }_{(+_1 +_4 )}
{\hat \Psi }_{(1,1;1,1)} \biggr]
\end{eqnarray}

{}From (\ref{eq34}) it follows that some of the horizontal fields in
(\ref{eq28}) (No 3, 4) remain massless at the tree-level.
This is a remarkable prediction: fields (\ref{eq28}) interact with the ordinary
chiral
matter fields only through the $U(1)_H$ and $SU(3)_H$
gauge boson and therefore are very interesting in the context of the
experimental searches for the new gauge bosons.

\bigskip
The Higgs fields  (see table 3, 13) could give the following
$(NS)^3$ contributions to the renormalizable superpotential:
\begin{eqnarray}
W_3=\sqrt{2}g\biggl\{
{\hat\Phi }_{ (5,1;1,3)}  {\hat\Phi }_{ (\bar 5,1;\bar 5,1)}
{\hat\Phi }_{ (1,1;5,\bar 3)} +
{\hat\Phi }_{ (5,1;1,3)}  {\hat\Phi }_{ (1,\bar 3;1,\bar 3)}
{\hat\Phi }_{ (\bar 5,3;1,1)} && \nonumber\\
+ {\hat\Phi }_{ (1,3;5,1)}  {\hat\Phi }_{ (\bar 5,1;\bar 5,1)}
{\hat\Phi }_{ (5,\bar 3;1,1)} +
{\hat\Phi }_{ (1,3;5,1)}  {\hat\Phi }_{ (1,\bar 3;1,\bar 3)}
{\hat\Phi }_{ (1,1;\bar 5,3)} +\mbox{conj.}
\biggr\}&&
\end{eqnarray}

So, $W_1 +W_2 +W_3$ is the most general renormalizable superpotential
which includes all non-zero three-point (F-type) expectation values
of the vertex operators for corresponding 2-dimensional conformal
model.
\bigskip

Now we consider the non-renormalizable four-point vertex operator of the form
$(R)^4$, which is the only nonvanishing type of N=4- point operators
(see the definitions in paper Kalara et al \cite {20'}).
We take that
$\frac{1}{\sqrt{2}}(\chi_n \pm i\chi_{n+1})=e^{\pm iH_k}$ ,
where $k= 1 +(n+1)/2$ ;
$H_{k'}$ and $H_{\bar k'}$ -- similarly   for $y_n$ and $\omega_n$.
Following to \cite {20'} the multipliers  for $\chi$ have been written
explicitely and
 for $y (\omega)$ we have also:
$\Sigma^{\pm}_{k'}\equiv e^{\pm\frac{i}{2}H_{k'}}$.

We need to calculate the correlator:
$\langle V^f_{1(-1/2)} V^f_{2(-1/2)} V^b_{3(-1)} V^b_{4(0)}\rangle $.
The 4-th vertex have to be written in noncanonical form (in picture 0).
So for R-case we need change  vertex $V^b_R$ from picture --1 to picture 0.
According to \cite {20'} the formula of  changing pictures
from  $q$ to $q+1$ is:
\begin{equation}
V_{q+1}(z)=\lim_{w\to z} e^c(w) T_F(w) V_q(z) \label{lim}
\end{equation}
The contribution is going only from  $T_F^{-1}$ and for complex case
one can obtain:
\begin{equation}
T_F^{-1}=\frac{i}{2\sqrt{2}} \sum_k e^{-iH_k}
\biggl[ (1-i)e^{iH_{k'}} e^{iH_{\bar k'}}
+(1+i)e^{iH_{k'}} e^{-iH_{\bar k'}} +(1+i)e^{-iH_{k'}} e^{iH_{\bar k'}}
+(1-i)e^{-iH_{k'}} e^{-iH_{\bar k'}} \biggr]
\end{equation}

Making use  (\ref{lim}) :
$$V^b_{R(-1)}=e^{-c}e^{i/2\ H_k}e^{i/2\ H_l}
\Sigma^{\pm}_{k'}\Sigma^{\pm}_{l'}
e^{i/2\ KX} \bar G e^{i/2\ K\bar X}$$
one can get:
\begin{eqnarray}
&&V^b_{R(0)}=\frac{i}{2\sqrt{2}}\biggl\{ e^{-i/2\ H_k} e^{i/2\ H_l}
\biggl( \Sigma_{k'}^{\mp} [(1\pm i)e^{iH_{\bar k'}}
+ (1\mp i)e^{-iH_{\bar k'}}]\biggr) \Sigma^{\pm}_{l'} \nonumber\\
&&+ e^{i/2\ H_k} e^{-i/2\ H_l} \Sigma^{\pm}_{k'}
\biggl( \Sigma_{l'}^{\mp} [(1\pm i)e^{iH_{\bar l'}}
+ (1\mp i)e^{-iH_{\bar l'}}]\biggr)  \biggr\}
\times e^{i/2\ KX} \bar G e^{i/2\ K\bar X}
\end{eqnarray}
Another vertex correllator constructions of $R^4$ one can see in \cite {20'}.
Now we can lead to the result of our study of the  $R^4$ term in model 1:

\begin{eqnarray}
(1,3,1,1)_{Mat.} & \sim e^{-i/2\ H_2}\Sigma^+_3\Sigma^+_4
 &  (f.-1/2) \nonumber\\
(+_1,-_3)(1,\bar 3,1,1)_{No 4} & \sim e^{-i/2\ H_6}\Sigma^+_5\Sigma^+_7
 &  (f.-1/2) \nonumber\\
(-_1,-_4)_{No 6} & \sim e^{i/2\ H_2}e^{i/2\ H_{10}}\Sigma^+_4\Sigma^-_5
 &  (b.-1) \nonumber\\
(+_3, +_4)_{No 6} & \sim e^{-i/2\ H_6}e^{i/2\ H_{10}}\Sigma^+_3\Sigma^+_7
[(1-i)e^{iH_9} + (1+i)e^{-iH_9}] & \nonumber\\
& + \underline{e^{i/2\ H_6}e^{-i/2\ H_{10}}\Sigma^-_3\Sigma^-_7}
[(1+i)e^{iH_4} + \underline{(1-i)e^{-iH_4}}] & (b.0) \nonumber
\end{eqnarray}

All others terms of $R^4$ contain the  correlators like,
which are equal to zero:
$\langle e^{-i/2\ H_k} e^{-i/2\ H_k} \rangle \equiv 0$.
One can see that in model 1 there is only one nonvanishing term:

\begin{eqnarray}
W_4= {\hat\Psi}_{(1,3;1,1)}  {\hat\Phi}^H_{2\ (+,-)(1,\bar 3;1,1)}
{\hat\sigma}_{1\ (-,-)} {\hat\sigma}_{3\ (+,+)}.
\end{eqnarray}
Calculable coefficient from $W_4$  have been omitted.

\bigskip
For consideration of the
five-point non-renormalizable function we may take into account only
the following candidates of the  $(NS)\times (R)^4$
vertex configurations:
\begin{eqnarray}
W_5=&&{\hat\Phi }_{ (5,1;1,3)} {\hat\Psi }_{(10,3;1,1)}
{\hat\Psi }_{(10,1;1,1)}
\times [{\hat\Psi }^H_{ (1,1;1,3)} {\hat\Psi }^H_{ (1,\bar 3;1,3)}\
\nonumber\\
+&& {\hat\Phi }^H_{ (1,1;1,\bar 3)} {\hat\Phi }_{ (1,\bar 3;1,1)}
+ {\hat\Phi }^H_{ (1,\bar 3;1,\bar 3)} {\hat\Phi }^H_{ (1,1;1,1)}]
\end{eqnarray}
The left N=2 superconformal invariance demands that these terms are
equal to zero.

Finally, we remark that the Higgs sector of our GUST allows conservation of
the $G_H$ gauge family symmetry down to the low energies ($\sim {\cal O}(1TeV)$
\cite{9'}). Thus in this energy region we can expect new interesting physics
(new gauge bosons, new chiral matter fermions, superweak-like CP--violation
in $K$,- $B$,- $D$-meson decays with ${\delta}_{KM} < {10}^{-4}$ \cite{9'}).

\begin{table}[h']
\caption{\bf Basis of the boundary conditions for all world-sheet fermions.
Model~5.}
\label{bas5}
\footnotesize
\begin{center}
\begin{tabular}{|c||c|ccc||c|cc|}
\hline
Vectors &${\psi}_{1,2} $ & ${\chi}_{1,..,6}$ & ${y}_{1,...,6}$ &
${\omega}_{1,...,6}$
& ${\bar \varphi}_{1,...,12}$ &
${\Psi}_{1,...,8} $ &
${\Phi}_{1,...,8}$ \\ \hline
\hline
$b_1$ & $1 1 $ & $1 1 1 1 1 1$ &
$1 1 1 1 1 1 $ & $1 1 1 1 1 1$ &  $1^{12} $ & ${\hat 1}^8 $ & ${\hat 1}^8$ \\
$b_2 = S $ & $1 1$ &
$1 1 0 0 0 0 $ & $ 0 0 1 1 0 0 $ & $ 0 0 0 0 1 1 $ &
$ 0^{12} $ & $ {\hat 0}^8 $ & $ {\hat 0}^8 $ \\
$b_3$ & $0 0$ & $ 0 0 1 1 1 1 $ & $0 0 0 0 1 1 $ & $ 0 0 1 1 0 0 $ &
$1^8  0^4 $ & $ {\hat 1}^8 $ &  $ {\hat 0}^8$ \\
$b_4$ & $1 1$ & $1 1 1 1 1 1$ &
$0 0 0 0 0 0$ &  $0 0 0 0 0 0 $ &  $ 0^{12} $ &
${\hat{1/2}}^8 $ & ${\hat 0}^8$  \\
$  b_5 $ & $ 1 1 $ & $ 0 0  1 1 0 0 $ &
$0 0 0 0 0 0 $ &  $1 1  0 0 1 1$ &  $ 1^{6} 0^2 1^2 0^2 $ &
${\hat{ 1/4}}^5  {\hat{-3/4}}^3 $ & $ {\hat{-1/4}}^5\ {\hat{3/4}}^3  $  \\
$  b_6 $ & $ 0 0 $ & $ 0 0 1 1 0 0 $ &
$ 1 0  0 0 0 0 $ &  $1 0  1 1 0 0$ &  $ 1^4 0^4 1 0 1 0 $ &
$ {\hat 1}^8  $ & $ {\hat 1}^8  $  \\
\hline \hline
\end {tabular}
\end{center}
\normalsize
\end{table}

\begin{table}[h']
\caption{\bf The choice of the GSO basis $\gamma [b_i, b_j]$. Model~5.
($i$ numbers rows and $j$ -- columns)}
\label{gso5}
\footnotesize
\begin{center}
\begin{tabular}{|c||c|c|c|c|c|c|}
\hline
& $b_1$ & $b_2$ & $b_3$ & $b_4$ & $b_5$ & $b_6$\\ \hline
\hline
$b_1$ & $0$ &    $1$ & $0$ & $1$ &    $0$ & $1$\\
$b_2$ & $1$ &    $1$ & $1$ & $1$ &    $1$ & $1$\\
$b_3$ & $0$ &    $1$ & $1$ & $1$ &  $-1/2$ & $1$\\
$b_4$ & $1$ &    $0$ & $0$ &$1/2$&  $1/4$ & $0$\\
$b_5$ & $0$ &    $0$ & $0$ & $1$ &    $0$ & $1$\\
$b_6$ & $1$ &    $1$ & $1$ &$1/2$&  $1/2$ & $0$\\
\hline \hline
\end {tabular}
\end{center}
\normalsize
\end{table}

\begin{table}[t']
\caption{\bf The list of quantum numbers of the states. Model~5.}
\label{spectr5}
\footnotesize
\noindent \begin{tabular}{|c|c||c|cccc|cccc|} \hline
N$^o$ &$ b_1 , b_2 , b_3 , b_4 , b_5 , b_6 $&
$ SO(4)\times U(1)^2_{1,2}$&$ U(5)^I $&$ U(3)^I $&$ U(5)^{II} $&$
U(3)^{II} $&$ {\tilde Y}_5^I $&$ {\tilde Y}_3^I $&$ {\tilde Y}_5^{II} $&$
{\tilde Y}_3^{II}$ \\ \hline \hline
1 & RNS &&5&$\bar 3$&1&1&--1&--1&0&0 \\
  &     &&1&1&5&$\bar 3$&0&0&--1&--1 \\
  &     &$+1_1\ +1_2$&1&1&1&1&0&0&0&0 \\
  &     &$+1_1\ -1_2$&1&1&1&1&0&0&0&0 \\
  &     &&1&1&1&1&0&0&0&0 \\
  &     &&1&1&1&1&0&0&0&0 \\
  &0\ 0\ 0\ 2\ 2(6)\ 0&&5&1&5&1&--1&0&--1&0 \\
${\hat \Phi }$
  &&&1&3&1&3&0&1&0&1 \\
  &&&5&1&1&3&--1&0&0&1 \\
  &&&1&3&5&1&0&1&--1&0 \\ \hline \hline
2 &0\ 0\ 0\ 1\ 0\ 0&&1&3&1&1&5/2&--1/2&0&0 \\
  &&&$\bar 5$&3&1&1&--3/2&--1/2&0&0 \\
  &&&10&1&1&1&1/2&3/2&0&0 \\
${\hat \Psi }^I$
  &0\ 0\ 0\ 3\ 0\ 0&&1&1&1&1&5/2&3/2&0&0 \\
  &&&$\bar 5$&1&1&1&--3/2&3/2&0&0 \\
  &&&10&3&1&1&1/2&--1/2&0&0 \\ \hline
3 &0\ 0\ 0\ 3\ 2\ 0&&1&1&1&$\bar 3$&0&0&--5/2&1/2 \\
  &&&1&1&5&$\bar 3$&0&0&3/2&1/2 \\
  &&&1&1&$\bar {10}$&1&0&0&--1/2&--3/2 \\
${\hat \Psi }^{II}$
  &0\ 0\ 0\ 1\ 6\ 0&&1&1&1&1&0&0&--5/2&--3/2 \\
  &&&1&1&5&1&0&0&3/2&--3/2 \\
  &&&1&1&$\bar{10}$&$\bar{3}$&0&0&--1/2&1/2 \\ \hline
4 &000231\ (111071)&$1,\ \pm 1/2_1$&1&1&1&3&0&--3/2&0&--1/2 \\
  &000271\ (111031)&$1,\ \pm 1/2_1$&1&$\bar 3$&1&1&0&1/2&0&3/2 \\
${\hat \Psi }^H$
  &000031\ (111271)&$1,\ \pm 1/2_1$&1&$\bar 3$&1&3&0&1/2&0&--1/2 \\
  &000071\ (111231)&$1,\ \pm 1/2_1$&1&1&1&1&0&--3/2&0& 3/2 \\ \hline

5 &$\frac{\mbox{000070\ (111230)}}{\mbox{000010\ (111250)}}$
&$\pm_{SO(4)},\ \mp 1/2_1$&1&1&1&1&$\pm$5/4&$\pm$3/4
&$\mp$5/4&$\mp$3/4 \\
${\hat \sigma }$
&$\frac{\mbox{010310\ (101150)}}{\mbox{010170\ (101330)}}$
&$\pm_{SO(4)},\ \pm 1/2_1$&1&1&1&1&$\pm$5/4&$\pm$3/4
&$\pm$5/4&$\pm$3/4 \\
\hline
\end{tabular}
\normalsize
\end{table}

\noindent
\begin{figure}[p]
\caption{Supersymmetry breaking.}
\label{fig2}
\bigskip
\centering\begin{tabular}{ccccccc} \bf \large
 N=2  SUSY : & \it V&=&(1,$\frac{1}{2}$)&+&($\frac{1}{2}$,0)&$SU(8)$ \\
$\Downarrow$ & & & & & & $\Downarrow$ \\
\bf\large N=1  SUSY : &  $V_{N=2}$ &$\rightarrow$&$V_{N=1}$&+&$S_{N=1}$&
  $\qquad SU(5)\times SU(3)\times U(1)$ \\
\end{tabular}

\bigskip

\raggedleft\begin{tabular}{|p{35mm}||c|c|c|c||}
\hline
   & \large \bf \hspace*{9mm}J=1\hspace*{9mm}
   & \large \bf \hspace*{8.5mm}J=1/2\hspace*{8mm}
   & \large \bf \hspace*{4.5mm}J=1/2\hspace*{4mm}
   & \large \bf \hspace*{4mm}J=0\hspace*{4mm} \\
\hline
{\large \bf $E_{\it vac}=[-1/2;-1]$  NS~sector } & (63) & --- & --- & (63) \\
\hline
{\large \bf $E_{\it vac}=[0;-1]$ SUSY~sector } & --- & $ (63)\times 2 $ & $
(63)\times 2 $ & --- \\
\hline \hline
\multicolumn{1}{|c|}{ } &
\multicolumn{4}{|c||}{ \large\bf Gauge multiplets } \\
\hline \hline
\end{tabular}

\bigskip
\begin{flushleft}
\hspace{60mm} $ {\bf \Downarrow} \quad b_5 $ GSO projection
\end{flushleft}

\bigskip

\raggedleft\begin{tabular}{|p{35mm}||c|c||c|c||}
\hline
   & \large \bf J=1
   & \large \bf J=1/2
   & \large \bf J=1/2
   & \large \bf J=0 \\
\hline
{\large \bf $E_{\it vac}=[-1/2;-1]$  NS~sector } &
 \small (24,1)+(1,1)+(1,8) & --- & --- & \small (5,\=3)+(\=5,3) \\ \hline
{\large \bf $E_{\it vac}=[0;-1]$ SUSY~sector } &
 --- & \scriptsize ((24,1)+(1,1)+(1,8))$\times$ 2  &
       \scriptsize ((5,\=3)+(\=5,3))$\times$ 2  & --- \\
\hline \hline
\multicolumn{1}{|c|}{} &
\multicolumn{2}{|c|}{ \large\bf Gauge multiplets } &
\multicolumn{2}{|c||}{ \large\bf Higgs multiplets } \\
\hline \hline
\end{tabular}

\end{figure}

\newpage

\subsection*{Acknowledgments}

 One of us (G.V.) is pleased
to thank INFN for financial support and the staff of
the Physics Department in Padova,  especially, Professor C. Voci
 for warm hospitality during his stay in Padova.

It is a great pleasure
for him to thank   Professor  G. Costa,
Professor  L. Fellin,
 Professor  A. Masiero,
 Professor M. Tonin and   Professor C. Voci for useful
discussions and for the help and support. Many interesting discussions with
colleagues from Theory  Department in Padova INFN Sezione, especially, to
Professor  S. Sartory,D. Degrassi  during this
work are also acknowledged.

G.V. would like to express his gratitude
to  Professor P. Drigo and their colleagues of the Medical
School of the University of Padova.

The research described in this publication was made possible in part
by Grants No RMP000, RMP300
from the International Science Foundation and
by Grant No 95-02-06121a from the Russian Foundation for Fundamental
Researches.

\newpage

\section{Appendix A. The N=1 SUSY character of the $SU(3)_H$ gauge family
symmetry}

We will consider the supersymmetric version of the Standard Model
extended by the family (horizontal) gauge symmetry.  The supersymmetric
Lagrangian of strong, electroweak and horizontal interactions,
based on the $SU(3)_C\times SU(2)_L \times U(1)_Y \times SU(3)_H$...
(and the Abelian gauge factor also can be taken into consideration),
has the standard general form \cite{9'}.
The superfields  of the considered model are presented
in Table \ref{tabl0}.

In models with a global supersymmetry it is impossible
to have   simultaneously a SUSY breaking and a vanishing cosmological term.
The reason is the semipositive definition  of the scalar potential in the rigid
supersymmetry approach
(in particular, in the case of a broken $SUSY$ we have $V_{min}>0$ ).
The problem of supersymmetry breaking, with the  cosmological
term $\Lambda =0$ vanishing, is solved in the framework of the $N=1$ $SUGRA$
models. This may be done under an appropriate choice of the
Kaehler potential, in particular, in the frames of "mini-maxi"- or
"maxi" type models \cite{31'}.
In such approaches, the spontaneous breaking of the local $SUSY$
is due to the possibility to get nonvanishing $VEV$s for the scalar fields from
the "hidden" sector of $SUGRA$ \cite{31'}. The appearance
in the observable sector
of the so-called soft
breaking terms  comes as a consequence of this effect.

In the "flat" limit, i.e.
neglecting gravity, one is left with standard supersymmetric lagrangian
and soft SUSY
breaking terms, which on the scales $\mu << M_{Pl}$ have the form:
\begin{eqnarray}
 {\cal L}_{SB}&=&\frac{1}{2}\sum_{i}m^{2}_{i}|\phi_{i}|^{2}
 +\frac{1}{2}m_{1}^{2}Tr|h|^{2}  +\frac{1}{2}m_{2}^{2}Tr|H|^{2}+\nonumber\\
 &+&\frac{1}{2}\mu_{1}^{2}|\eta |^{2}+\frac{1}{2}\mu_{2}^{2}|\xi|^{2}
 +\frac{1}{2}M^{2}Tr|\Phi |^{2}+ \label{2.2}\\
&+&\frac{1}{2}\sum_{k}M_{k}\lambda^{a}_{k}\lambda^{a}_{k}+h.c.
+ \mbox {trilinear terms},\nonumber
 \end{eqnarray}
 where $H_1=H$, $H_2=h$ and  $i$ runs over all the scalar matter fields $\;
\tilde Q$,
 ${\tilde u}^c$, ${\tilde d}^c$,$\tilde L$, ${\tilde e}^c$,
 ${\tilde \nu}^c$ and $k$ - runs over all the gauge groups:
  $SU(3)_H$, $SU(3)_C,\; SU(2)_L,\; U(1)_Y$. At the energies close
 to the Planck scale all the masses, as well as the gauge coupling, are
 correspondingly equal (this is true if the analytic
 kinetic function satisfies  $f_{\alpha\beta}\sim\delta_{\alpha\beta}$ )
\cite{31'},
 but at low energies they have different  values depending on the
corresponding  renormgroup equation (RGE). The squares of some masses  may
be negative,
 which permits the spontaneous gauge symmetry breaking.

Considering the SUSY version of the $SU(3)_H$-model,
it is natural to ask:
why do  we need to supersymmetrize  the model? Basing on our
present-day knowledge
of the nature of supersymmetry \cite {31',32'}, the answer will be:

(a) First, it  is necessary to preserve the hierarchy of the scales:
$M_{EW} <M_{SUSY}< M_H < \cdots ? \cdots < M_{GUT} $
Breaking the horizontal gauge symmetry, one has  to preserve
 SUSY on that scale.
 Another sample of hierarchy to be considered is :
$M_{EW}<M_{SUSY}\sim M_H$.
In this case, the scale $M_H$ should be rather low ($M_H\leq$ a few TeV).

(b) To use the SUSY $U(1)_{R}$ degrees of freedom
for constructing the
superpotential and forbidding undesired Yukawa couplings.

(c) Super-Higgs mechanism - it is possible to describe Higgs bosons
by means of massive gauge superfields \cite{32'}.

Since the expected scale of the horizontal symmetry breaking is sufficiently
large:
$M_H>>M_{EW}$ , $M_H>>M_{SUSY}$ (where $M_{EW}$ is the scale of the electroweak
symmetry breaking,
and $M_{SUSY}$ is the value of the splitting into ordinary particles and their
superpartners),
it is reasonable to search for the SUSY-preserving stationary vacuum solutions.

Let us construct the gauge invariant superpotential $P$.
With the fields given in Table \ref{tabl0},
 the most general superpotential will
have the form
\begin{eqnarray}
P&=& \lambda_0 \biggl[\; {{1}\over {3}}Tr\bigl({\hat \Phi}^3 \bigr)+
{{1}\over {2}}
M_ITr\bigl({\hat \Phi}^2 \bigr)\;\biggr]+
\lambda_1 \biggl[\; \eta {\hat \Phi }\xi
 +M'\eta \xi\; \biggr]
+ \lambda_2 Tr\bigl(\hat h {\hat \Phi}\hat H \bigr)+\nonumber\\
&+& \mbox {~(Yukawa couplings)} +
\mbox {~( Majorana terms $\nu^c$ ),} \label{2.3}
\end{eqnarray}
where  Yukawa Couplings could be constructed, for example,
using the Higgs fields, H and h, transforming under $SU(3)_H \times SU(2_L)$,
like (8,2):
\begin{eqnarray}
P_Y=\lambda_3 Q{\hat H}d^c + \lambda_4 L{\hat H}e^c +
\lambda_5 Q{\hat h}u^c.
\end{eqnarray}
Also, one can consider another types of superpotential $P_Y$, using
the  Higgs fields from Table \ref{tabl0}.

\begin{table}[t]
\caption{The Higgs Superfields with their
 $SU(3)_H, SU(3)_C,\ SU(2)_L,\ U(1)_Y$ (and possible
$U(1)_H$- factor) Quantum Numbers}
\label{tabl0}
\begin{center}
\begin{tabular}{||c|ccccc|} \hline
&H&C&L&Y&$Y_H$ \\ \hline
$\Phi$& 8 & 1 & 1 & 0 & 0 \\
H & 8 & 1 & 2 & $-1/2$ & $-y_{H1}$ \\
h & 8 & 1 & 2 & 1/2 &$y_{H1}$ \\
$\xi$ & $\bar 3$ & 1 & 1 & 0 & 0 \\
$\eta$ & 3 & 1 & 1 & 0 & 0 \\
$Y$&$\bar 3$&  1&  2&  1/2 &$ -y_{H2}$ \\
$X$&$ 3$&  1&  2&$ -1/2$& $y_{H2}$   \\
$\kappa_1$&1&  1&  1 (2) &  0 (1/2)&$ -y_{H3}$ \\
$\kappa_2$&      1 & 1 & 1 (2) &  0 (-1/2) & $y_{H3}$    \\ \hline
\end{tabular}
\end{center}
\end{table}
Note, the fields $\Phi$, H, h - can be obtained
on the level 2 of Kac-Moody algebra g or effectively on the level 1
of algebras $g^I, g^{II}$
after "integration" over heavy fields, when $G^I \times G^{II} \rightarrow
G^{symm}$ (see section 5). Higgs fields $X$ and $Y$ are very important in
models with forth SU(3$)_H$-singlet generation.
In the construction of the stationary solutions,
only the following contributions of the scalar potential are taken into
account:
\begin{eqnarray}
V &=& \sum_i |F_i|^2+\sum_{a} |D^{a}|^2=V_F+V_D \geq 0 \label{2.5}\\
\mbox {where~~~~}V_F &=& \sum {\biggl|{{\partial P_F}\over {\partial
F_i}}\biggr|}^2=
{\biggl|{{\partial P_F}\over {\partial F_{\Phi^a}}}\biggr|}^2+
{\biggl|{{\partial P_F}\over {\partial F_{\xi_i}}}\biggr|}^2+
{\biggl|{{\partial P_F}\over {\partial F_{\eta_i}}}\biggr|}^2\label{2.6}
\end{eqnarray}

The case ${\langle }V{\rangle }=0$ of supersymmetric vacuum can be realized
within different gauge
scenarios \cite {9'}. By switching on the SUGRA, the vanishing scalar potential
is
no more required to conserve the supersymmetry with the necessity.
The different gauge breaking scenarios  do not result in obligatory
vacuum degeneracy, as in the case of the global SUSY version.
Let us write down each of the terms of formula (\ref{2.6}):
\begin{eqnarray}
P_F(\Phi ,\xi ,\eta )&=& \lambda_0{\biggl[\;{\frac{i}{4\times 3}}\;f^{abc}
\Phi^a \Phi^b \Phi^c+{\frac{1}{4\times 3}}\;d^{abc}\Phi^a \Phi^b
\Phi^c+{\frac{1}
{4}}\;M_I\Phi^c \Phi^c\; \biggr]}_F+\nonumber\\
&+& \lambda_1 \biggl[\; \eta_i \;{(T^c)_j^i}\;
\xi^j \Phi^c+M'\eta_i \xi^i\; \biggr]_F+ \label{2.7}\\
&+& \lambda_2 \biggl[\;{\frac{i}{4}}f^{abc}\; h^a_i \Phi^b H^c_j \epsilon^{ij}+
\frac{d^{abc}}{4} \; h^a_i \Phi^b H^c_j \epsilon^{ij}\; \biggr]_F+h.c.\nonumber
\end{eqnarray}
The contribution of $D$-terms into the scalar potential will be :
\begin{eqnarray}
V_D&=&g_H^2|\eta^+ T^a\eta -\xi^+ T^a\xi +i/2\ f^{abc} \Phi^b{\Phi^c}^++
i/2\ f^{abc} h^b{h^c}^++i/2\ f^{abc} H^b{H^c}^+|^2 \nonumber\\
&+&g_2^2|h^+ \tau^i /2\ h +\ H^+ \tau^i /2\ H|^2+
(g')^2 | 1/2\ h^+ h -\ 1/2\  H^+ H |^2 \label{2.8}
\end{eqnarray}
The SUSY-preserving condition for scalar potential (\ref{2.5}) is
determined by the flat $F_i-$
and $D^{a}$ directions: ${\langle }F_i{\rangle }_0={\langle }D^{a}{\rangle
}_0=0$.
It is possible to remove the degeneracy of the
supersymmetric vacuum solutions taking into account the interaction with
supergravity, which was endeavored in SUSY GUT's, e.g. in the $SU(5)$ one
\cite{31'}
$(SU(5)\rightarrow SU(5),$ $SU(4)\times U(1),$ $SU(3) \times SU(2) \times
U(1))$.

The horizontal symmetry spontaneous breaking to
the intermediate subgroups in the
first three cases of \cite{9'}  can be realized,
using the scalar components of the
chiral complex superfields $\Phi$,
which are singlet under the standard gauge group.
The $\Phi$-superfield transforms as the adjoint representation of $SU(3)_H$.
The
intermediate scale $M_I$ can be sufficiently large: $M_I > 10^5-10^6$GeV.
 The complete breaking of the remnant symmetry group $V_H$ on the scale
$M_H$ will  occur
due to the nonvanishing VEV's of the scalars from the chiral
superfields $\eta (3_H)$
and $\xi(\bar 3_H)$. The $V_{min}$, again, corresponds to the flat directions:
${\langle }F_{\eta,\xi}{\rangle }_0=0$. The version (iv) corresponds to the
minimum of the scalar
potential in the case when  ${\langle }\Phi{\rangle }_0=0$.

As for the electroweak breaking, it is due to the VEV's of the fields $h$ and
$H$,
providing masses for quarks and leptons. Note that VEV's of the fields $h$ and
$H$
must be of the order of $M_W$ as they determine the quark and lepton mass
matrices. On the
other hand, the masses of physical Higgs fields $h$ and $H$, which mix
 generations, must be some orders higher than $M_W$, so that
 not to contradict the
experimental restrictions on FCNC. As a careful search for the Higgs
potential shows, this is the picture that can be attained.

\newpage

\section{Appendix B. The Superstring theory scale  of unification and
the estimates on the horizontal coupling constant and the Weinberg angle.}

Really, the estimates on the $M_{H_0}$- scale depend on the
value of the family gauge coupling.
These estimates can be made in GUST using the string scale
\begin{eqnarray}
M_{SU} \approx 0,73\,g_{string} \times {10}^{18} GeV
\label{Mstring}
\end{eqnarray}
and the renormalization group equations (RGE) for the gauge couplings,
${\alpha}_{em}$, ${\alpha}_{3}$, ${\alpha}_{2}$, to the low energies :
\begin{eqnarray}
{\alpha}_{em}(M_Z) &\approx & 1/128\ ,\nonumber\\
{\alpha}_{3}(M_Z) &\approx & 0.11\ ,\nonumber\\
{\sin}^2{\theta}_{W}(M_Z) &\approx & 0.233\ . \label{experiment}
\end{eqnarray}
The string unification scale could be contrasted with the
$SU(3^c)\times SU(2) \times U(1)$ naive unification scale,
$M_{GUT} \approx {10}^{16} GeV$, obtained by running the SM particles and their
SUSY-partners to high energies. The simplest solution to this problem
is the introduction  of the new heavy particles with SM quantum
numbers, which can  exist in string spectra  \cite{20'}.

However there are some other ways to explain the difference between
scales of string ($M_{SU}$) and ordinary ($M_{GUT}$) unifications.
If one uses the breaking scheme $G^I\times G^{II}\,\rightarrow G^{sym}$
( where $G^{I,II}=U(5)\times U(3)_H \subset E_8$ ) described above, then
unification scale $M_{GUT} \sim 10^{16}\,$GeV is the scale of breaking
the $G\times G$ group, and string unification do supply the equality
of coupling constant $G\times G$ on the string scale
$M_{SU}\sim 10^{18}\,$GeV.
Otherwise, we can have an addition scale of the symmetry breaking
$M_{sym} > M_{GUT}$.
In any case on the scale of breaking $G\times G\,\rightarrow G^{sym}$
the gauge coupling constants satisfy the equation
\begin{equation}
g^2_{sym}=1/4\,(g_I^2 + g_{II}^2 )\ .
\label{symscale}
\end{equation}
Thus in this scheme knowing of scales $M_{SU}$ and $M_U$ gives
us a principal possibility to trace the evolution of coupling
constant of the original  group $G\times G$  to the low energy
and estimate the value of horizontal gauge constant $g_{3H}$.

The coincidence of $\sin^2 \theta_W$ with experiment will
show how realistic this model is.

Let us consider some relations which determine the value
of $\sin^2\theta_W$ for different unification groups and
for different ways of the breaking.

Firstly let us consider the case of
$SO(10)\times U(3)\longrightarrow SU(5)\times U(1)\times
[SU(3)\times U(1)]_H$ breaking, which can be illustrated
by Model~2. In this case matter fields are generated by
world-sheet fermions with periodic boundary conditions.
Consequently all representations of matter fields can be
considered as the result of destruction of $\underline 16$ and
$\underline{\bar{16}}$ representations of the $SO(10)$ group.

If we write the general expansion for a world-sheet
fermion in the form
\begin{equation}
f(\sigma ,\tau) =\sum_{n=0}^{\infty}\left[
b^+_{n+\frac{1-\alpha}{2}}\exp [-i(n+\frac{1-\alpha}{2})
(\sigma +\tau )]
+d_{n+\frac{1+\alpha}{2}}\exp [i(n+\frac{1+\alpha}{2})
(\sigma +\tau )]\right]
\end{equation}
where the quantization conditions are given by the
anti-commutation relations
$$\{ b^+_a\,,\,b_b \} =\{ d^+_a\,,\,d_b \} =\delta_{ab}$$
then the representation $\underline{\bar{16}}$ of $SO(10)$
in terms of the creation ($b^{i+}_0$) and annihilation
($b^i_0$) operators will have the form
\begin{equation}
\underline{\bar{16}}=\underline 1 +\underline{\bar{10}}
+\underline 5 =(1+ b^{i+}_0 b^{j+}_0
+ b^{i+}_0 b^{j+}_0 b^{k+}_0 b^{l+}_0)|0\rangle\ ,
\end{equation}
where $i,j,k,l=1,\cdots 5$.

The Clifford algebra is realized via the $\gamma$-matrix for
$SO(10)$ group
$\gamma_k=(b_k +b_k^+)$ and $\gamma_{5+k}=-i(b_k -b_k^+)$.
Generators of the $U(5)$ subgroup can be written in terms $b_0^i$
as $T[U(5)]=1/2\,[b_i,b_j^+]$. Then the operator of the
$U(1)_5$ hypercharge is
\begin{equation}
Y_5=1/2\,\sum_i [b_i,b_i^+]=5/2 -\sum_i b_i^+,b_i\ .\label{Y5}
\end{equation}
But this generator is not normalized, since
$Y_5(\underline 1 ,\,\underline{\bar{10}} ,\,\underline 5)
=5/2,\ 1/2,\ -3/2$ corresponding, and Tr$_{\underline{16}}Y_5^2=20$.

In our scheme the electromagnetic charge is
\begin{equation}
Q_{EM}=T_5-2/5\,Y_5\ ,
\end{equation}
where $T_5=\mbox{diag}(1/15\,,\ 1/15\,,\ 1/15\,,\ 2/5\,,\ -3/5\,)$.
For representation $\underline 5$ of the $SU(5)$ this means that
\begin{eqnarray}
Q_{EM}(\underline 5)&=&[\mbox{diag}(0^3\,,\ 1/2\,,\,-1/2)
+\mbox{diag}(2/30^3\,,\,-3/30^2)]-2/5\cdot (-3/2)=\nonumber \\
&=&\mbox{diag}(0^3\,,\ 1/2\,,\,-1/2)
+1/2\,[\mbox{diag}(2/15^3\,,\,-3/15^2)+6/5]= \label{Q} \\
&=&t_3+1/2\ [{\tilde t}_0-4/5\ Y_5]
=t_3+1/2\,\mbox{diag}(4/3^3\,,\ 1^2)
=t_3 +1/2\,y\ ,\nonumber
\end{eqnarray}
where $y$ is the electroweak hypercharge.

Now let us write down the principal equation for coupling constants
\begin{equation}
g_5t_0A_{\mu}+(kg_5)YA'_{\mu}=g_1yB_{\mu}+g'_1y'B'_{\mu}\ .
\label{coupl}
\end{equation}
In this equation $(kg_5)$ is $U(1)_5$ coupling constant on
the scale where $U(5)$ is breaking down (on the
$SO(10)\rightarrow U(5)$ scale $k=1$); operators
$t_0\sim {\tilde t}_0$ , $Y\sim Y_5$ and have equal norm;
$A$ and $B$ are gauge fields.

Diagonal generators can be written in terms of creation-annihilation
operators as
\begin{equation}
\mbox{diag}(A_i)=\sum_{i=1}^5 A_i(1-b^+_i b_i)=-\sum A_i b^+_i b_i\ .
\label{diag}
\end{equation}
Consequently Tr$_{\underline{16}}{\tilde{t_0}}^2 =8/15$.
If we shall normalize generators as
Tr$_{\underline{16}}t_0^2=\mbox{Tr}_{\underline{16}}Y^2=8$,
then $Y(\underline 5)=\sqrt{2/5}\,Y_5(\underline 5)=-3/\sqrt{10}$
and $t_0=1/\sqrt{15}\ \mbox{diag}(2^3\,,\,-3^2)$.
Now after rewriting the equation (\ref{coupl}) separately
for three up components and two down components, and substitution
$B_{\mu}=cA_{\mu}+sA'_{\mu}$ , $B'_{\mu}=-sA_{\mu}+cA'_{\mu}$
where $c^2+s^2=1$, we find from equation (\ref{coupl})
\begin{equation}
\sin^2 \theta_W =\frac{g_1^2}{g_1^2+g_5^2}
=\left.\frac{15k^2}{16k^2 +24}\right|_{k^2=1}=\frac{3}{8} \label{sin}
\end{equation}

\medskip

Now let us consider the breaking $E_6\rightarrow U(5)\times U(1)$,
which corresponds to models like Model~4.
The expansion of matter representation $\underline{27}$ of the
$E_6$ group under the group $SU(5)\times U(1)_5$ is
\begin{eqnarray}
&&\underline{27}=({\underline{\bar 5}}_{3/2}+{\underline{10}}_{-1/2}+
{\underline{1}}_{-5/2})+({\underline{5}}_{1}
+{\underline{\bar 5}}_{-1})+{\underline{1}}_{0}= \nonumber\\
&&=[(b_0^{i+}+b_0^{i+}b_0^{j+}b_0^{k+}+
b_0^{i+}b_0^{j+}b_0^{k+}b_0^{l+}b_0^{m+})
+(d_{1/2}^{i+}+b_{1/2}^{i+})+1]|0\rangle\ .
\end{eqnarray}
The generalization of the formula (\ref{Y5}) for the case when
representation contains states from different sectors with
different boundary conditions is
\begin{equation}
Y_5=\sum_i \left( \frac{\alpha_i}{2} +\sum_E^{\infty}
[d^+_E(f_i)d_E(f_i) -b^+_E(f_i^{\ast})b_E(f_i^{\ast})]\right)
\label{Y_5}
\end{equation}
and analogically for formula (\ref{diag})
$$\mbox{diag}(A_i)=\sum_i [A_i\cdot\sum_E^{\infty}(
d^+_E(f_i)d_E(f_i)-b^+_E(f_i^{\ast})b_E(f_i^{\ast}))]\ ,
\qquad(\sum A_i =0).$$

Now we have Tr$_{\underline{27}}Y_5^2=30$ and
Tr$_{\underline{27}}{\tilde t}_0^2=4/5$. By comparison with
preceding case  both norms are 1.5 times greater,
hence formula (\ref{sin}) is true for this case too.
But now the $B'_{\mu}$ is some linear combination of gauge
fields.

\medskip

Further, let us consider the Model~1. This case corresponds
to breaking $SO(16)\rightarrow U(8)\rightarrow U(5)\times U(3)$.
The matter fields arise from sectors with $\alpha =\pm 1/2$
and correspond to chips of the $SU(8)$ representations
\begin{displaymath}
\left. \begin{array}{cccccc}
8 & \rightarrow & [(1,\ 3)] & +(5,\ 1) &  &  \\
56 & \rightarrow & [(1,\ 1) & +(10,\ 3)]
& +(\bar{10},\ 1) & +(5,\ \bar 3)  \\
\bar{56} & \rightarrow & [(10,\ 1) & +(\bar 5 ,\ 3)]
& +(1,\ 1) & +(\bar{10},\ \bar 3)  \\
\bar 8 & \rightarrow & [(\bar 5 ,\ 1)] & +(1,\ \bar 3) &  &
\end{array} \right\} \sim 128_{SO(16)},
\end{displaymath}
where only the fields in square brackets survive after the GSO projection.

For these model it is necessary to  change
$Y_5\rightarrow {\tilde Y}_5=-1/4\,(Y_5+5Y_3)$ in formula (\ref{Q}).
Now we can calculate the norms of ${\tilde t}_0$ and ${\tilde Y}_5$
operators for this model.
$$\mbox{Tr}_{128}{\tilde Y}_5^2=160=20\times 8$$
$$\mbox{Tr}_{128}{\tilde t}_0^2=\frac{64}{15}=\frac{8}{15}\times 8$$
Hence we find again the formula (\ref{sin}),
but $A'_{\mu}$  is linear combination of gauge fields, which
corresponds to ${\tilde Y}_5$ hypercharge and $kg_5$ is his
coupling constant.

\medskip

The analysis of RG--equations allows to
state that horizontal coupling constant $g_{3H}$ does not exceed
electro-weak one $g_2$.

For example, if below the $M_{GUT}$ scale in non-horizontal sector
we have effectively the standard model with four generations and
two Higgs doublets (like Model 1, 2), then the evolution of gauge coupling
constants is described by equations
\begin{eqnarray}
&&\alpha^{-1}_S(\mu )=\alpha^{-1}_5(M_{GUT} ) +8\pi b_3\ln (\mu /M_{GUT}) \\
&&\alpha^{-1}(\mu )\sin^2\theta_W =\alpha^{-1}_5(M_{GUT} ) +8\pi b_2\ln (\mu
/M_{GUT}) \\
&&\frac{15}{k^2+24}\alpha^{-1}(\mu )\cos^2\theta_W =
(k^2\alpha_5(M_{GUT} ))^{-1} +8\pi b_1\ln (\mu /M_{GUT}) \ ,
\end{eqnarray}
where
$$b_3=\frac{1}{16\pi^2}\ ,\qquad b_2=-\frac{3}{16\pi^2}\ ,\qquad
b_1=-\frac{21}{40\pi^2}\ .$$
{}From these equations and from (\ref{experiment}) we can find
\begin{equation}
M_{GUT}=1.3\cdot 10^{16}\mbox{GeV}\ ,\qquad k^2=0.9\ ,
\qquad \alpha_5^{-1}=14\ .
\end{equation}

Now we can get the relation between $g_{str.}\equiv g$ and $M_{sym.}$
from RG equations for gauge running constants $g_5^{sym.}\equiv g_5$,
$g_5^I$ and $g_5^{II}$ on $M_{GUT}$ -- $M_{SU}$ scale.
For example, for Model~1
$$ b_5^{sym.}=-\frac{3}{4\pi^2}\ ,\qquad
b_5^{I}=-\frac{5}{16\pi^2}\ ,\qquad
b_5^{II}=\frac{3}{16\pi^2}\ ,$$
and from RGE we find the following relation
\begin{equation}
\frac{g_5^2}{4\pi^2 +6g_5^2\ln (M_{GUT}/M_{sym})} = g^2
\frac{8\pi^2-g^2\ln (M_{sym}/M_{SU})}{
[8\pi^2-5g^2\ln (M_{sym}/M_{SU})]\cdot [8\pi^2+3g^2\ln (M_{sym}/M_{SU})]}\ .
\end{equation}
According to this equation we obtain that if $M_{SU}\approx 10^{18}$ Gev
and the scale of breaking down to symmetric subgroup changes
in region $M_{sym}\div 1.5\cdot 10^{16}$ GeV --- $10^{18}$ GeV
then $g_{str.} \sim O(1)$.
Note that these values agree
with formula (\ref{Mstring}).

Using RG equations for the running constant $g_{3H}$
and the value of the string coupling constant $g_{str.}$
we can estimate a value of
the horizontal coupling constant at low energies.
For Model~1 we have
$$ b_{3H}^{sym.}=-\frac{5}{2\pi^2}\ ,\qquad
b_{3H}^{I}=-\frac{21}{16\pi^2}\ ,\qquad
b_5^{II}=-\frac{13}{16\pi^2}\ ,$$
and taking into account the relation (\ref{symscale}), we find
from RGE for $g_{3H}$ that
$$g_{3H}^2 \biggl( O(1\mbox{TeV})\biggr)\approx 0.05\ , $$
and this value depends very slightly on the scale $M_{sym.}$.
However, note that for all our  estimation the presence of
the breaking $G\times G$ group to
diagonal subgroup $G_{sym}$ played the crucial role.

The above calculations show that for evaluation of intensity
of a processes with a gauge horizontal bosons at low energies
we can use inequality
$${{\alpha}_{3H}(\mu)}\,\leq \,{{\alpha}_2(\mu)}\ .$$

\newpage
\section{Appendix C. Rules for constructing consistent string
 models out of free world-sheet fermions.}

 The partition function of the theory is a sum over terms
 corresponding to world-sheets of different genus $g$.
 For consistens of the theory we must require that partition
 function to be invariant under modular transformation, which
 is reparametrizations not continuously connected to the identity.
 For this we must sum over the different possible boundary conditions
 for the world-sheet fermions with appropriate weights \cite{SW}.

 If the fermions are parallel transported around a nontrivial loop
 of a genus-$g$ world-sheet $M_g$, they must transform into themselves:
 \begin{equation}
 \chi^I\longrightarrow L_g(\alpha )^I_J\chi^J \label{tr}
 \end{equation}
 and similar for the right-moving fermions.
 The only constraints on $L_g(\alpha )$ and $R_g(\alpha )$ are that
 it be orthogonal matrix representation of $\pi_1 (M_g)$ to leave
 the energy-momentum current invariant and supercharge (\ref{sch})
 invariant up to a sign. It means that
 \begin{eqnarray}
 && \psi^{\mu}\longrightarrow -\delta_{\alpha}\psi^{\mu}\ ,\qquad
 \delta_{\alpha}=\pm 1\ ,\label{bc1} \\
 && L^I_{gI'}L^J_{gJ'}L^K_{gK'} f_{IJK}=-\delta_{\alpha}f_{I'J'K'}
 \label{bc2}
 \end{eqnarray}
 and consequently $-\delta_{\alpha}L_g(\alpha )$ is an automorphism
 of the Lie algebra of $G$.

 Farther, the following restrictions on $L_g(\alpha )$ and
 $R_g(\alpha )$ are imposed:

 $(a)$ $L_g(\alpha )$ and $R_g(\alpha )$ are abelian matrix
 representations of $\pi_1(M_g)$. Thus all of the $L_g(\alpha )$
 and all of the $R_g(\alpha )$ can be simultaneously diagonalized in
 some basis.

 $(b)$ There is commutativity between the boundary conditions on
 surfaces of different genus.

 When all of the $L(\alpha )$ and $R(\alpha )$ have been
 simultaneously diagonalized the transformations like (\ref{tr})
 can be writen as
 \begin{equation}
 f\longrightarrow -\exp (i\pi\alpha_f)f\ .
 \end{equation}
 Here and in eqs. (\ref{bc1}), (\ref{bc2}) the minus signs
 are conventional.

 The boundary conditions (\ref{tr}), (\ref{bc1}) are specified
 in this basis by a vector of phases
 \begin{equation}
 \alpha=[\alpha(f_1^L),\cdots,\alpha(f_k^L)\ |\
 \alpha(f_1^R),\cdots,\alpha(f_l^R)]\ .
 \end{equation}
 For complex fermions and $d=4$, $k=10$ and $l=22$.
 The phases in this formula are redused mod(2) and are chosen
 to be in the interval $(-1,+1]$.

 Modular transformations mix spin-structures amongst one another
 within a surface of a given genus. Thus, requiring the modular
 invariance of the partition function imposes constraints
 on the coefficients ${\cal C}
 \left[
 \begin{array}{c}
 {\alpha}_1\ \cdots\ \alpha_g \\
 {\beta}_1\ \cdots\ \beta_g
 \end{array}\right]$
 (weights in the partition function sum, for example see eq.(\ref{Z})
 which in turn impose constraints on what spin-structures are allowed
 in a consistent theory. In according to the assumptions $(a)$ and $(b)$
 these coefficients factorize:
 \begin{equation}
 {\cal C}
 \left[
 \begin{array}{c}
 {\alpha}_1\ \cdots\ \alpha_g \\
 {\beta}_1\ \cdots\ \beta_g
 \end{array}\right]={\cal C}
 \left[
 \begin{array}{c}
 {\alpha}_1 \\
 {\beta}_1
 \end{array}\right]{\cal C}
 \left[
 \begin{array}{c}
 {\alpha}_2 \\
 {\beta}_2
 \end{array}\right] \cdots {\cal C}
 \left[
 \begin{array}{c}
 {\alpha}_g \\
 {\beta}_g
 \end{array}\right]
 \end{equation}
 The requirement of modular invariance of the partition function
 thus gives rise to constraints on the one-loop coefficiennts
 ${\cal C}$ and hence on the possible spin structures
 $(\alpha , \beta)$ on the torus.

 For rational phases $\alpha(f)$ (we consider only this case)
 the possible boundary conditions $\alpha$ comprise a finite additive
 group $\Xi=\sum_{i=1}^k\oplus Z_{N_i}$ which is generated by
 a basis $(b_1,\cdots ,b_k)$, where $N_i$ is the smallest integer
 for which $N_i b_i =0\,\mbox{mod}(2)$.
 A multiplication of two vectors from $\Xi$ is defined by
 \begin{equation}
 \alpha\cdot\beta=(\alpha^i_L\beta^i_L-\alpha^j_R\beta^j_R)_{complex}
 +1/2\,(\alpha^k_L\beta^k_L-\alpha^l_R\beta^l_R)_{real}\ .
 \end{equation}
 \medskip
 The basis satisfies the following conditions derived in \cite{19'}:\\
 \medskip
 (A1) The basis $(b_1,\cdots ,b_k)$ is chosen to be canonical:
 $$\sum m_i b_i=0 \Longleftrightarrow m_i=0\,\mbox{mod}(N_i)
 \quad\forall i\ .$$
 Then an arbitrary vector $\alpha$ from $\Xi$ is a linear
 combination $\alpha=\sum a_i b_i$.\\
 \medskip
 (A2) The vector $b_1$ satisfies $1/2\,N_1 b_1={\bf 1}$.
 This is clearly satisfied by $b_1={\bf 1}$.\\
 \medskip
 (A3) $N_{ij}b_i\cdot b_j=0\,\mbox{mod}(4)$ where $N_{ij}$ is the
 least common multiple of $N_i$ and $N_j$.\\
 \medskip
 (A4) $N_i b_i^2=0\,\mbox{mod}(4)$; however, if $N_i$ is even we
 must have $N_i b_i^2=0\,\mbox{mod}(8)$.\\
 \medskip
 (A5) The number of real fermions that are simultaneously periodic
 under any three boundary conditions $b_i$, $b_j$, $b_k$
 is even, where $i$, $j$, $k$ are not necessarily distinct.
 This implies that the number of periodic real fermions in any
 $b_i$ be even.\\
 \medskip
 (A6) The boundary condition matrix corresponding to each $b_i$
 is an automorphism of the Lie algebra that defines the supercharge
 (\ref{sch}). All such automorphisms must commute with one another,
 since they must simultaneously diagonalizable.
 \medskip

 For each group of boundary conditions $\Xi$ there are a number
 of consistent choices for coefficients ${\cal C}[\cdots ]$,
 which determine from requirement of invariant under modular
 transformation. The number of such theories corresponds to the
 number of different choices of ${\cal C}
 \left[
 \begin{array}{c}
 b_i \\ b_j
 \end{array}\right]$.
 This set must satisfy equations:\\
 \medskip
 (B1)$\qquad\qquad\qquad {\cal C}
 \left[
 \begin{array}{c}
 b_i \\ b_j
 \end{array}\right] =\delta_{b_i} e^{2\pi in/N_j}
 =\delta_{b_j} e^{i\pi(b_i\cdot b_j)/2} e^{2\pi im/N_i}\ .$\\
 \medskip
 (B2)$\qquad\qquad\qquad {\cal C}
 \left[
 \begin{array}{c}
 b_1 \\ b_1
 \end{array}\right] =\pm e^{i\pi b_1^2/4}\ .$\\
 \medskip
 The values of ${\cal C}
 \left[
 \begin{array}{c}
 \alpha \\ \beta
 \end{array}\right]$ for arbitrary $\alpha,\,\beta\in\Xi$ can be
 obtained by means of the following rules:\\
 \medskip
 (B3)$\qquad\qquad\qquad {\cal C}
 \left[
 \begin{array}{c}
 \alpha \\ \alpha
 \end{array}\right] =e^{i\pi (\alpha\cdot\alpha +1\cdot 1)/4}{\cal C}
 {\left[
 \begin{array}{c}
 \alpha \\ b_1
 \end{array}\right]}^{N_1/2}\ .$\\
 \medskip
 (B4)$\qquad\qquad\qquad {\cal C}
 \left[
 \begin{array}{c}
 \alpha \\ \beta
 \end{array}\right] =e^{i\pi (\alpha\cdot\beta )/2} {\cal C}
 {\left[
 \begin{array}{c}
 \beta \\ \alpha
 \end{array}\right]}^{\ast}\ .$\\
 \medskip
 (B5)$\qquad\qquad\qquad {\cal C}
 \left[
 \begin{array}{c}
 \alpha \\ \beta +\gamma
 \end{array}\right] =\delta_{\alpha}{\cal C}
 \left[
 \begin{array}{c}
 \alpha \\ \beta
 \end{array}\right] {\cal C}
 \left[
 \begin{array}{c}
 \alpha \\ \gamma
 \end{array}\right]\ .$
 \medskip

 The relative normalization of all the ${\cal C}[\cdots ]$ is fixed
 in these expressions conventionally to be ${\cal C}
 \left[
 \begin{array}{c}
 0 \\ 0
 \end{array}\right] \equiv 1\ .$
 \medskip

 For each $\alpha\in\Xi$ there is a corresponding Hilbert space of
 string states ${\cal H}_{\alpha}$ that potentially contribute to the
 one-loop partition function. If we wrrite
 $\alpha =(\alpha_L\,|\,\alpha_R)$, then the states in
 ${\cal H}_{\alpha}$ are those that satisfy the Virasoro condition:
 \begin{equation}
 M_L^2=-c_L +1/8\,\alpha_L\cdot\alpha_L+\sum_{L-mov.}(frequencies)
 =-c_R +1/8\,\alpha_R\cdot\alpha_R +\sum_{R-mov.}(freq.)=M_R^2\,.
 \end{equation}
 Here $c_L=1/2$ and $c_R=1$ in the heterotic case.
 In ${\cal H}_{\alpha}$ sector the fermion $f$ $(f^{\ast})$ has
 oscillator frequencies
 \begin{equation}
 \frac{1\pm\alpha (f)}{2} +\mbox{integer}\ .
 \end{equation}

 The only states $|s{\rangle }$ in ${\cal H}_{\alpha}$ that contribute to the
 partition function are those that satisfy the generalized GSO conditions
 \begin{equation}
 \left\{ e^{i\pi (b_i\cdot F_{\alpha})}-\delta_{\alpha} {\cal C}
 {\left[
 \begin{array}{c}
 \alpha \\ b_i
 \end{array}\right]}^{\ast}\right\} |s{\rangle }=0
 \end{equation}
 for all $b_i$. Where $F_{\alpha}(f)$ is a fermion number operator.
 If $\alpha$ contains periodic fermions then $|0{\rangle }_{\alpha}$ is
 degenerate, transforming as a representation of an $SO(2n)$
 Clifford algebra.

\end{document}